\documentclass[oldversion]{aa} 
\usepackage{graphicx}
\usepackage{aalongtable}
\usepackage{txfonts}
\usepackage{rotating}
\usepackage{lscape}
\newcommand{\gsim}{\;\lower.6ex\hbox{$\sim$}\kern-7.75pt\raise.65ex\hbox{$>$}\;}
\newcommand{\lsim}{\;\lower.6ex\hbox{$\sim$}\kern-7.75pt\raise.65ex\hbox{$<$}\;}

\begin{document}
\title{Inconsistent metallicity spreads in first generation stars of globular
clusters from high resolution spectroscopy and HST photometry
 }

\author{
Eugenio Carretta\inst{1}
\and
Angela Bragaglia\inst{1}
}

\authorrunning{Carretta and Bragaglia}
\titlerunning{Discrepancies in metallicity spreads}

\offprints{E. Carretta, eugenio.carretta@inaf.it}

\institute{
INAF-Osservatorio di Astrofisica e Scienza dello Spazio di Bologna, Via P. Gobetti
 93/3, I-40129 Bologna, Italy}

\date{}

\abstract{An open issue about multiple stellar populations in globular clusters
(GCs) is the possible existence of metallicity spreads in first generation (FG)
stars. Recent estimates based on HST pseudo-colours map (PCM) derived unlikely
large spreads in [Fe/H] from spreads in the colour $col=m_{F275W}-m_{F814W}$.
The inferred metallicity spreads for many GCs are comparable or even larger than
those observed in dwarf galaxies. This result is clearly unexpected and at
odds with the birth time of stars in dwarf galaxies, spanning several
billion years, as opposed to very short formation times of the stellar component
in GCs (a few million years). The contradiction is corroborated by the
comparison of the widths of red giant branches in both classes of objects.
Moreover, the so called spreads in FG stars  estimated from the PCMs are always
larger than the intrinsic metallicity spreads derived from spectroscopy. We
used 30 pairs of FG stars with similar parameters in 12 GCs to highlight that a
constant displacement in $\Delta col$ corresponds to variable differences in
[Fe/H] up to 0.2 dex, depending on the GC. Providing for the first time
quantitative measurements of the extension in $\Delta col$ of the sequences of
FG and SG stars, we found no relation between metallicity spreads previously
derived and extension of FG stars. We found that the length of the FG region
correlates with the average global metallicity of GCs, and not with the observed
metallicity spreads. The extension of FG stars also correlates with the
extension of SG stars, and the global mass of the GCs. Our findings seriously
challenge the scenario claiming more inhomogeneous mixing among FG stars,
invalidating previous speculations in the literature.
}
\keywords{Stars: abundances -- Stars: atmospheres --
Stars: Population II -- Galaxy: globular clusters: general }

\maketitle

\section{Introduction}

A still unsolved issue on multiple populations (MPs) in globular clusters (GCs)
concerns the possible existence of metallicity spreads among first generation
(FG) stars. These stars formed in the primary star formation phase at the very
beginning of the lifetime of these old stellar systems, shortly after the Big
Bang (e.g., Carretta et al. 2000, VandenBerg et al. 2013). As hinted by their
narrow photometric sequences in the colour-magnitude diagrams (CMDs), GCs do not
have an extended star formation. In particular, FG stars have an abundance
pattern analogous to field halo stars of similar metallicity, bearing traces
only of supernovae (SNe) nucleosynthesis, that is enhanced levels of
$\alpha-$elements typical of contributions from massive stars exploding as
core-collapse SNe. Type Ia SNe, with their longer evolving times, do not provide
a noticeable contribution to the chemistry of most GCs, as proven by the
homogeneity of Fe-peak species in the majority of GCs (better than 12\% or less
than 0.05 dex,  Carretta et al. 2009a; see also, e.g., Bailin and von Klar 2022,
Willman and Strader 2012). The most massive objects among FG stars provide the
nuclearly processed matter, enriched in proton-capture elements from H-burning
at high temperature, that together with varying amounts of pristine gas produce 
the chemical composition typical of second generation (SG) stars (e.g., Langer et
al. 2003, Gratton et al. 2001), although the  exact nature of the polluters is
still debated (e.g., the reviews by Gratton et al. 2004, 2012, 2019, and Bastian
and Lardo 2018).

First generation stars can be easily separated from SG stars directly, deriving
their abundances with spectroscopy. The chief disadvantage is represented by the
time consuming observations, albeit mitigated by the use of efficient
spectrographs with high multiplexing gain at large telescopes. The tagging of
MPs can also be indirectly performed by photometric means, exploiting the effect
of different abundances involved in the H-burning on filters at the bandpasses
appropriate to sample spectral regions where main molecular bands  associated to
CNO products are located. With the benefit of sampling many more stars in a
short time, photometric methods have the main handicap to follow primarily the N
abundance, obtained using filters encompassing the ultraviolet or blue regions
where NH or CN features are located (e.g., Calamida et al. 2007, Sbordone et al.
2011, Carretta et al. 2011a, Lardo et al. 2011, Lee et al. 2009,  Milone et al.
2017, Johnson et al. 2023). Hence, the photometric approach essentially measures
the N content in different populations and is prone to some ambiguity, since N
can be generated by the conversion of C for temperatures above 10 MK, but also
from O when the ON branch of the CNO cycle is activated at higher temperatures. 

Discrepancies between the spectroscopic and the photometric approaches in
tagging MPs are examined in Carretta and Bragaglia (2024; hereinafter CB24),
using pseudo-colour maps (PCMs) and high resolution spectroscopy for 22 GCs. We
showed that PCMs are able to provide a coarse division of MPs, with FG and SG
stars usually separated into two groups, with an efficient trade off between
limited observing time and number of surveyed stars. However, we
found that photometric methods tend to overestimate the fraction of FG stars, 
with the number of mismatches depending on the GC mass (CB24). This happens
both with ground-based and space-based photometry.

The results by CB24 suggest that the photometric approach is sound, in broad
terms, but maybe still not completely understood. Many of the conclusions based
on PCMs seem still based on empirical evidence and some inferences may be
misleading.

In the present paper we use the sample of PCMs by CB24, the only published so
far, to explore in details the claims of the possible existence of a metallicity
spread among FG stars. Our purposes are to ascertain whether a spread in [Fe/H]
is truly connected to the extension of the region of FG stars in the PCM,
whether the spread in the region of SG stars also indicates a metallicity
dispersion in this population, and finally to disclose potential inconsistencies
between spectroscopy and HST photometry in this issue regarding MPs. 

For clarity, in Table~\ref{t:notation}, adapted from CB24, we reproduce the
simplified notation to describe the PCMs, whose detailed description can be
found in Milone et al. (2017: M17). Here we added the definition of $RG1$ as the
region in the PCM populated by FG stars, and analogously $RG2$ as the locus in
the PCM with the preponderant presence of SG stars.

The problem of possible metallicity spreads among GC stars seems to be
primarily based on indirect inferences. None of the candidate polluters proposed
as sources of matter enriched in proton-capture elements produces also
variations in metallicity, so the [Fe/H] content of MPs is expected to be very
homogeneous, as supported by the cosmic scatter of iron in GCs (e.g., Carretta et
al. 2009a). However, from the PCMs of 55 GCs, still unpublished, M17
state that the extension of the $RG1$ (see
Tab.~\ref{t:notation}) is too large if compared to the photometric errors
obtained by bootstrap technique. Milone et al. (2017) claim that this
comparison implies that FG stars may have a dishomogeneous composition. Since the
x-coordinate  ($\Delta col$) in the PCMs is mainly sensitive to variations in
effective temperature of stars (through the differential use of a long baseline
colour), changes in He content or in metallicity were proposed as responsible
for the observed spread. Since FG stars should not have an enhanced
He level due to H-burning in massive polluters, Milone et al. argue that the
extension of $RG1$ along $\Delta col$ is entirely due to metallicity variations.

Legnardi et al. (2022: L22) extend the work of M17, who only considered red
giant branch (RGB) stars, to the main sequence (MS) of two GCs (NGC~6362,
NGC~6838),  also observed with HST in the same filters. They find essentially
the same results for RGB and MS stars, that is a separation between $RG1$ and
$RG2$ (although clearer on the RGB due to the much smaller photometric errors)
and an elongation of the $RG1$ distribution in $\Delta col$ larger than that of
$RG2$ and of estimated errors. They further check the possibility that such a
spread could be due to the presence of binaries or a spread in helium. As for
M17, the latter occurrence is difficult to reconcile with the fact that these
stars represent the primordial population of the GC, without any chemical
variations due to polluters. Based on simulations, they discard the two
explanations and attribute the spread in $\Delta col$ solely to a spread in
metallicity. They then proceed to translate the width of FG stars on the RGB
measured by M17, i.e., $W1g = W_{F275W,F814W}^{1G}$ to a spread in metallicity,
using the colour-metallicity relations of Dotter et al. (2008). This procedure
results in spreads from about 0.05 to 0.3 dex (their Table~3).

In the present paper we first highlight the inconsistency of metallicity spreads
in FG stars as  derived in L22 from the PCMs with the observed
dispersions in [Fe/H] from high resolution spectroscopy (Section 2).  We then
use a large set of stars identified with spectroscopic criteria as FG stars in
20 GCs (CB24) to analyse the discrepancies between spectroscopy and HST
photometry and to investigate the dependence of the coordinate $\Delta col$ on
[Fe/H], providing empirical constraints to the true rate of changes in both
metallicity and  $\Delta col$ (Section 3). In Section 4 we quantify  for the
first time the real extension of the sequence of FG stars in the PCM. These
precise measurements are compared with global parameters of GCs to study
dependences and possible correlations. Section 5 is devoted to scrutinising the
existence of possibly neglected metallicity spreads among SG stars. Finally,
in Section 6 we present our conclusions and further implications of our
findings. 

\begin{table}
\centering
\caption[]{Notation adopted in the present paper}
\begin{tabular}{l}
\hline

$col = mag_{F275W}-mag_{F814W}$ \\

$col3 = (mag_{F275W}-mag_{F336W}) - (mag_{F336W}-mag_{F438W})$ \\

 $\Delta col = W_{col} ( (col -xr)/(xr-xb) )$  \\
 
 $\Delta col3 = W_{col3} ( (col3 -yr)/(yr-yb) ) $  
(see Section 2) \\

 $W_{col}$ and $W_{col3}$ = widths of the RGB in $col$ and $col3$\\

PCM = pseudo-colour map = $\Delta col$ in abscissa and $\Delta col3$ in ordinate \\

$RG1$ = region in the PCM populated by FG stars only\\

$RG2$ = region in the PCM populated by SG (or non-FG) stars\\

len$RG1$ = extension of the $RG1$ region in $\Delta col$\\

iqrSG = extension of the $RG2$ region in $\Delta col$\\

\hline
\end{tabular}
\label{t:notation}
\end{table}

\section{Metallicity spreads of GC FG stars in context}

From the extension in $\Delta col$ of the $RG1$ in the unpublished PCMs of GCs, 
L22 advocate that metallicity spreads must be present among the stars born in the
primary star formation within GCs. The magnitude of the spread derived 
for many GCs is significantly overestimated using their approach. The 
values of the the metallicity spread attributed by L22 to FG stars are generally
too large for stellar systems long known to be homogeneous in heavy elements, as
supported by their narrow photometric sequences in the CMDs.

\begin{figure}
\centering
\includegraphics[scale=0.40]{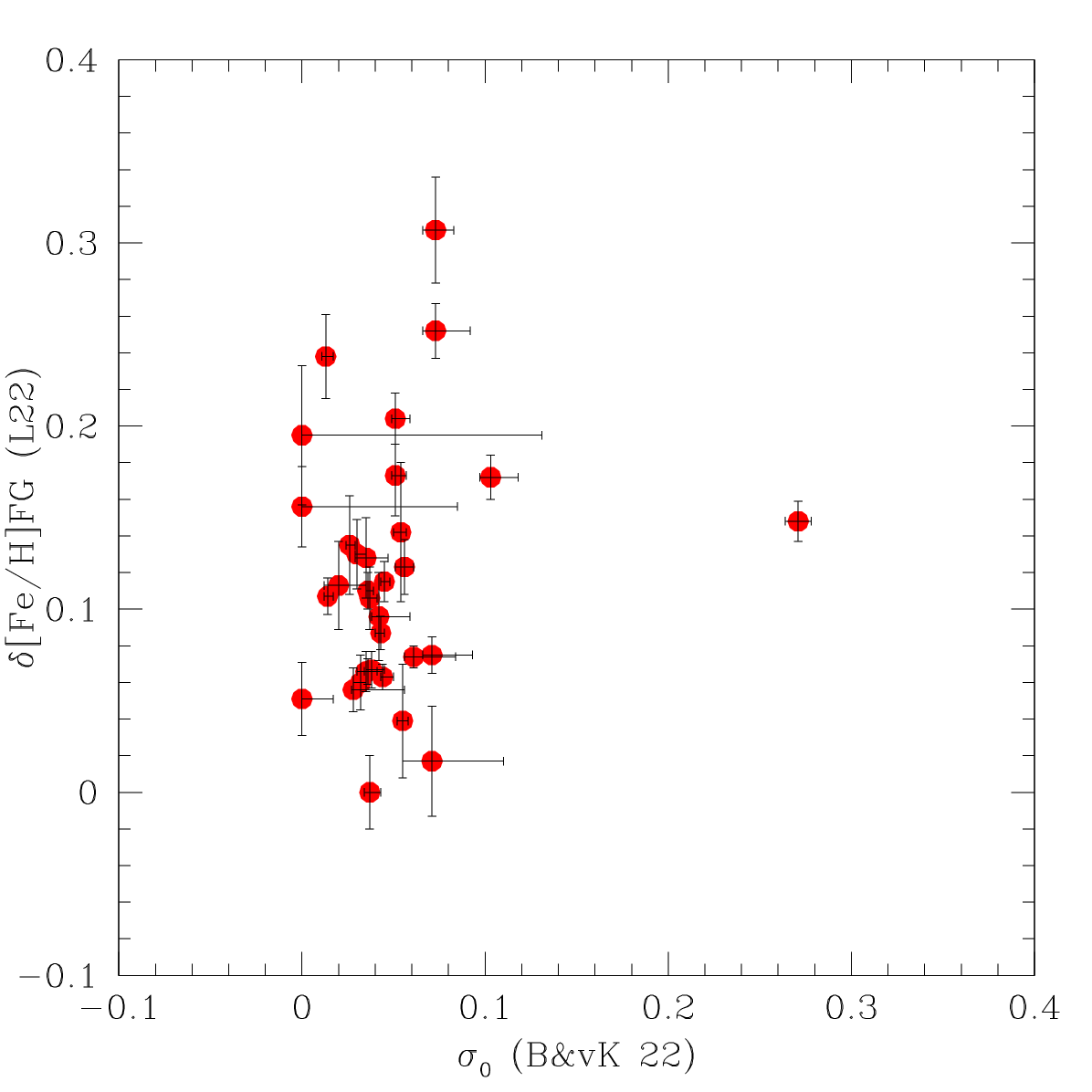}
\caption{Metallicity spreads for FG stars derived from PCMs by L22  as a
function of the intrinsic dispersion in [Fe/H] calculated in Bailin and von Klar
(2022) for 33 GCs in common.}
\label{f:solol22b22}
\end{figure}

To better visualise this inconsistency, in Fig.~\ref{f:solol22b22} we compare the
estimates by L22, $\delta$[Fe/H](FG), to the intrinsic
metallicity dispersions obtained by Bailin and von Klar (2022). They update the
estimates of Bailin (2019) of the internal dispersion $\sigma_0$ from high
resolution spectroscopy of RGB stars, taking into account both random and
systematic errors associated to the abundance analysis. The values by Bailin and
von Klar are referred to the total internal spread in iron abundance, without
distinction between FG and SG stars.

With the notable exception of $\omega$ Cen, the intrinsic iron spreads in GCs
are usually small, mostly below 0.1 dex. In contrast, FG spreads derived in L22
show a long tail extending to dispersions in [Fe/H] as large as 0.2-0.3 dex, the
record being NGC~5024, with a spread of 0.307 dex.

\begin{figure}
\centering
\includegraphics[scale=0.40]{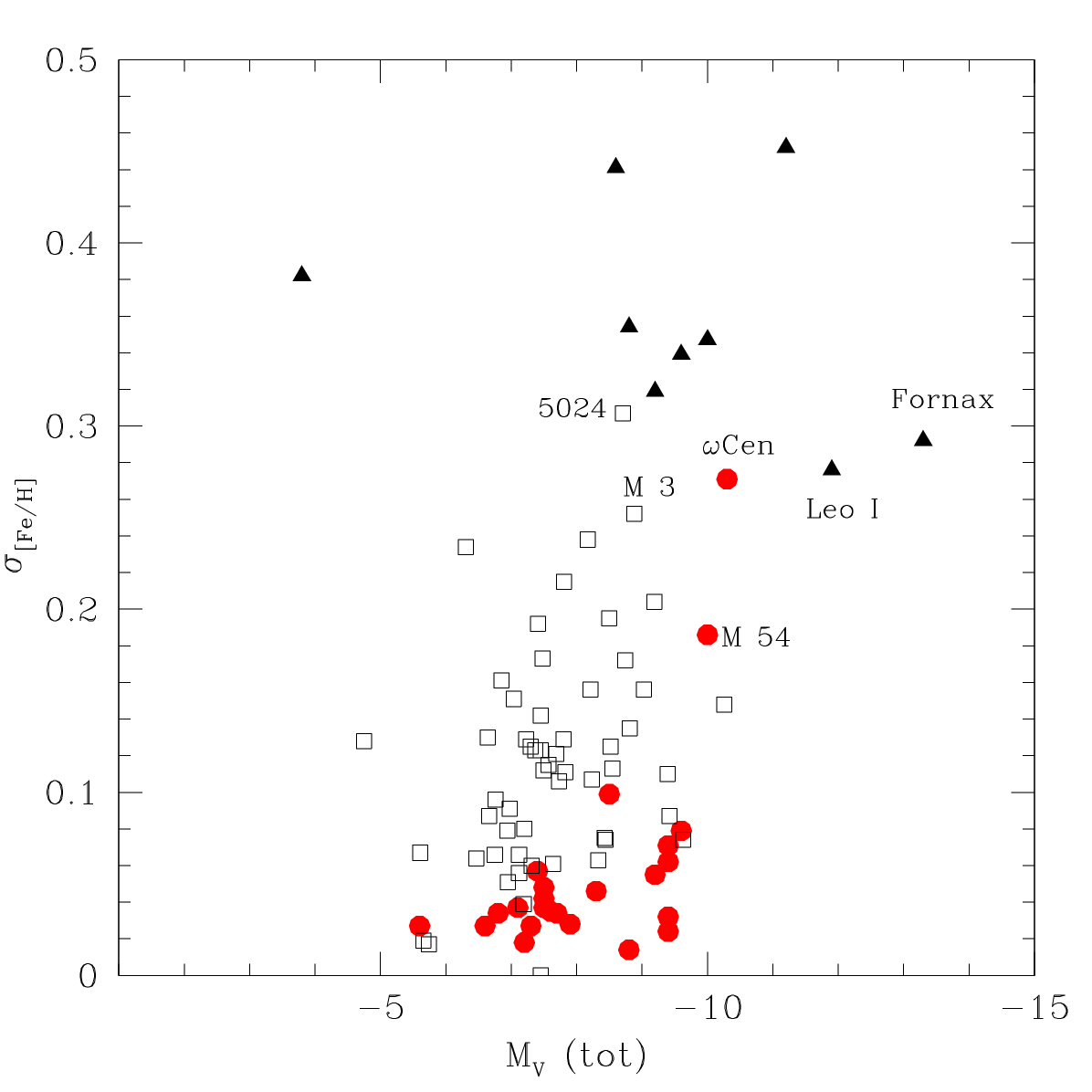}
\caption{Spread in [Fe/H] for FG stars in 55 GCs (black open squares) as a
function of the GC present-day mass as expressed by the total absolute magnitude
$M_V$ (Harris 2010). Metallicity spreads are derived from PCMs by L22. Filled red
circles and black triangles are spectroscopic [Fe/H] spreads
$\sigma_{\rm [Fe/H]}$  of GCs and dwarf galaxies, respectively, from Willman and
Strader (2012). A few interesting objects are labelled in the figure.}
\label{f:dawillman012}
\end{figure}

Only 33 GCs are in common between the studies of L22 and Bailin and von Klar 
(2022). Furthermore, for some GCs the number of stars with abundance analysis
from high resolution spectroscopy is limited. A more complete and instructive view for
the soundness of the estimates by L22 is provided in Fig.~\ref{f:dawillman012},
where the spreads in [Fe/H] for FG stars in all the 55 GCs analysed in L22 (open
black squares) are plotted as  a function of the total absolute magnitude of the
GCs, a proxy for the present-day total stellar mass, from Harris (2010). In the
same figure a  variety of other stellar systems from the compilation by Willman
and Strader (2012) are also reported. To explore a meaningful definition for
``galaxy", they used a Bayesian Markov Chain Monte Carlo technique to derive
consistent intrinsic  metallicity spreads for 24 GCs and 16 dwarf galaxies and
satellites of the Milky Way. The GCs considered are mostly from our homogeneous
FLAMES survey of large sample of stars in GCs (Carretta et al. 2009b), plus
other literature data for M~22 and $\omega$ Cen. Complete references for the
sample of dwarfs can be found in Willman and Strader (2012).

From Fig.~\ref{f:dawillman012}, several conclusions can be drawn.
First, also in this figure the vast majority of [Fe/H] spreads advocated for a
limited fraction of stars (the FG stars) in each GC are much larger than the
estimates obtained from spectroscopic observations of the GCs as  whole systems.
Second, it is not clear whether a trend as a function of mass is present for the
values by L22, whereas an increase of the global spread for GCs of increasingly
larger mass is rather evident. Third, and most astonishing, a few estimates by
L22 are comparable to or even exceed the metallicity dispersion observed in much
more massive dwarf galaxies.

In particular, this is true for the values derived by L22 for NGC~5024 (M~53)
and NGC~5272 (M~3), with $\delta$[Fe/H](FG)=0.307 and 0.252 dex, respectively.
These are exceptionally large amounts, since both M~53 (Boberg et al. 2016) and M~3
(Cohen and Melendez 2005, Sneden et al. 2004) are GCs known to be extremely
homogeneous in the abundance of heavy elements. According to L22, NGC~5024 would
host among its FG stars a metallicity spread larger that the one detected in
$\omega$ Cen\footnote{Notice that L22 provide for $\omega$ Cen a mere spread of
0.148 dex. This GC is then dropped from the sample in Legnardi et al. (2024),
together with a few others GCs, due to its complexity}. In addition, several
GCs in  Fig.~\ref{f:dawillman012}, including M~53 and M~3,  have a spread of
iron in FG larger than the metallicity dispersion in M~54 (NGC~6715), the
massive GC in the core of the tidally  disrupting dwarf spheroidal galaxy
Sagittarius of which M~54 probably represents the nuclear star cluster (Siegel
et al. 2007, Carretta et al. 2010).

\begin{figure*}
\centering
\includegraphics[scale=0.30]{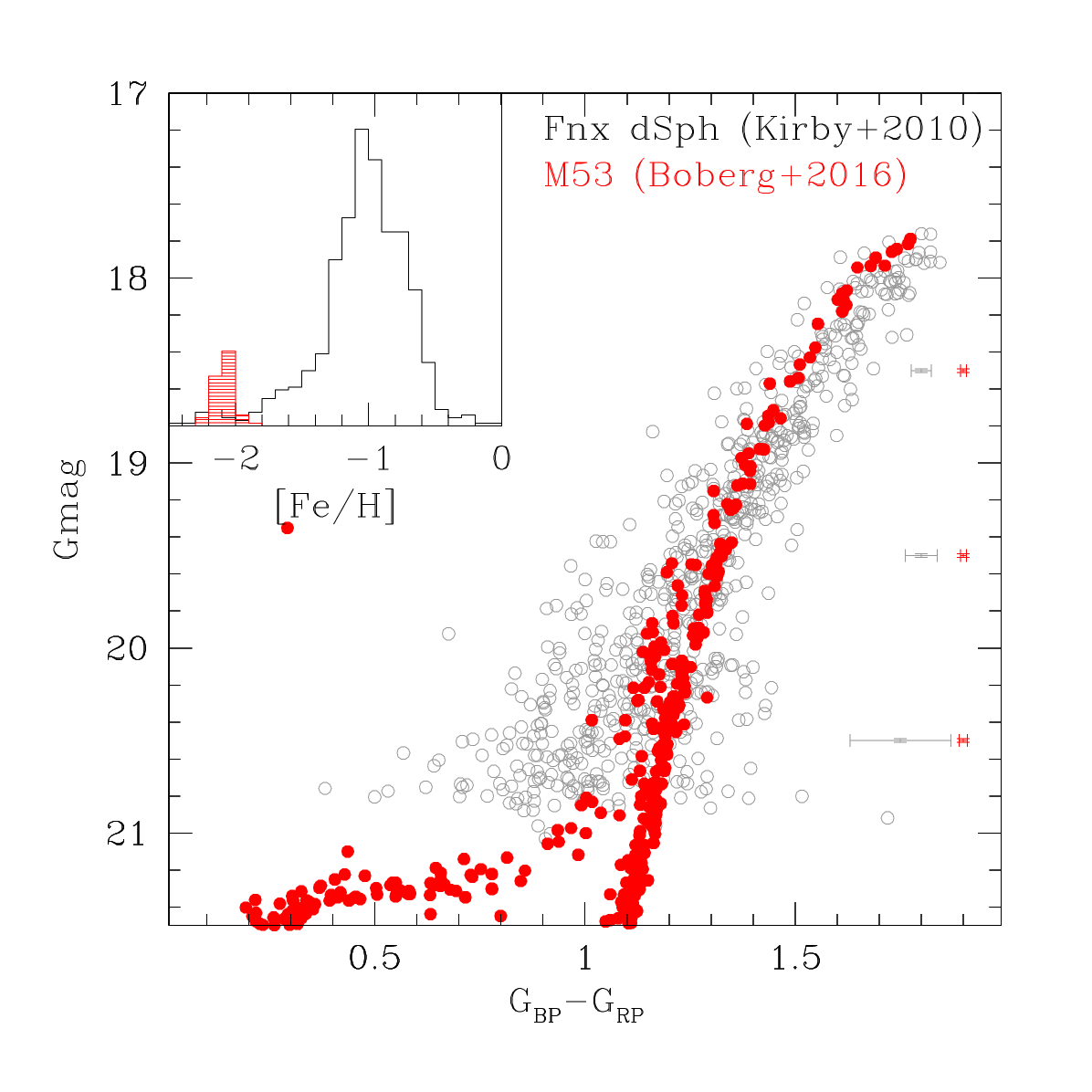}\includegraphics[scale=0.30]{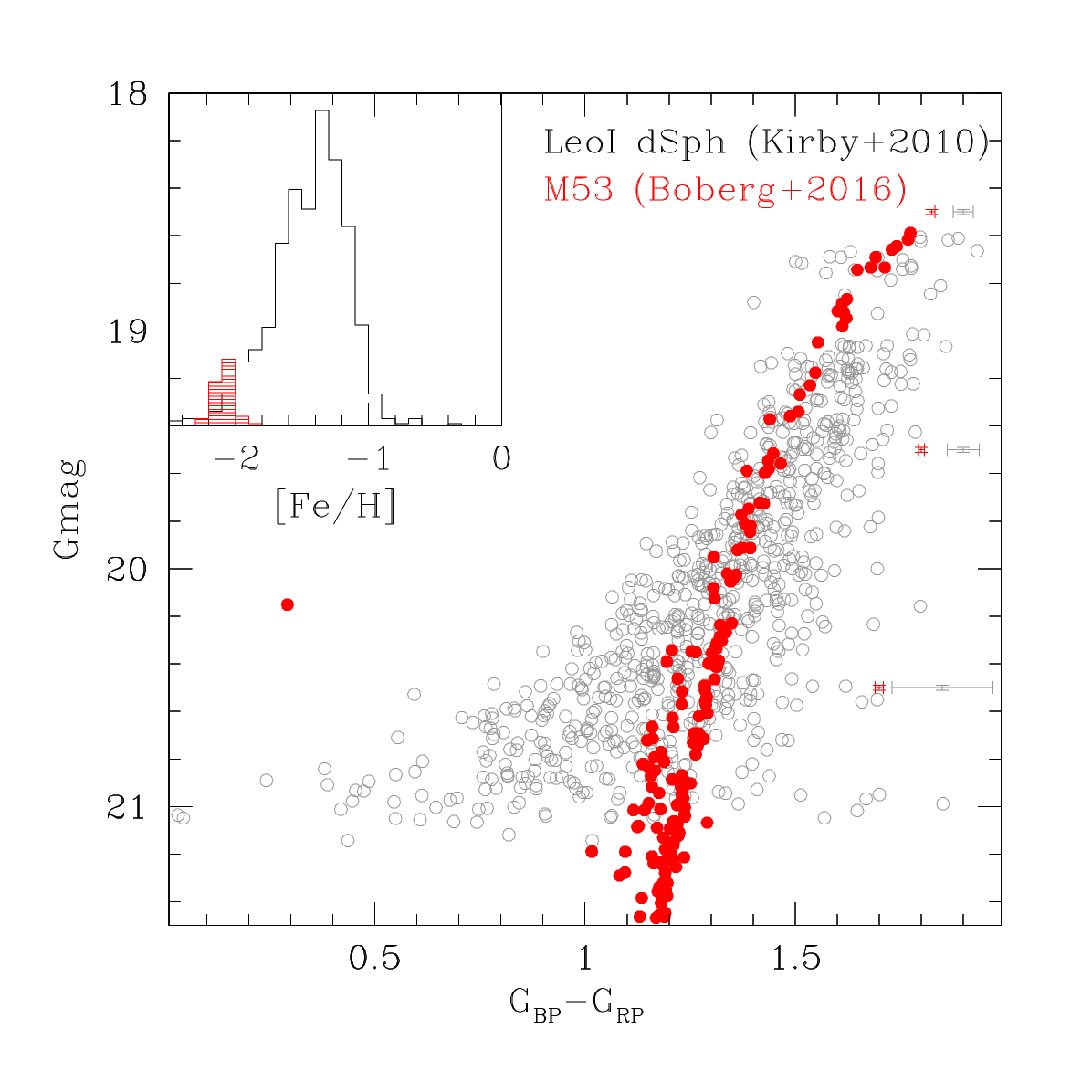}
\includegraphics[scale=0.30]{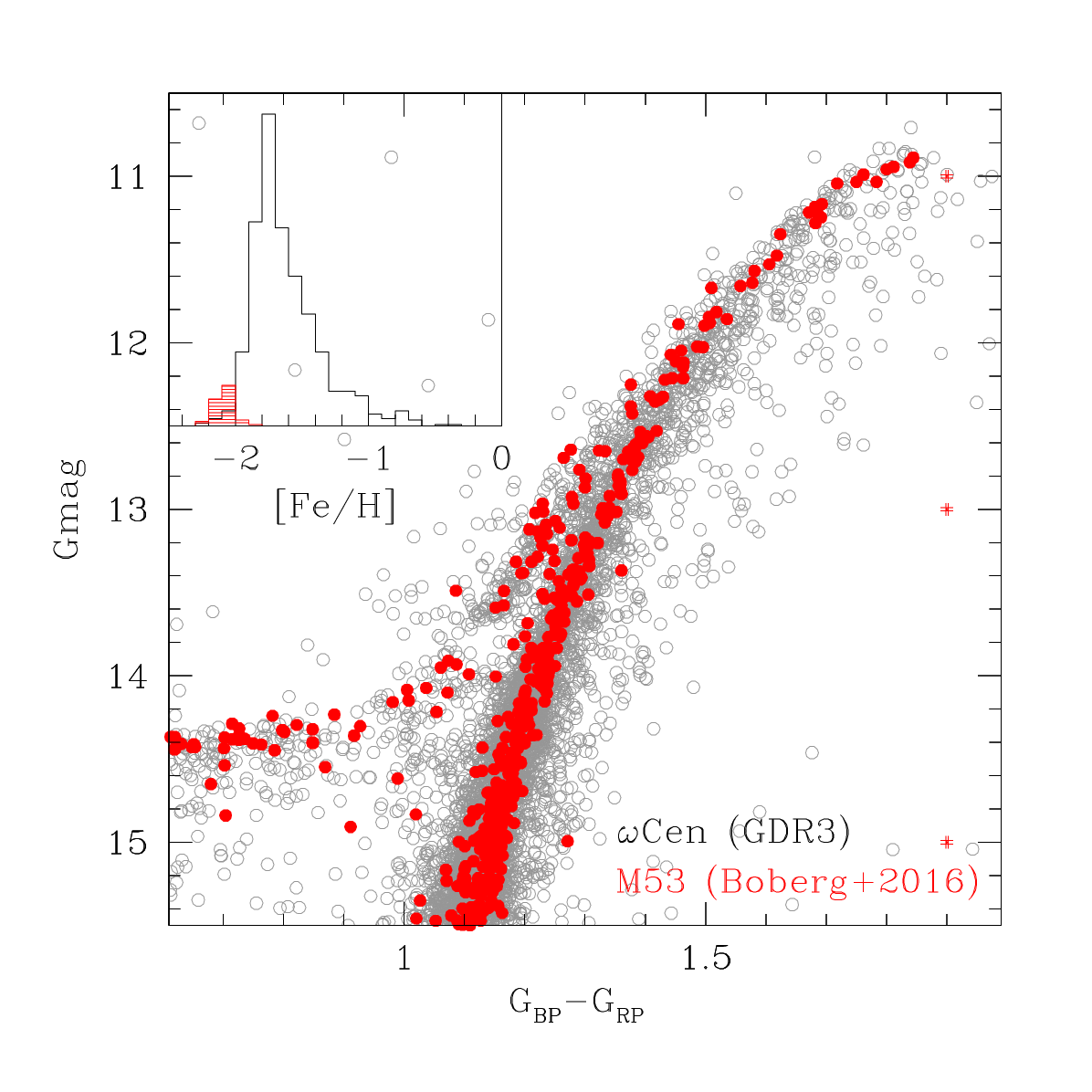}\includegraphics[scale=0.30]{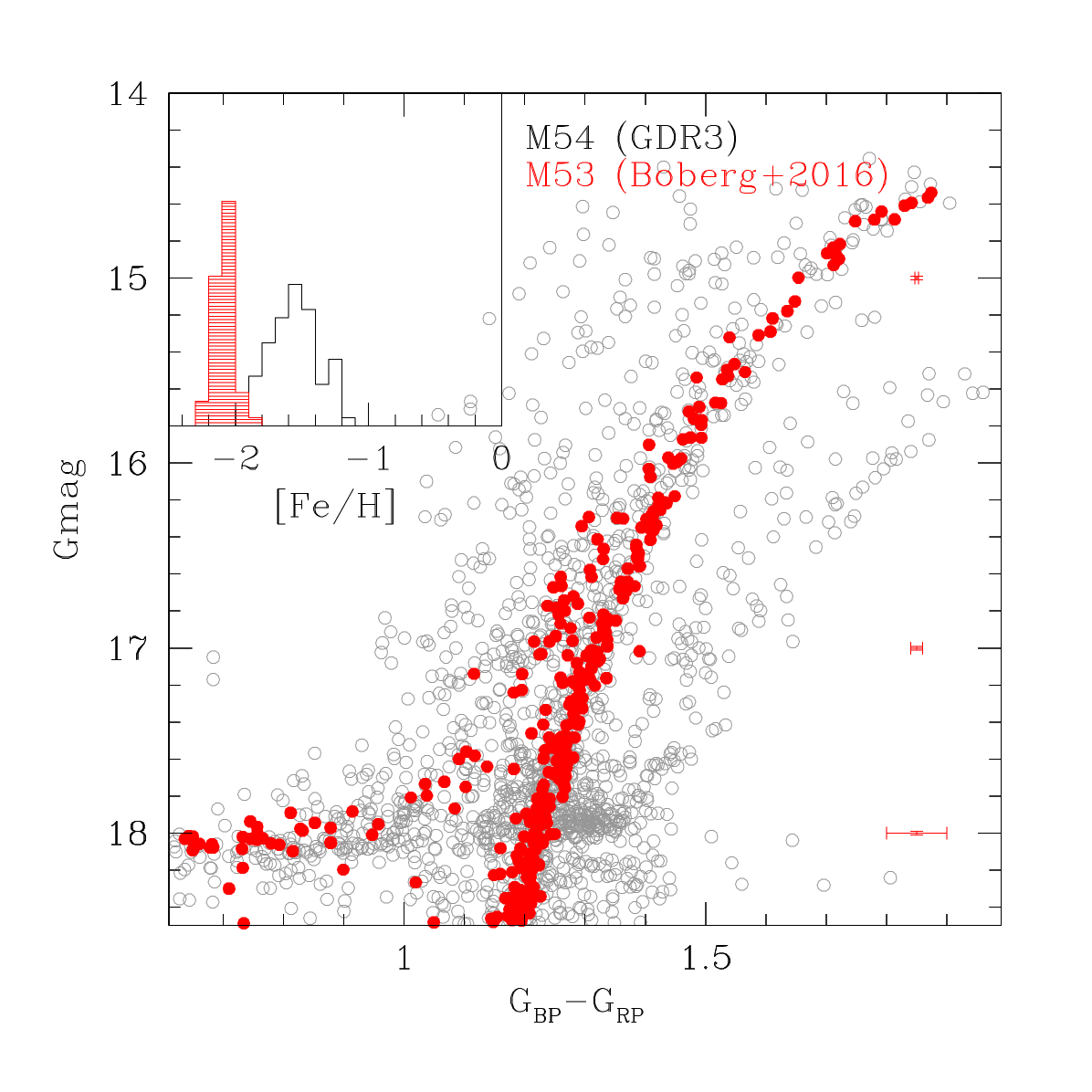}
\caption{Comparison of the RGBs in the CMDs of GCs and dwarf galaxies. Upper
panels: NGC~5024 (M~53, filled red points) is compared to the dwarf galaxies
Fornax and Leo I (empty circles). Lower panels: Comparison of NGC~5024 with two
massive GCs, $\omega$ Cen and M~54 (left and right panel, respectively). The
inset in each panel shows the metallicity distribution of the compared 
samples. Typical error bars in the Gaia photometric system are indicated.}
\label{f:dwarfs}
\end{figure*}

At a given absolute luminosity, the colour of the RGB has a negligible
dependence on the stellar mass, hence on the age of the stellar population, at
least for ages higher than a few Gyr (e.g., Salaris, Cassisi, Weiss 2002). On the
other hand, the position of a star across the RGB is strongly affected by the
value of the metallicity. By comparing the colour spreads of the RGBs of the old 
stellar populations in the systems discussed above, one can grasp an immediate
quantitative idea on the soundness and ranking of the different estimates.
This comparison is done in Fig.~\ref{f:dwarfs}, using for simplicity the Gaia
bandpasses to overcome problems related to the variety of photometric systems
used for GCs and dwarfs. Metal rich stars would be located on the red side of
the RGB, whereas metal-poor stars would populate the bluer region, regardless of
the selected bandpass.

To obtain the best sample of stars in NGC~5024, we considered the photometric
constraints described in Leitinger et al. (2023). They use the UBVRI photometry
by Stetson et al. (2019) and the resulting sample of stars was then
cross-matched with the database of H.
Baumgardt\footnote{\tt \tiny https://people.smp.uq.edu.au/HolgerBaumgardt/globular/parameter.html}
of member stars in GCs. Red points in Fig.~\ref{f:dwarfs} are
stars in NGC~5024 with membership probability $\geq 0.9$. Arbitrary shifts in
colour G$_{\rm BP}$-G$_{\rm RP}$ and in G$_{\rm mag}$ were applied to make the
RGB tip approximatively coincide for all the compared systems, as we were simply
interested in comparing the size of spread in colour as proxy of the metallicity
spread.

We took the metallicity information for the dSph's Leo~I and Fornax from Kirby
et al. (2010), who used Keck/DEIMOS medium-resolution spectroscopy combined with
spectral synthesis to derive them. That catalogue was cross-matched to Gaia data
release 3 (GDR3, Gaia Collaboration et al. 2023) to obtain the magnitudes in the
Gaia bandpasses for direct comparison with the GC CMDs. The sample of GC stars
in $\omega$~Cen (NGC~5139) and NGC~6715 (M54) are from  Stetson et al. (2019)
and have also been cross-matched with GDR3.

The [Fe/H] spread in FG stars for NGC~5024 from L22 (0.307 dex) is slightly larger than
the value reported by Willman and Strader (2012) for Fornax that, unlike GCs, is
known to have experienced extended star formation lasting several Gyrs, as
demonstrated from the width of the metallicity distributions and the age spreads
deduced from its broadband colours (combined with spectroscopic metal
abundances, e.g., Battaglia et al. 2006).

In the CMD of NGC~5024, stars on the asymptotic giant branch are well
recognisable, and the stars unambigously on the RGB form a very narrow sequence,
as expected for a stellar population with the same age and composition, 
whereas old populations on the RGB of dSphs show a much larger spread in colour,
more consistent with measured spectroscopic spreads of [Fe/H] of about 0.3 dex
(Kirby et al. 2010, Willman and Strader 2012).
The luminosity-Z spread relation observed in dSphs (e.g., Kirby et al. 2011)
shows that dSphs were able to sustain a star formation of long duration thanks
to the ability to retain the gas, the more effectively the more luminous is the
galaxy. Conversely, the formation of GCs was an extremely fast event: none of
the proposed scenarios (e.g., Bastian and Lardo 2018, Gratton et al. 2019)
involves star formation histories lasting more than a few tens of Myr, after
which only passive evolution sets on. This time limit is  established relying on
the evolving times of various plausible candidates proposed for the enrichment
in light, proton-capture elements characterising the MPs in GCs and the
simultaneous lack of variations in heavier elements in most GCs.

Also instructive is the comparison shown in the lower two panels of
Fig.~\ref{f:dwarfs}, where NGC~5024 is plotted against the RGB colour
distribution of the two most massive GCs in the Galaxy: $\omega$ Cen and M~54.
Both GCs have an intrinsic [Fe/H] spread, well documented from spectroscopic or
photometric measurements, and both are considered to be former nuclear star
clusters of dwarf galaxies (e.g., Bekki and Freeman 2003, Carretta et al. 2010,
see also the review by Gratton et al. 2004). Again, the spread in NGC~5024
estimated for FG stars by L22 seems to be completely at odds with the large
values measured in the two massive GCs, as the comparison of the CMDs
unambigously shows (see also Fig.~\ref{f:solol22b22}).

Typical uncertainties in magnitudes and colours are indicated by the error
bars in Fig.~\ref{f:dwarfs}, estimated from Table 3 of Gaia Collaboration et al.
(2018a) and figure A.7 of Gaia Collaboration et al. (2018b). While the faint end
of RGB in the more distant dwarf galaxies is close to the limits of precise Gaia
photometry, the comparison at the bright regime does not support a correpondence
between colour spreads and metallicity dispersions derived by L22. This is even
more evident when the comparison involves NGC~5024 and other GCs like $\omega$
Cen and M~54\footnote{The error bars close the the lower RGB region in the dwarf
galaxies, when properly focused deep photometry is used, decrease to values more
comparable to those of GCs.}.

Beside this direct comparison of observables on the RGB, there is also
support from theory to suggest that the metallicity spreads given in L22 are
probably overestimated and unrealistic. The proto globular cluster clouds out of
which $\omega$ Cen and M~54 formed were massive enough to satisfy the theoretical
prerequisites for showing clear evidence of self-enrichment. These requirements 
are well illustrated by Dopita and Smith (1986) and include primary star
formation with SN explosions and a potential well deep enough to trap the ejecta
that must be able to cool enough to be used later in the following phase of
secondary star formation.

For star formation to occur in gas enriched by SNe exploded in the primary star
formation, the ejecta must cool and remain trapped in a layer of gas both
gravitationally bound to the cloud and able to becoming self-gravitationally
unstable. These constraints imply that there is a limit in mass for the
progenitor of GCs to sustain some self-enrichment. Dopita and Smith (1986)
calculated that energetic SN ejecta can be slowed below the escape velocity of
the cloud, by explosions close to the mass center, only in clouds more massive
than $\sim 3 \times 10^6$ M$_\odot$. Explosions near the tidal radius would have
gas and metal enriched ejecta unavoidably lost to the intergalactic medium. At
the low extreme, clouds would be completely disrupted by SN explosions if less
massive than $\sim 10^5$ M$_\odot$. The bottom line is that only the most
massive clouds are destined to form the final, self-enriched GCs. 

\begin{figure}
\centering
\includegraphics[scale=0.40]{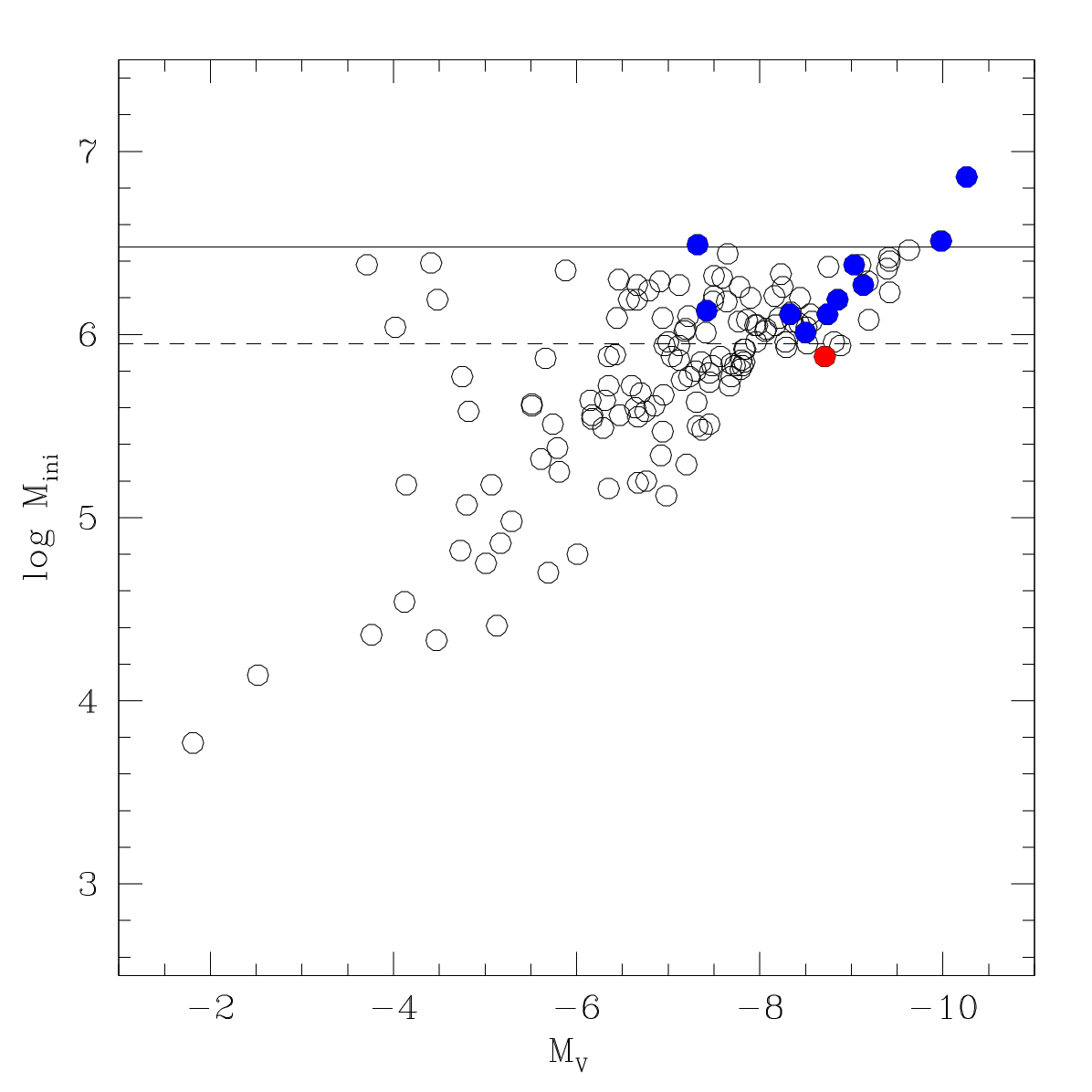}
\caption{Initial masses of GCs from Baumgardt et al. (2019) for the population
of GCs in the Milky Way plotted as a function of the present-day total absolute
luminosity. Solid and dashed lines indicate values $\sim 3 \times 10^6$ 
M$_\odot$ and $9 \times 10^5$ M$_\odot$. The red filled circle represents NGC~5024
and blue circles indicate GCs presently known to have in some extent a
metallicity dispersion.} 
\label{f:bhsb}
\end{figure}

A quantitative idea of the fraction of Galactic GCs potentially capable of
satisfying these requisites is given in Fig.~\ref{f:bhsb}. The estimates of
initial masses for the population of GCs in the Milky Way (Baumgardt et al.
2019) are plotted as a function of the present day masses (represented by the
total absolute magnitude). The solid line shows the lower mass limit for a proto
cluster cloud to efficiently retain SN ejecta according to the calculations by
Dopita and Smith (1986). Only three GCs (namely $\omega$ Cen, M~54, and Liller 1)
are expected to be above this threshold. NGC~5024 (red filled circle) lies
well below the required value, even at the estimated beginning of its evolution,
before its mass is next reduced by dynamical evolutionary effects (both internal
and external).

We know, of course, that actually a number of GCs present traces of pollution by
SNe, in various amounts. These are the GCs named iron complex clusters by
Johnson et al. (2015): M~2, NGC~1851, M~54, $\omega$ Cen, NGC~5286, NGC~5824,
NGC~6273, Terzan 5, M~22. A recent addition to this group is Liller 1 (Fanelli
et al. 2024). The iron complex GCs are indicated in Fig.~\ref{f:bhsb} as filled
blue points and the dashed line (at about $9 \times 10^5$ M$_\odot$) defines the
lower envelope for initial masses of these massive GCs. NGC~5024 lies just below
the group of iron complex GCs, but allowing for uncertainties in the estimates
of the initial masses of objects just slightly younger than the Universe, it 
could even match this less stringent constraint. However, the striking
difference in [Fe/H] spread between NGC~5024 and GCs with a spectroscopic spread
of 0.3 dex still lingers. Boberg et al. (2016) find a dispersion of only 0.07
dex associated to the mean metallicity of NGC~5024 from a large sample of stars.
This value is more similar to the one detected for NGC~1851 (Carretta et al.
2011b), a member of the iron complex GCs with a minimal inhomogeneity in [Fe/H].

The conclusion from the present section is that the metallicity spread for FG in
NGC~5024 (as well as for NGC~5272) computed by L22 from the PCM seems to be
largely overestimated. Neither simple observations based on the comparison of the RGB
colour spreads nor theoretical model constraints seem to support the notion of
iron spread so large as advocated by L22 for these two GCs.
Since the method used for NGC~5024 and NGC~5272 is not different from the one
applied to the other GCs, a logical inference is that probably the estimates
for the [Fe/H] spread in FG stars of all GCs may be an overestimate.

These greatly overstated values derived from the PCMs could probably explain 
the discrepancies between high resolution spectroscopy and HST photometry,
probed in the next Section.

\section{Discrepancies between spectroscopy and HST photometry}

The approach used by L22 to derive a putative metallicity spread among FG
stars is to link the extension of $RG1$ to the width W1g of FG stars
on the RGB in the CMD $m_{F814W}$ versus $col$ taken from M17. The underlying
principle is that stars of different metallicity occupy different sides of the
RGB. 

From Section 2 what does not compute is the sheer (and unbelievably large)
amount of the spreads attributed to FG stars, derived by L22 by translating W1g
into metallicity. This was done using a generic ``colour-metallicity relation"
from Dotter et al. (2008). If the  method used by L22 is the equivalent of
interpolating mean ridge lines of GCs at different metallicities, but instead
simply employing theoretical isochrones, it is susceptible to even larger errors.
Salaris et al. (2002) stated that at fixed magnitude on the  RGB a colour
difference in $V-I$ of about 0.05 or 0.10 mag translates into uncertainties as
large as $\sim 0.3-0.5$ dex when [Fe/H] is determined from the comparison of
observed and theoretical RGBs.

For the first time it is possible to test this scenario and obtain a more sound
estimate of the variation in [Fe/H] among FG stars thanks to the large (and
unique, so far) published database of stars used for the chemical tagging of the
PCMs in CB24. From the few hundreds stars with both HST photometry and high
resolution spectroscopy identified in 22 GCs, we considered 159 FG stars. These
stars lie on the $RG1$ and are classified as FG with spectroscopy, 
according to their Na abundance, since we showed that a number of stars tagged
using photometric criteria are actually mismatches, that is FG stars
misclassified as SG stars and vice versa (CB24). We prefer to rely on the
criterion based on Na, since the abundance from spectroscopy is the direct 
discriminant for MPs in GCs (see Carretta et al. 2009b).

\begin{table*}
\centering
\caption{Pairs of FG stars with similar atmospheric parameters}
\begin{tabular}{rrlrlrllrl}
\hline
pair &$\Delta col$ & $\Delta col3$ & [Fe/H] & [Na/Fe] &   ID     &  GC  &[Fe/H]$_{GC}$ &  Atm.parameters & ref. abundance analysis \\
\hline
  1 & -0.0591	   &  0.0218	   & -1.240 &  0.017  & R0000202 &  288 & -1.219 &  5073 2.69 -1.24 1.13 & Carretta et al. (2009b)\\   
  1 & -0.0477	   &  0.0725	   & -1.188 & -0.002  & R0001020 &  288 & -1.219 &  5103 2.68 -1.18 1.09 & \\  
  2 & -0.0690	   &  0.0064	   & -1.150 &  0.065  & R0006958 & 2808 & -1.128 &  4493 1.54 -1.15 1.52 & Carretta (2015) \\   
  2 & -0.1760	   &  0.0671	   & -1.111 &  0.078  & R0035565 & 2808 & -1.128 &  4461 1.45 -1.11 1.50 & \\   
  3 & -0.0690	   &  0.0064	   & -1.150 &  0.065  & R0006958 & 2808 & -1.128 &  4493 1.54 -1.15 1.52 & \\   
  3 & -0.1844	   &  0.0337	   & -1.161 & -0.032  & R0038704 & 2808 & -1.128 &  4502 1.52 -1.16 1.56 & \\   
  4 & -0.0852	   & -0.0153	   & -1.113 & -0.082  & R0037787 & 2808 & -1.128 &  4703 1.89 -1.11 1.35 & \\   
  4 & -0.2384	   &  0.0431	   & -1.168 & -0.089  & R0039175 & 2808 & -1.128 &  4712 1.88 -1.17 1.54 & \\   
  5 &  0.0954	   & -0.0221	   & -1.457 & -0.103  & R0000657 & 3201 & -1.495 &  4737 2.17 -1.46 1.67 & Carretta et al. (2009b)\\   
  5 & -0.0927	   &  0.0098	   & -1.432 & -0.196  & R0000897 & 3201 & -1.495 &  4733 2.18 -1.43 1.11 & \\  
  6 &  0.0954	   & -0.0221	   & -1.457 & -0.103  & R0000657 & 3201 & -1.495 &  4737 2.17 -1.46 1.67 & \\  
  6 & -0.4515	   &  0.1263	   & -1.562 & -0.122  & R0002678 & 3201 & -1.495 &  4761 2.26 -1.57 1.42 & \\  
  7 &  0.0270	   &  0.0140	   & -1.496 & -0.246  & R0000690 & 3201 & -1.512 &  4495 1.61 -1.50 1.59 & \\  
  7 &  0.0105	   & -0.0009	   & -1.440 & -0.129  & R0001697 & 3201 & -1.512 &  4496 1.61 -1.44 1.69 & \\  
  8 & -0.4515	   &  0.1263	   & -1.562 & -0.122  & R0002678 & 3201 & -1.495 &  4761 2.26 -1.57 1.42 & \\  
  8 & -0.0927	   &  0.0098	   & -1.432 & -0.196  & R0000897 & 3201 & -1.495 &  4733 2.18 -1.43 1.11 & \\  
  9 & -0.0221	   &  0.0164	   & -2.265 &  0.036  & R0000883 & 4590 & -2.227 &  4856 2.01 -2.26 0.41 & Carretta et al. (2009b)\\ 
  9 & -0.0522	   &  0.0234	   & -2.304 &  0.009  & R0001662 & 4590 & -2.227 &  4833 2.01 -2.30 0.20 & \\   
 10 &  0.0054	   & -0.0531	   & -2.019 &  0.022  & R0004864 & 4833 & -2.040 &  4829 2.05 -2.02 1.68 & Carretta et al. (2014)\\  
 10 & -0.0489	   & -0.0006	   & -2.038 &  0.031  & R0003934 & 4833 & -2.040 &  4852 2.09 -2.04 1.59 & \\   
 11 & -0.1730	   &  0.1328	   & -1.813 &  0.125  & R0012060 & 6093 & -1.792 &  4414 1.18 -1.81 1.39 & Carretta et al. (2015)\\  
 11 & -0.1604	   &  0.1297	   & -1.843 &  0.177  & R0012523 & 6093 & -1.791 &  4434 1.21 -1.84 1.58 & \\   
 12 & -0.1331	   &  0.0233	   & -1.804 &  0.038  & R0004959 & 6093 & -1.791 &  4935 2.23 -1.80 1.54 & \\   
 12 & -0.0828	   &  0.1008	   & -1.791 &  0.225  & R0003113 & 6093 & -1.791 &  4952 2.25 -1.79 1.23 & \\   
 13 & -0.0722	   &  0.0217	   & -1.784 &  0.170  & R0002307 & 6093 & -1.791 &  5056 2.46 -1.78 0.65 & \\   
 13 & -0.1606	   &  0.1118	   & -1.782 &  0.030  & R0012580 & 6093 & -1.791 &  5065 2.48 -1.78 1.82 & \\   
 14 & -0.0764	   &  0.0798	   & -1.763 &  0.154  & R0003844 & 6093 & -1.791 &  5111 2.58 -1.76 0.67 & \\   
 14 & -0.0642	   &  0.0636	   & -1.772 & -0.020  & R0001255 & 6093 & -1.791 &  5111 2.56 -1.77 1.96 & \\   
 15 & -0.0642	   &  0.0636	   & -1.772 & -0.020  & R0001255 & 6093 & -1.791 &  5111 2.56 -1.77 1.96 & \\   
 15 & -0.0932	   &  0.1011	   & -1.796 &  0.169  & R0010817 & 6093 & -1.791 &  5117 2.59 -1.80 1.75 & \\   
 16 & -0.0746	   & -0.0104	   & -1.446 & -0.043  & R0001634 & 6254 & -1.556 &  4938 2.37 -1.44 0.82 & Carretta et al. (2009b)\\  
 16 & -0.1384	   &  0.0208	   & -1.640 & -0.137  & R0001704 & 6254 & -1.556 &  4932 2.35 -1.64 1.12 & \\   
 17 &  0.0578	   &  0.0203	   & -1.549 & -0.128  & R0001056 & 6254 & -1.556 &  5011 2.50 -1.55 1.27 & \\   
 17 & -0.1435	   &  0.0441	   & -1.577 & -0.117  & R0004888 & 6254 & -1.556 &  5014 2.52 -1.58 1.85 & \\   
 18 & -0.1435	   &  0.0441	   & -1.577 & -0.117  & R0004888 & 6254 & -1.556 &  5014 2.52 -1.58 1.85 & \\   
 18 & -0.0438	   & -0.0074	   & -1.466 & -0.096  & R0001523 & 6254 & -1.556 &  5032 2.55 -1.47 1.54 & \\   
 19 & -0.0001	   &  0.0004	   & -2.058 & -0.001  & R0000841 & 6397 & -1.993 &  4931 2.42 -2.06 0.35 & Carretta et al. (2009b)\\   
 19 & -0.0252	   &  0.0128	   & -2.036 & -0.153  & R0000943 & 6397 & -1.993 &  4936 2.05 -2.04 1.38 & \\   
 20 & -0.0427	   &  0.0153	   & -1.880 & -0.170  & R0000162 & 6535 & -1.963 &  4901 2.58 -1.88 0.80 & Bragaglia et al. (2017)\\   
 20 & -0.0720	   &  0.0164	   & -1.950 & -0.059  & R0000450 & 6535 & -1.963 &  4886 2.55 -1.95 0.83 & \\    
 21 & -0.0427	   &  0.0153	   & -1.880 & -0.170  & R0000162 & 6535 & -1.963 &  4901 2.58 -1.88 0.80 & \\    
 21 & -0.0186	   & -0.0049	   & -1.970 & -0.017  & R0000199 & 6535 & -1.963 &  4926 2.29 -1.97 1.39 & \\    
 22 & -0.0829	   &  0.0260	   & -1.611 &  0.050  & R0006086 & 6752 & -1.555 &  4315 1.12 -1.61 1.50 & Carretta et al. (2009c)\\   
 22 & -0.0806	   &  0.0520	   & -1.548 &  0.126  & R0003489 & 6752 & -1.555 &  4338 1.19 -1.55 1.66 & \\   
 23 & -0.1417	   &  0.0232	   & -0.773 &  0.113  & R0000776 & 6838 & -0.808 &  4777 2.48 -0.77 1.49 & Carretta et al. (2009b)\\   
 23 & -0.0796	   &  0.0483	   & -0.862 &  0.233  & R0001634 & 6838 & -0.808 &  4766 2.47 -0.86 1.07 & \\   
 24 & -0.1417	   &  0.0232	   & -0.773 &  0.113  & R0000776 & 6838 & -0.808 &  4777 2.48 -0.77 1.49 & \\   
 24 & -0.0687	   &  0.0240	   & -0.776 &  0.137  & R0001344 & 6838 & -0.808 &  4785 2.49 -0.78 1.62 & \\   
 25 & -0.0687	   &  0.0240	   & -0.776 &  0.137  & R0001344 & 6838 & -0.808 &  4785 2.49 -0.78 1.62 & \\   
 25 & -0.0578	   &  0.0060	   & -0.787 &  0.233  & R0000187 & 6838 & -0.808 &  4800 2.47 -0.79 1.37 & \\   
 26 & -0.0578	   &  0.0060	   & -0.787 &  0.233  & R0000187 & 6838 & -0.808 &  4800 2.47 -0.79 1.37 & \\   
 26 & -0.0978	   &  0.0353	   & -0.837 &  0.240  & R0000458 & 6838 & -0.808 &  4814 2.51 -0.84 1.39 & \\   
 27 & -0.1979	   &  0.0756	   & -1.070 &  0.00   & R0000591 & 6121 & -1.070 &  4800 2.47 -1.07 1.37 & Marino et al. (2008)\\        
 27 & -0.0997	   &  0.0496	   & -1.050 &  0.00   & R0002084 & 6121 & -1.070 &  4800 2.57 -1.05 1.18 & \\         
 28 &  0.0676	   &  0.0119	   & -1.000 &  0.06   & R0000301 & 6121 & -1.070 &  4830 2.56 -1.00 1.25 & \\         
 28 & -0.0470	   &  0.0434	   & -1.090 &  0.13   & R0000346 & 6121 & -1.070 &  4830 2.52 -1.09 1.28 & \\         
 29 & -0.0470	   &  0.0434	   & -1.090 &  0.13   & R0000346 & 6121 & -1.070 &  4830 2.52 -1.09 1.28 & \\         
 29 & -0.0618	   & -0.0084	   & -1.080 &  0.07   & R0001488 & 6121 & -1.070 &  4850 2.54 -1.08 1.34 & \\         
 30 & -0.0618	   & -0.0084	   & -1.080 &  0.07   & R0001488 & 6121 & -1.070 &  4850 2.54 -1.08 1.34 & \\         
 30 & -0.1268	   &  0.0543	   & -1.090 &  0.02   & R0001736 & 6121 & -1.070 &  4850 2.63 -1.09 1.22 & \\         

\hline
\end{tabular}
\label{t:pairs}
\end{table*}

\begin{figure*}
\centering
\includegraphics[scale=0.23]{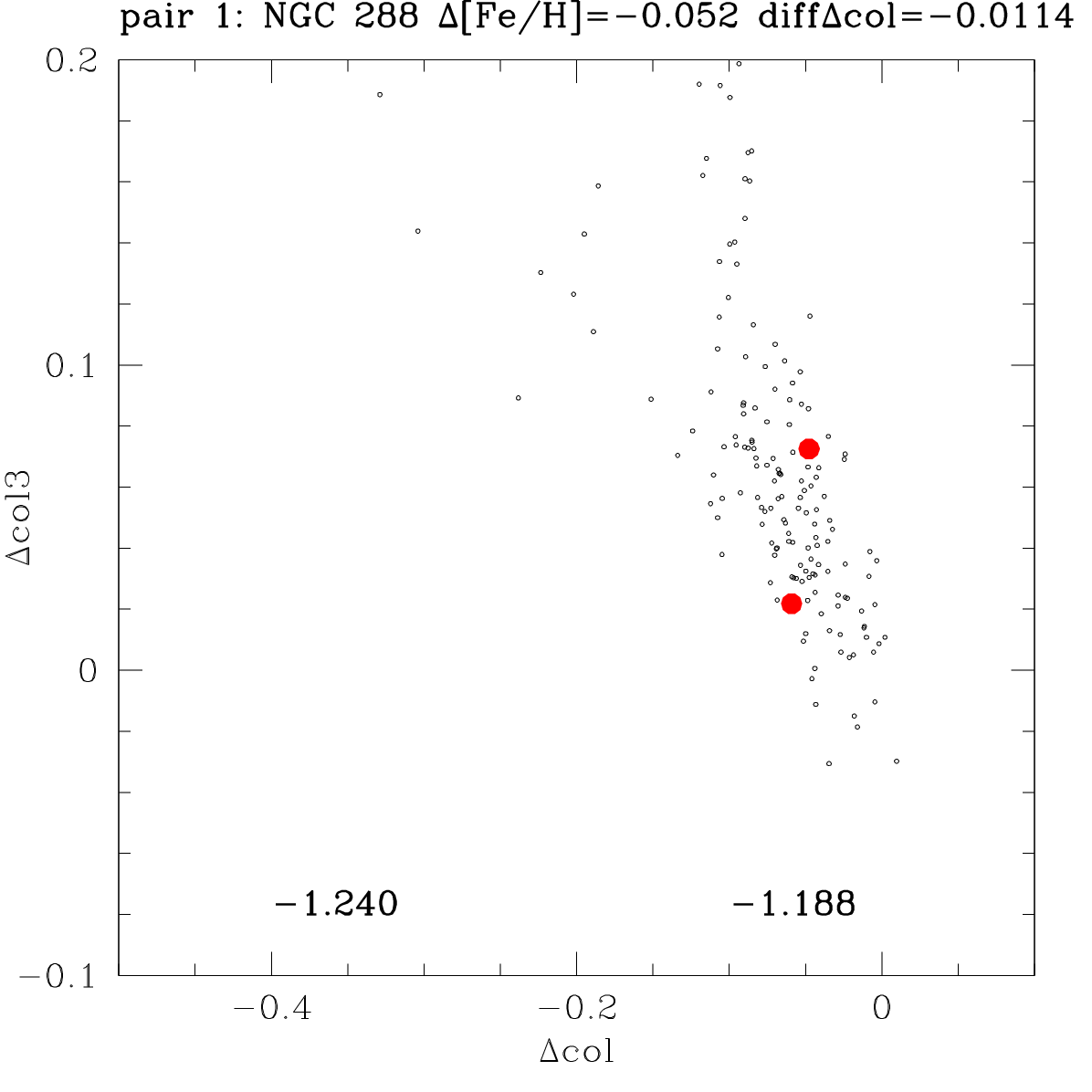}\includegraphics[scale=0.23]{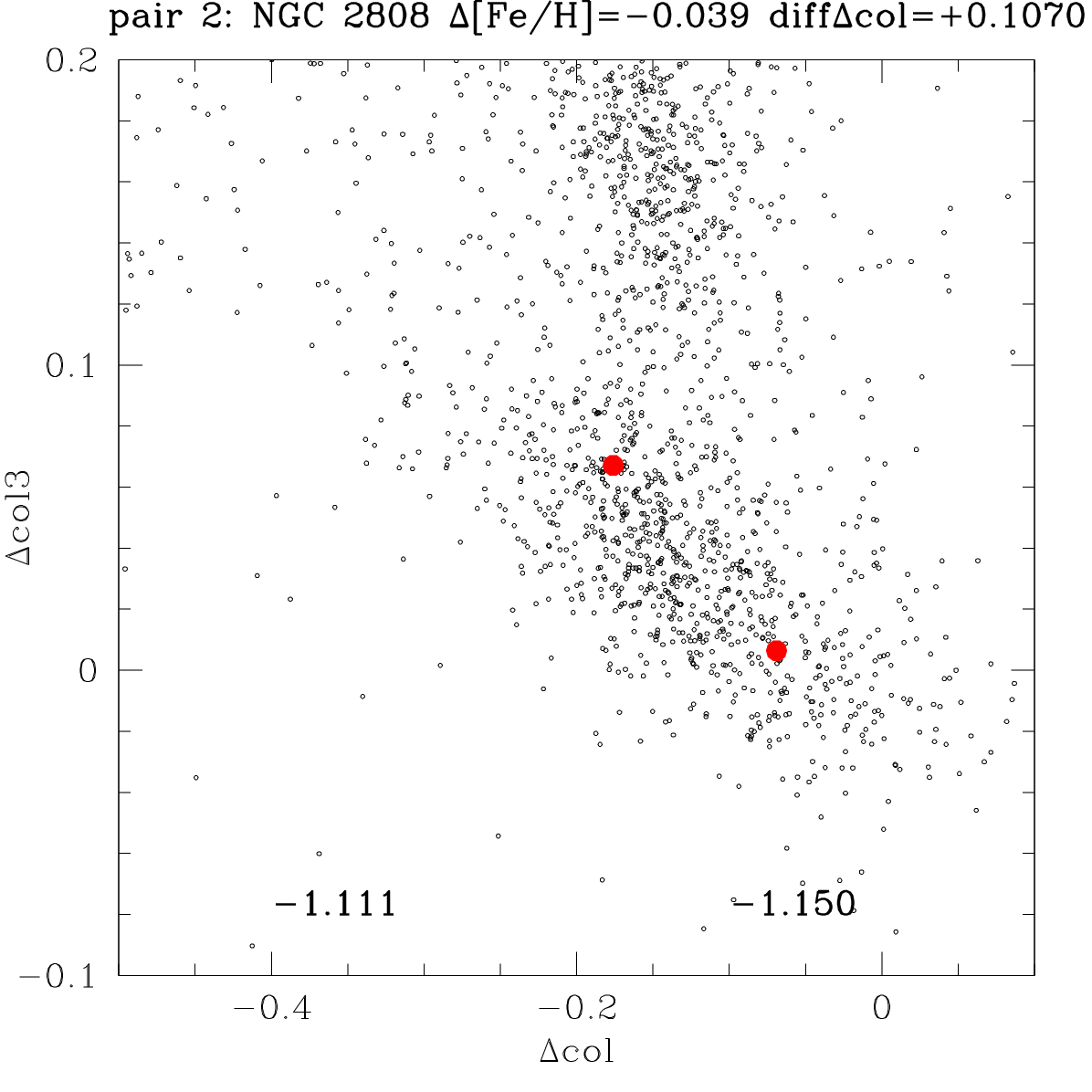}\includegraphics[scale=0.23]{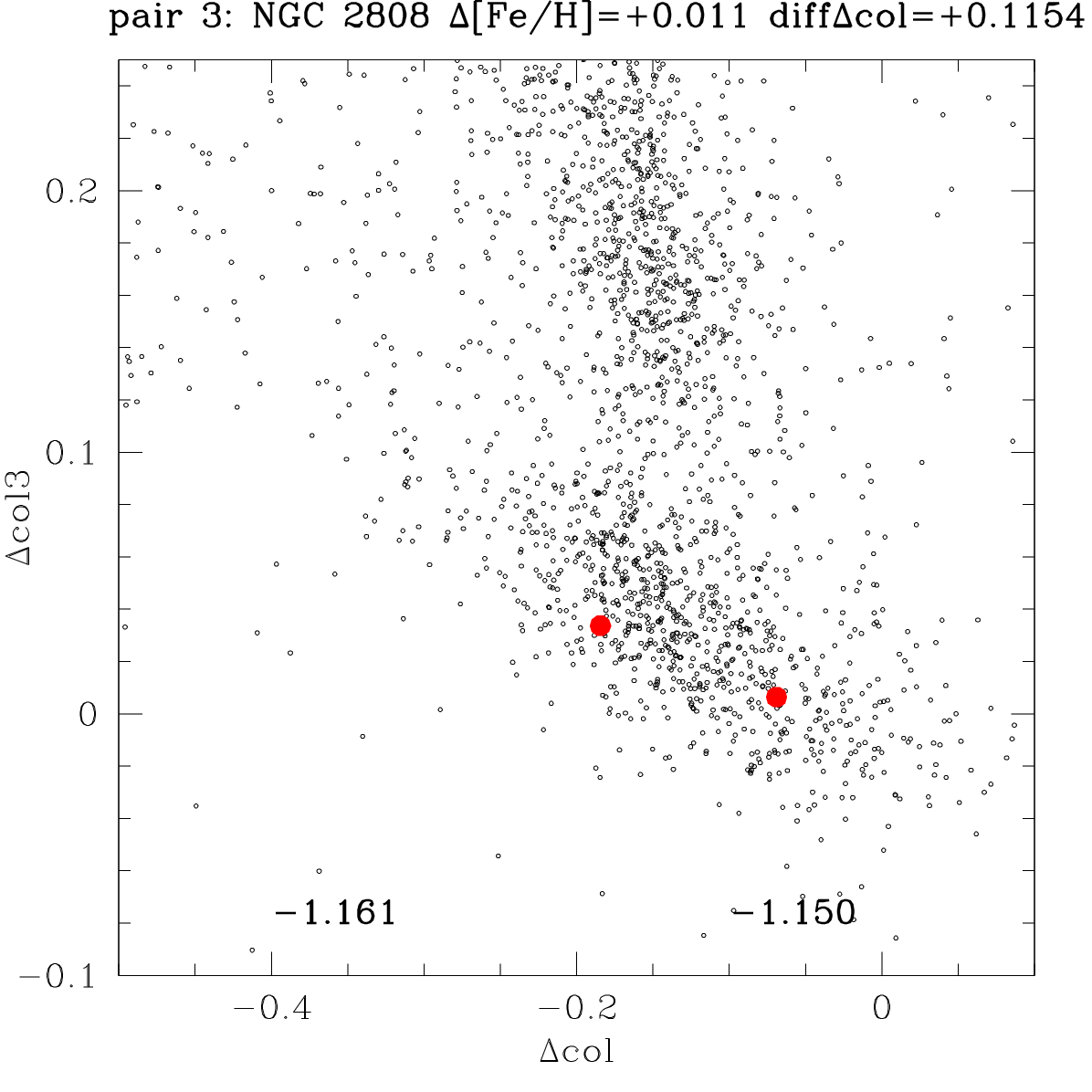}\includegraphics[scale=0.23]{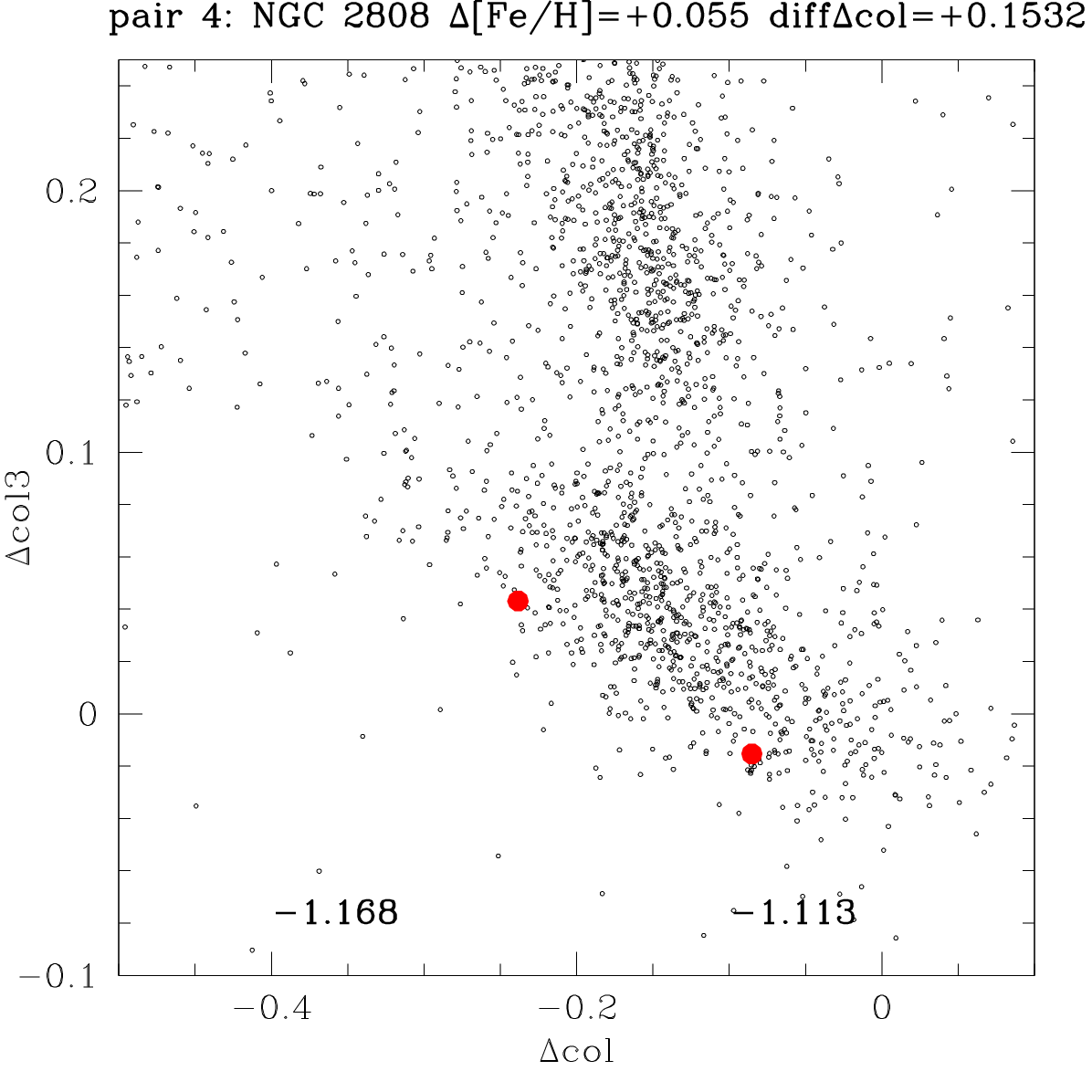}
\includegraphics[scale=0.23]{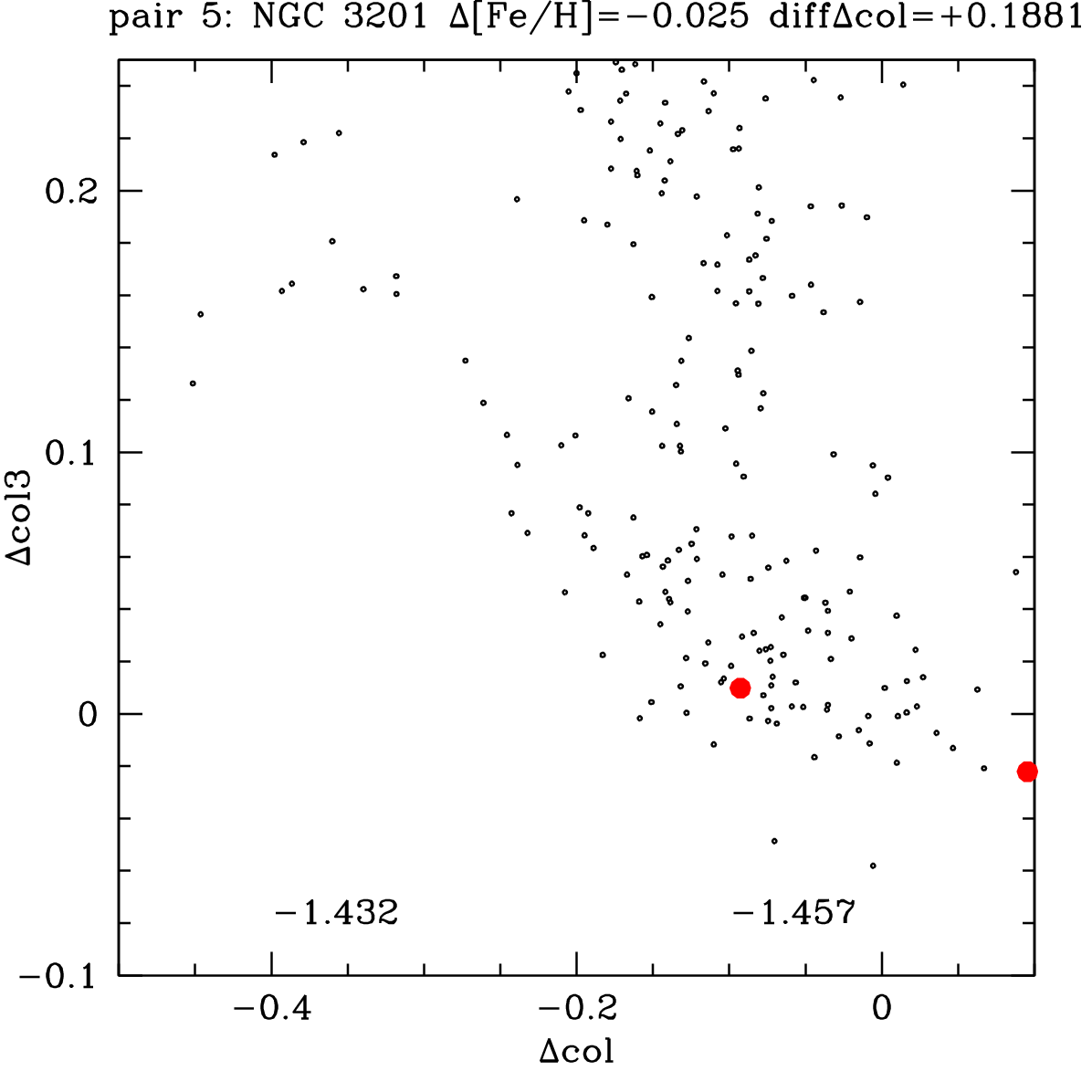}\includegraphics[scale=0.23]{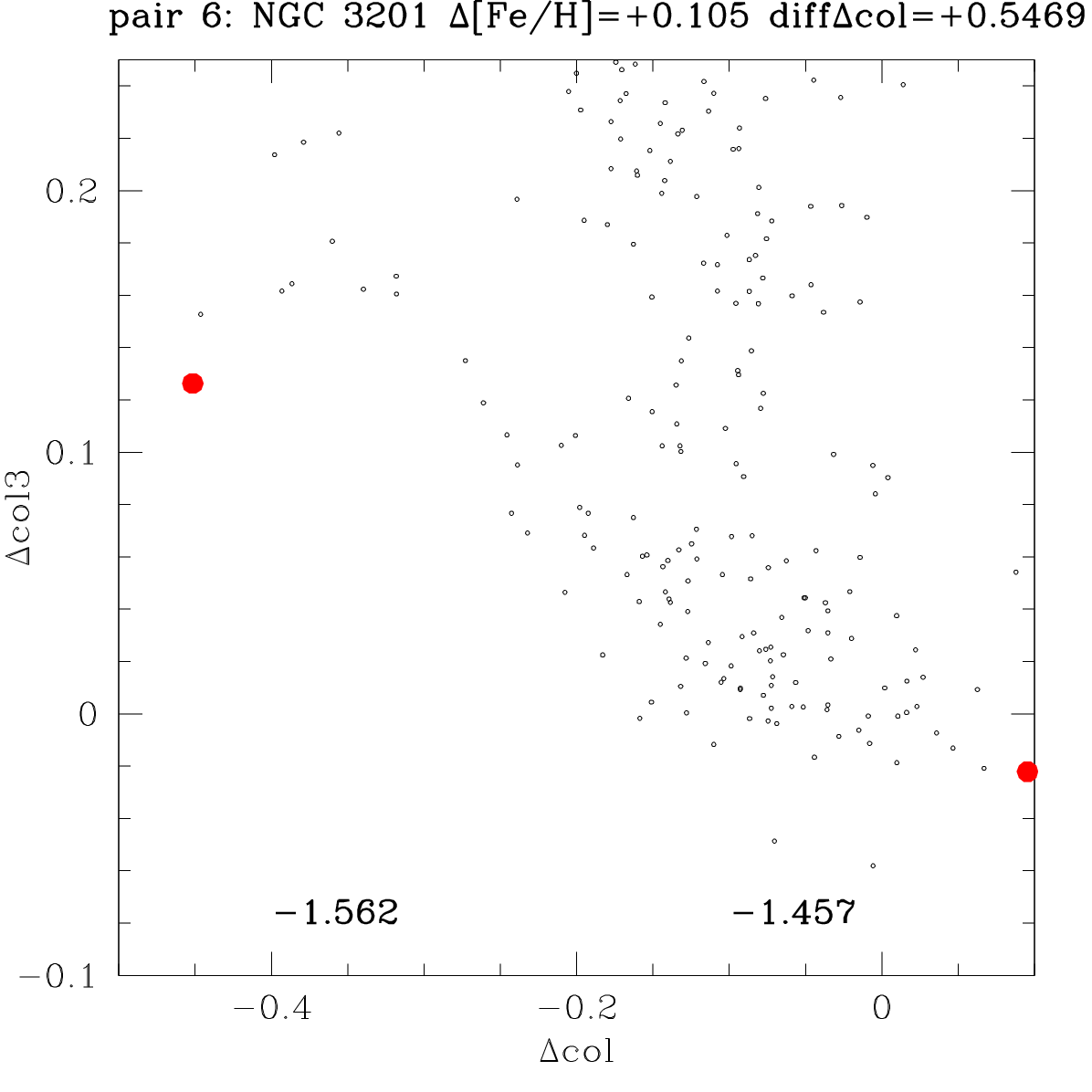}\includegraphics[scale=0.23]{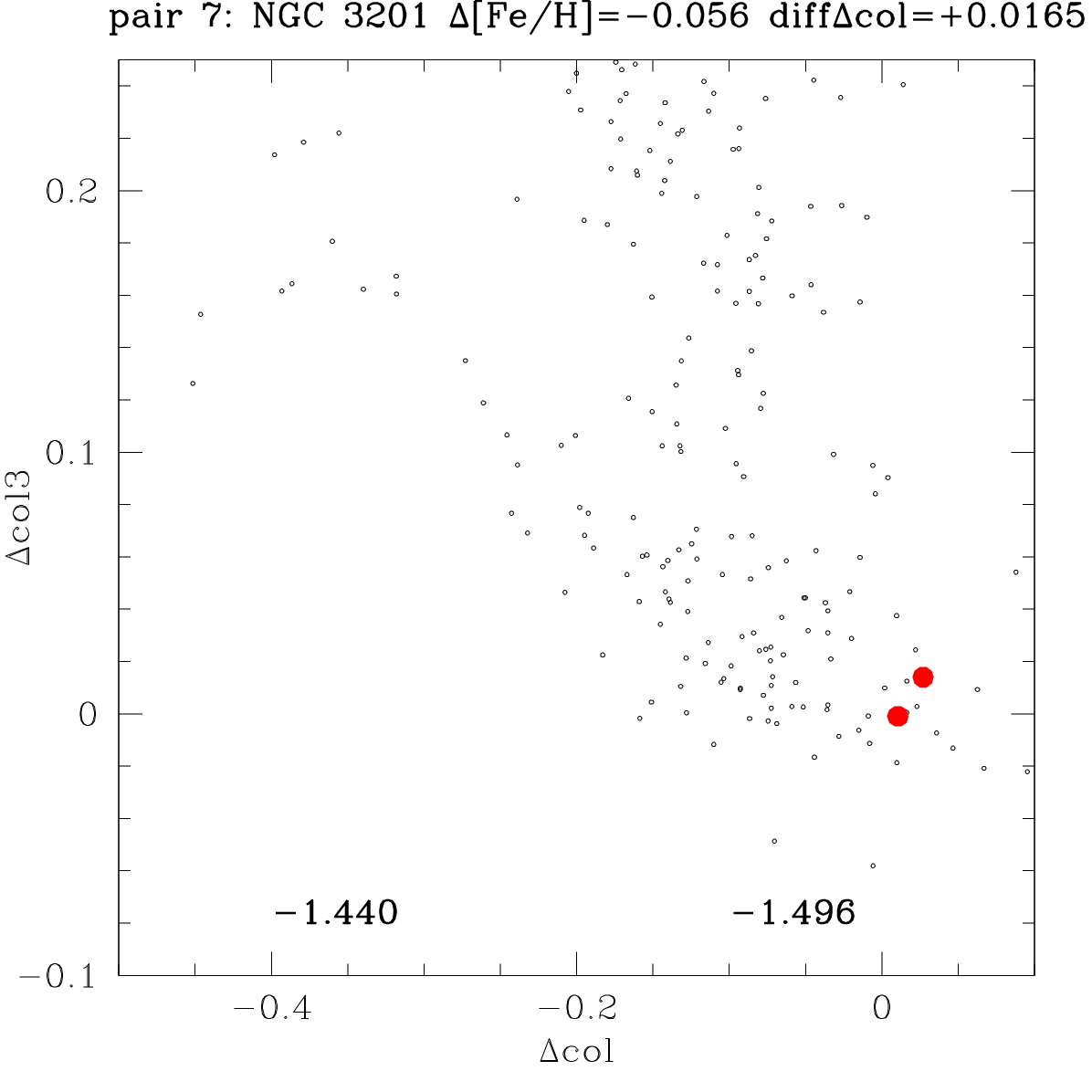}\includegraphics[scale=0.23]{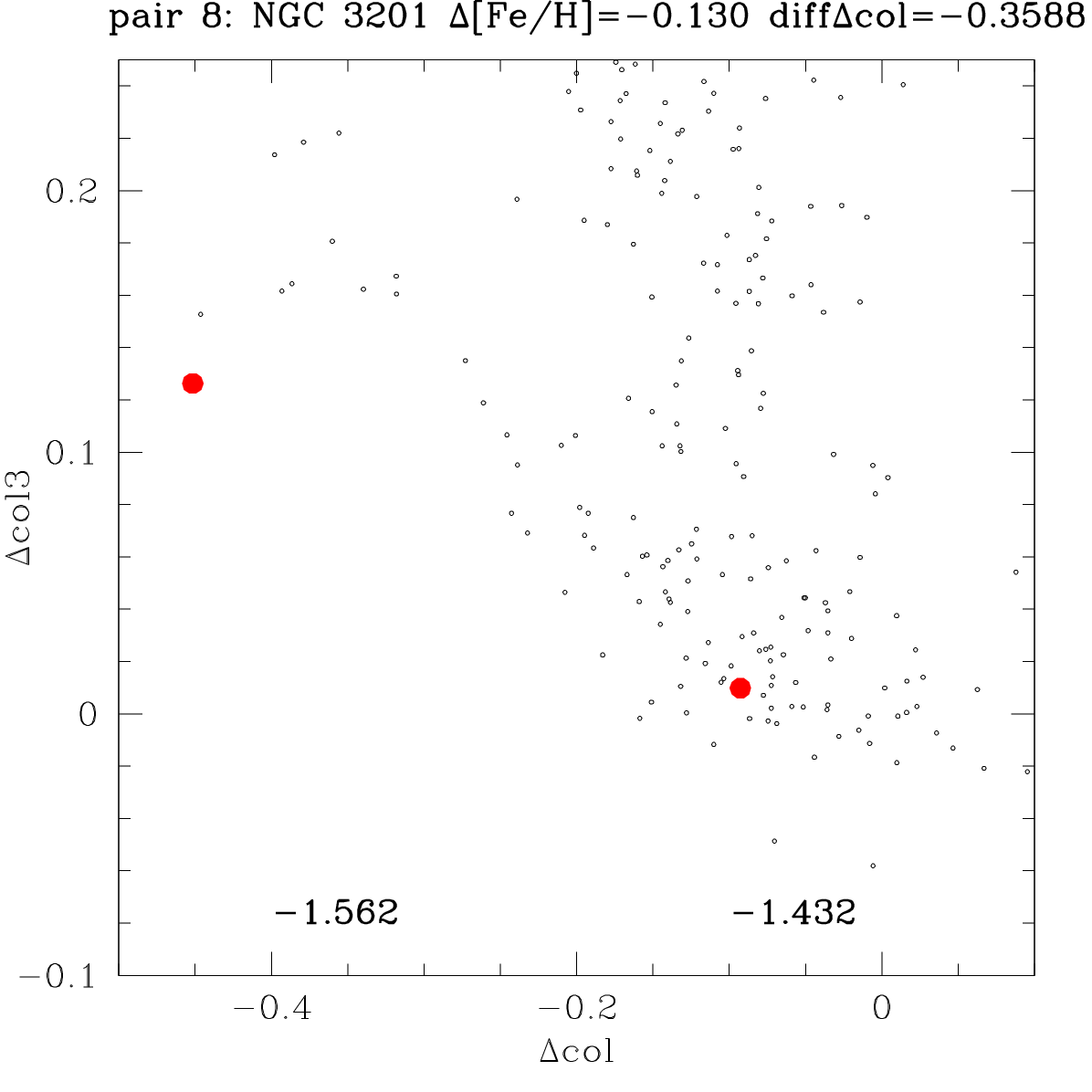}
\includegraphics[scale=0.23]{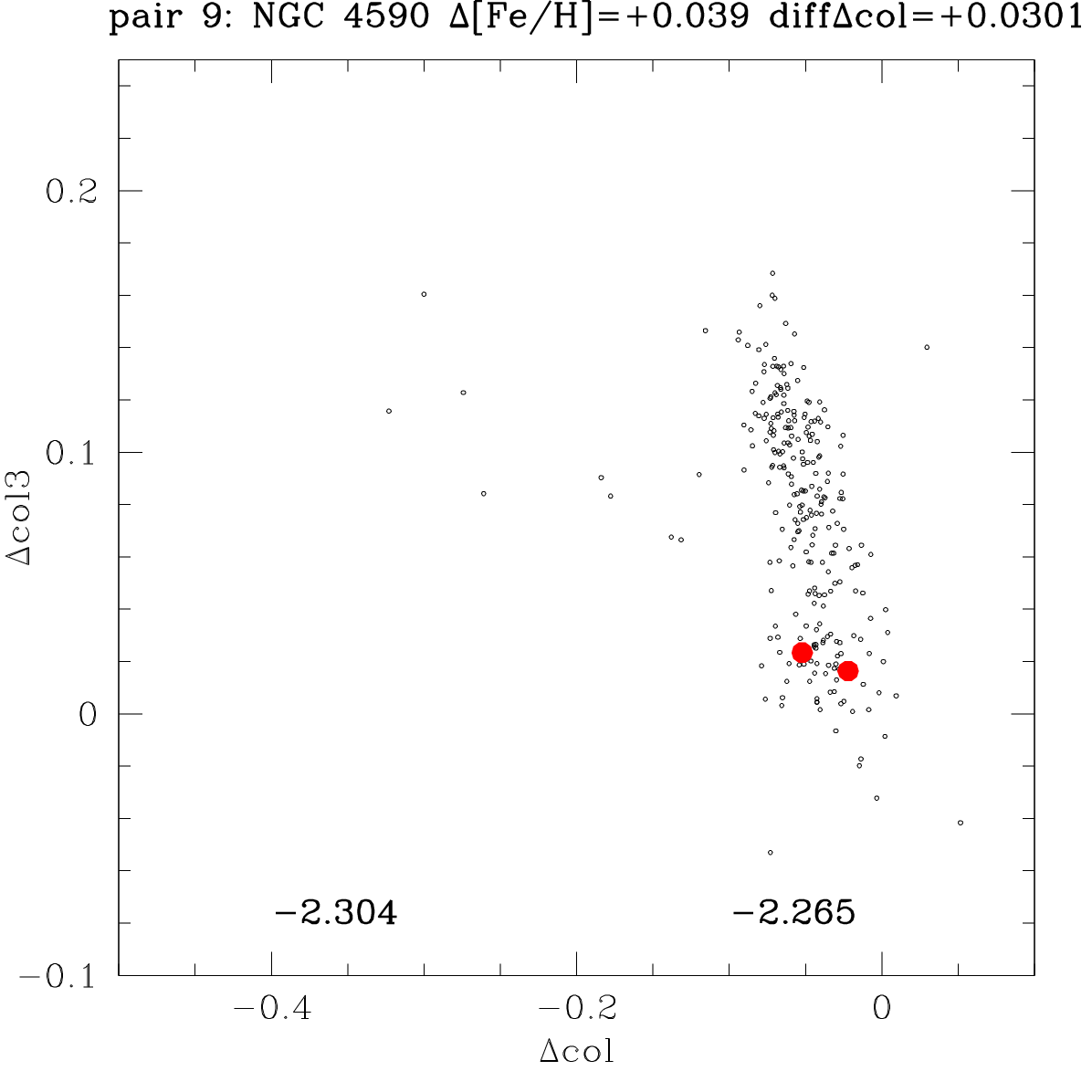}\includegraphics[scale=0.23]{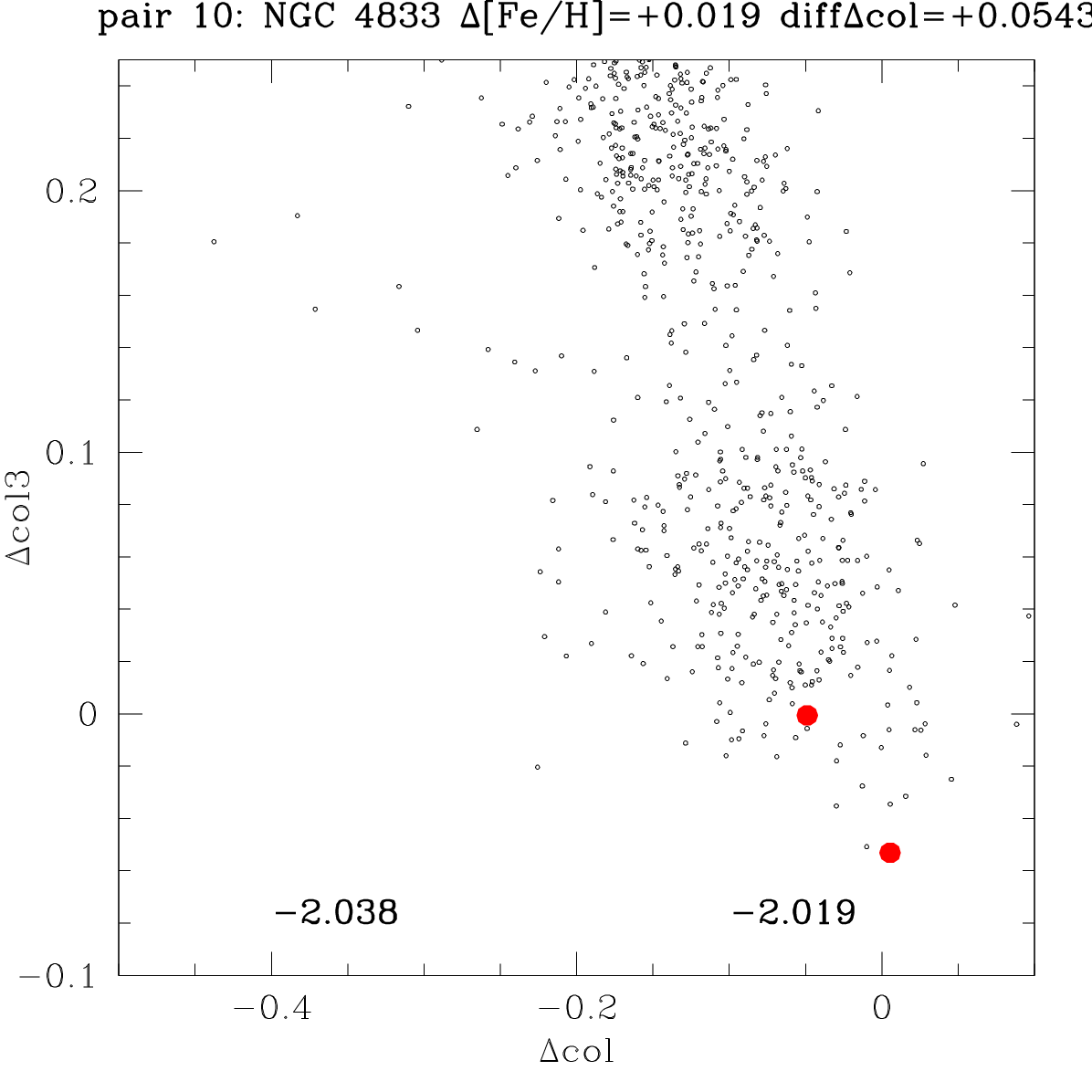}\includegraphics[scale=0.23]{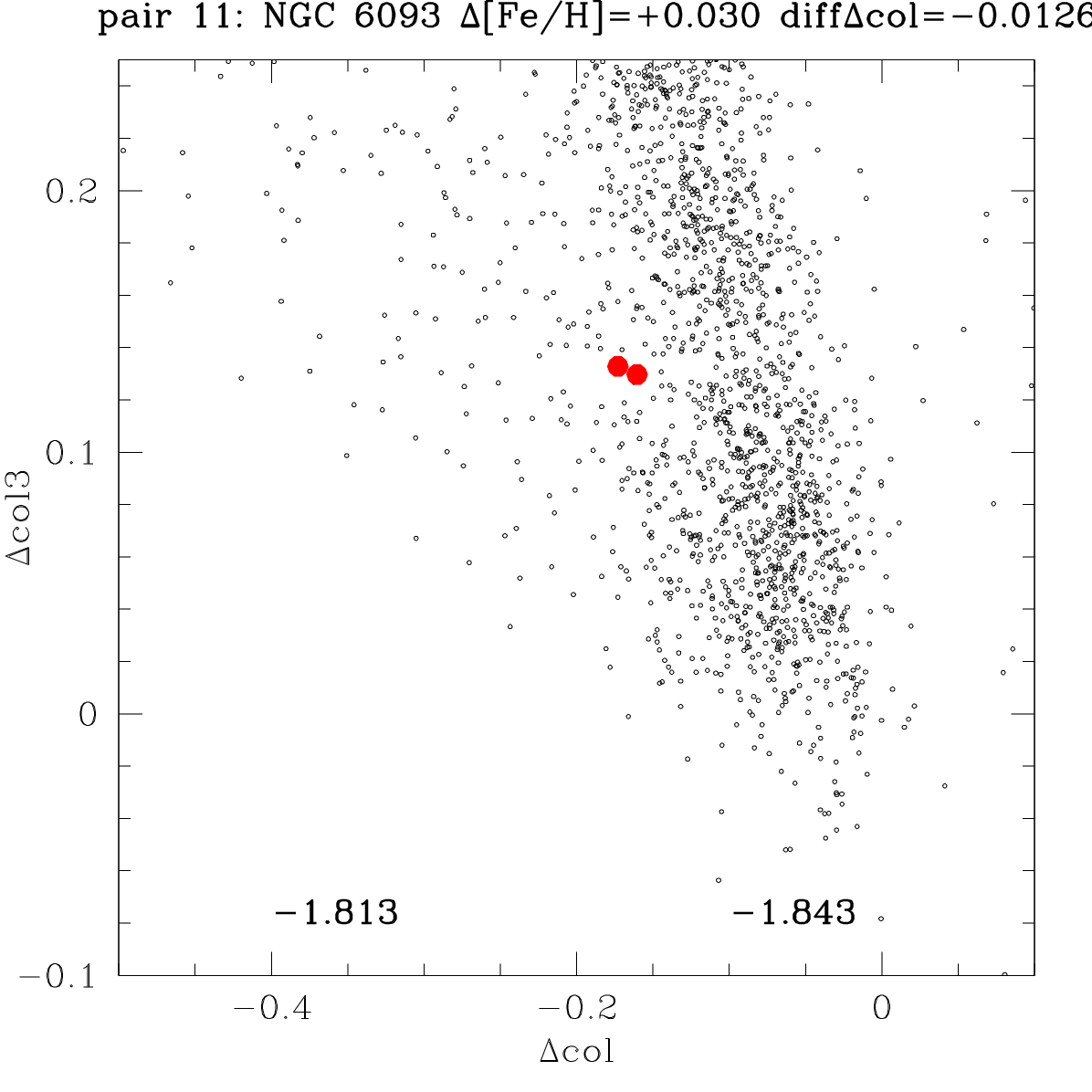}\includegraphics[scale=0.23]{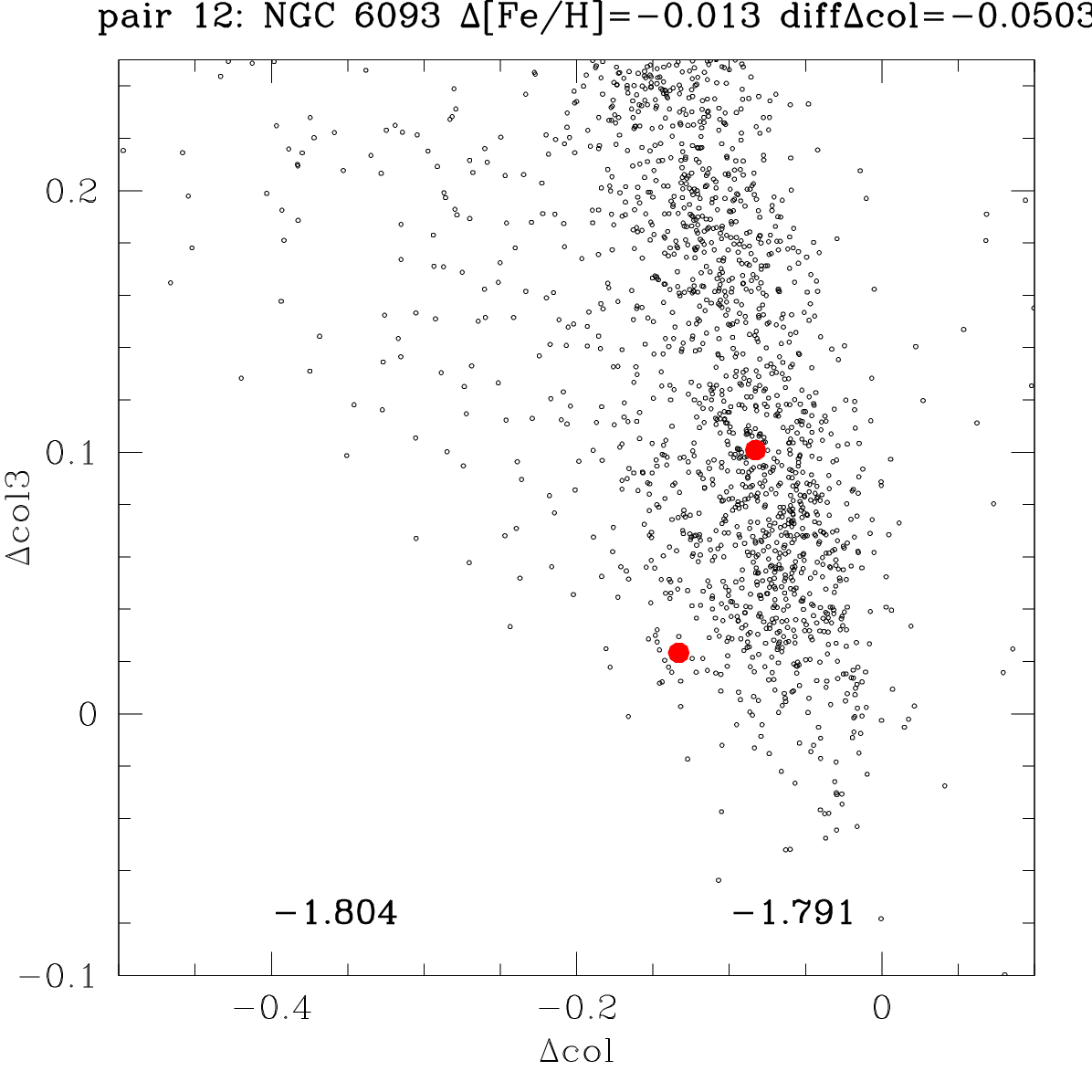}
\includegraphics[scale=0.23]{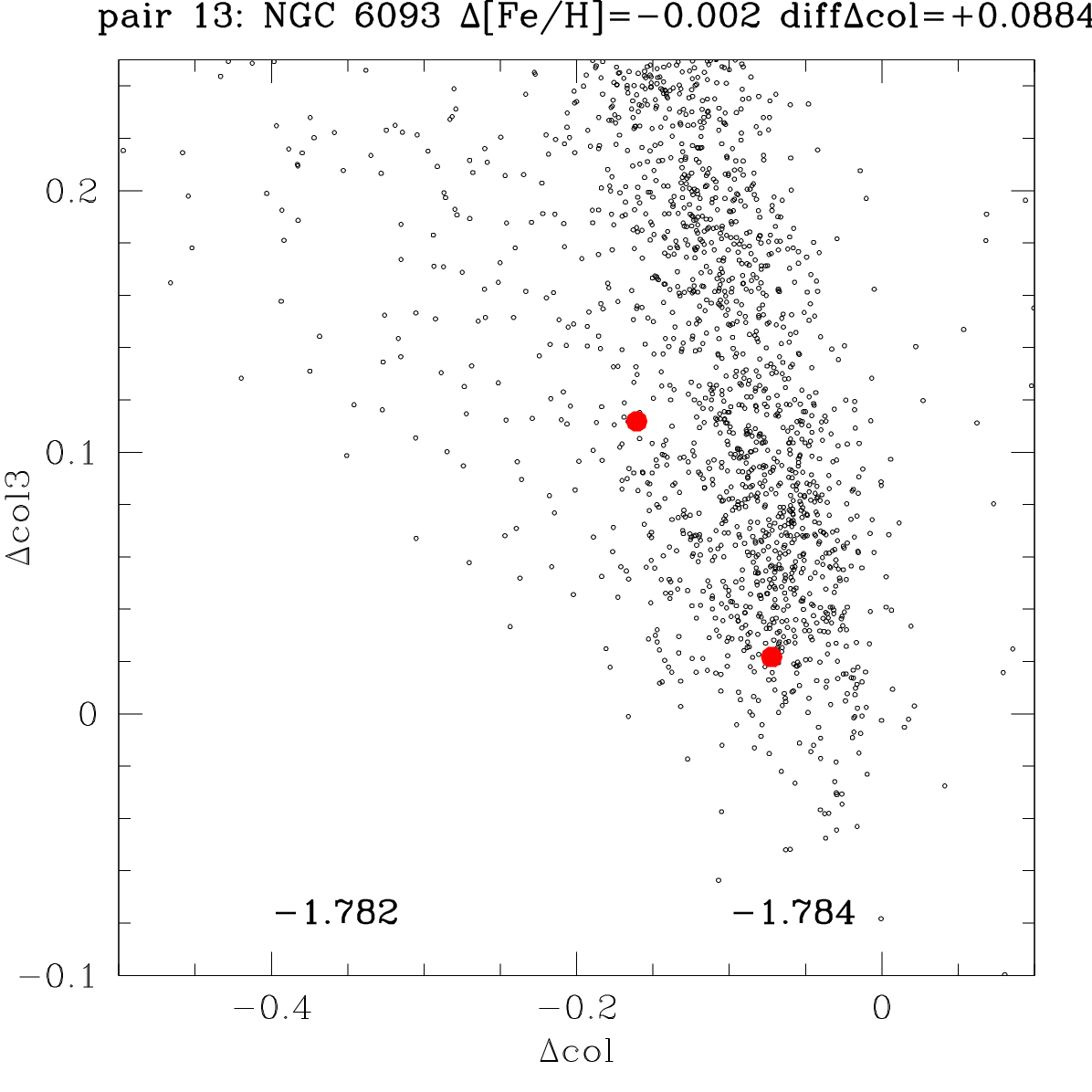}\includegraphics[scale=0.23]{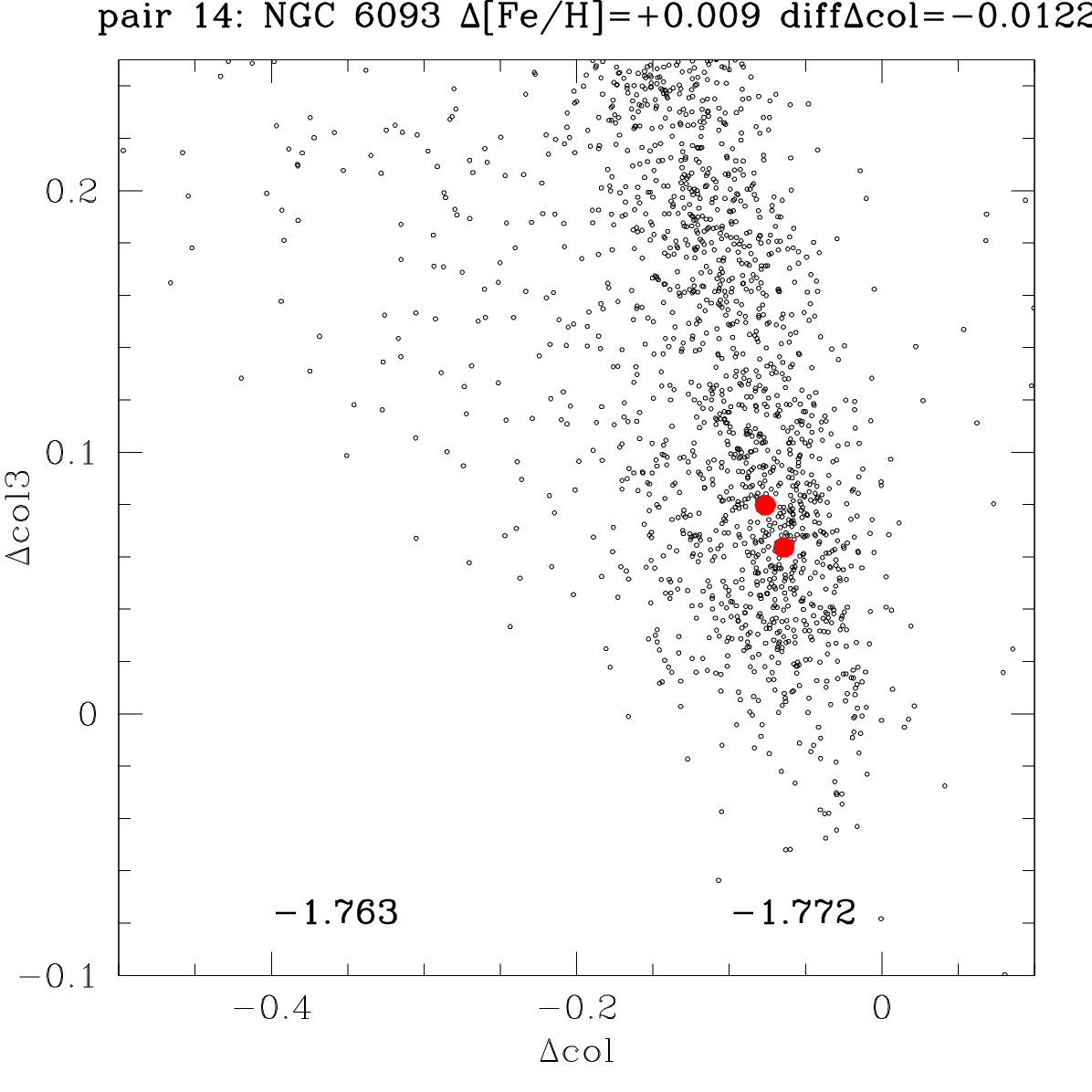}\includegraphics[scale=0.23]{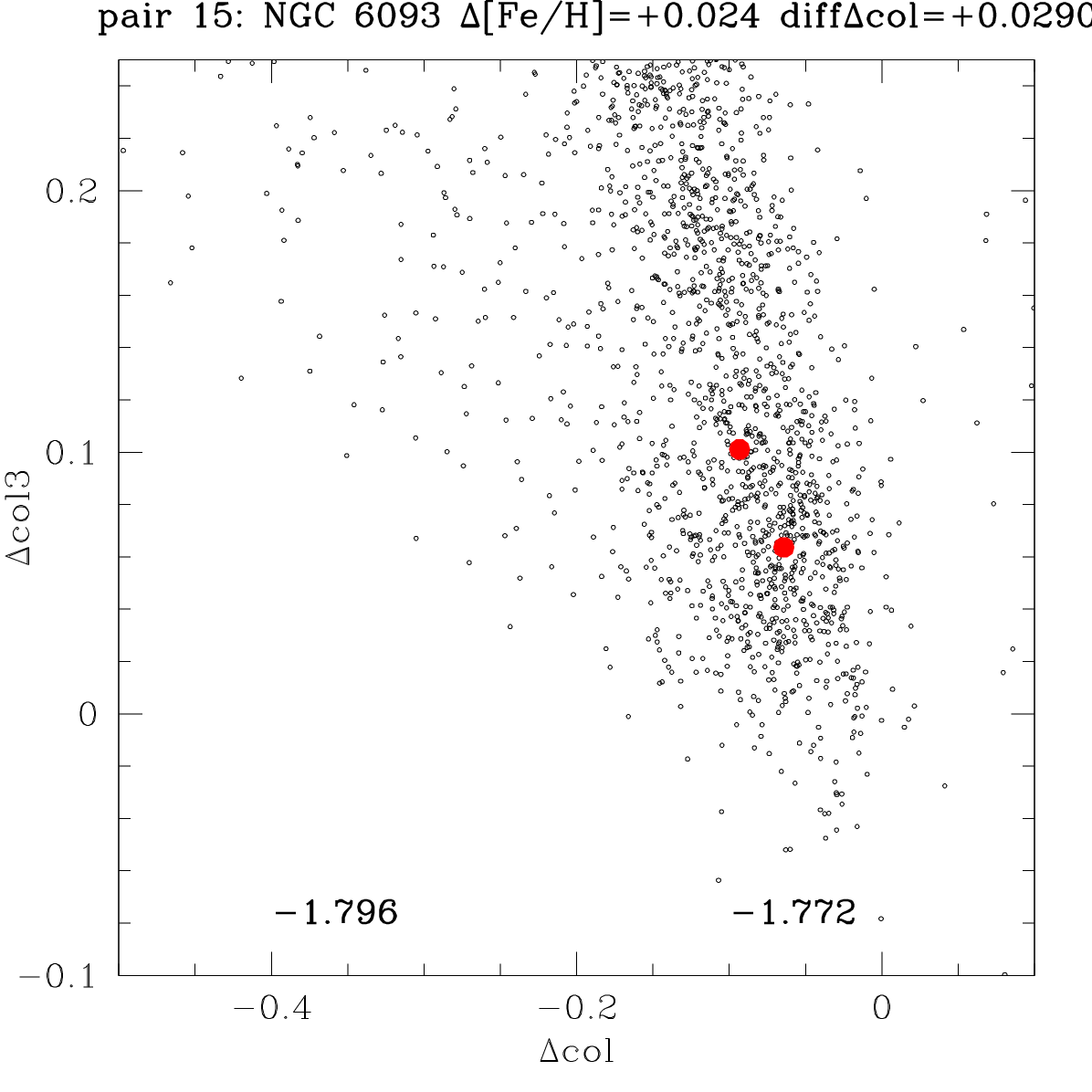}\includegraphics[scale=0.23]{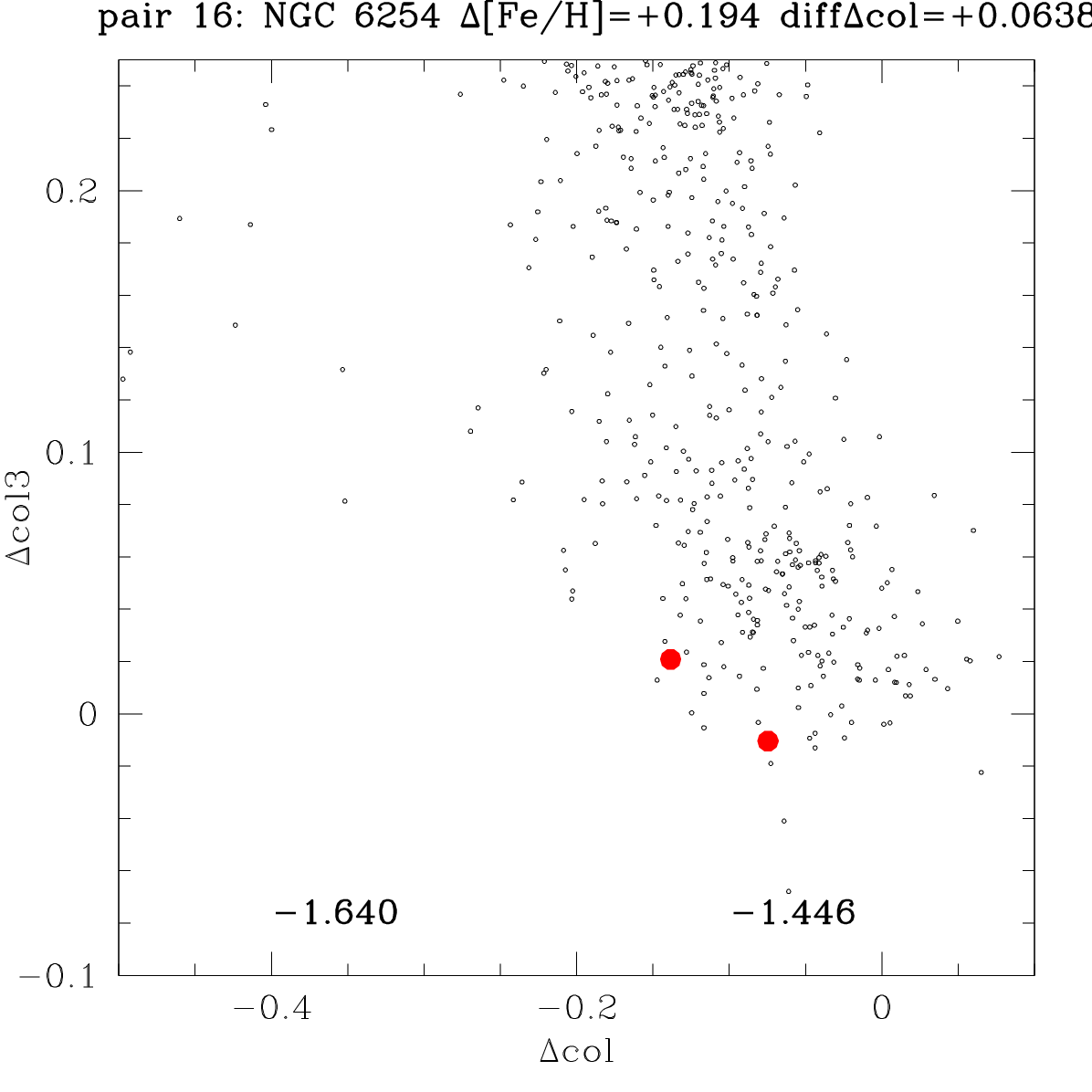}
\caption{Position of pairs of FG stars with similar atmospheric parameters
(large red filled symbols) in NGC~288, NGC~2808, NGC~3201, NGC~4590, NGC~4833,
NGC~6093, and NGC~6254}
\label{f:coppie1}
\end{figure*}

\begin{figure*}
\centering
\includegraphics[scale=0.23]{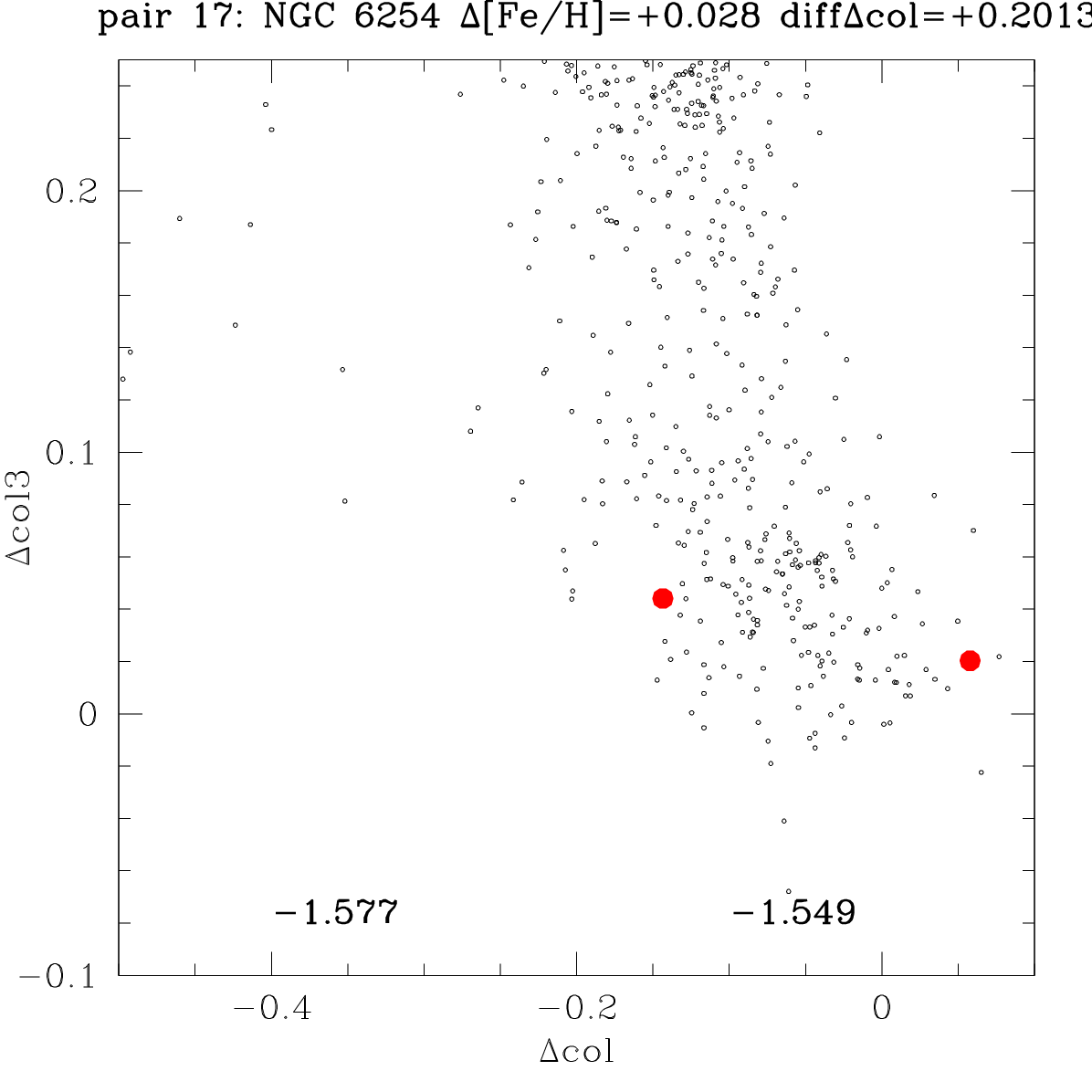}\includegraphics[scale=0.23]{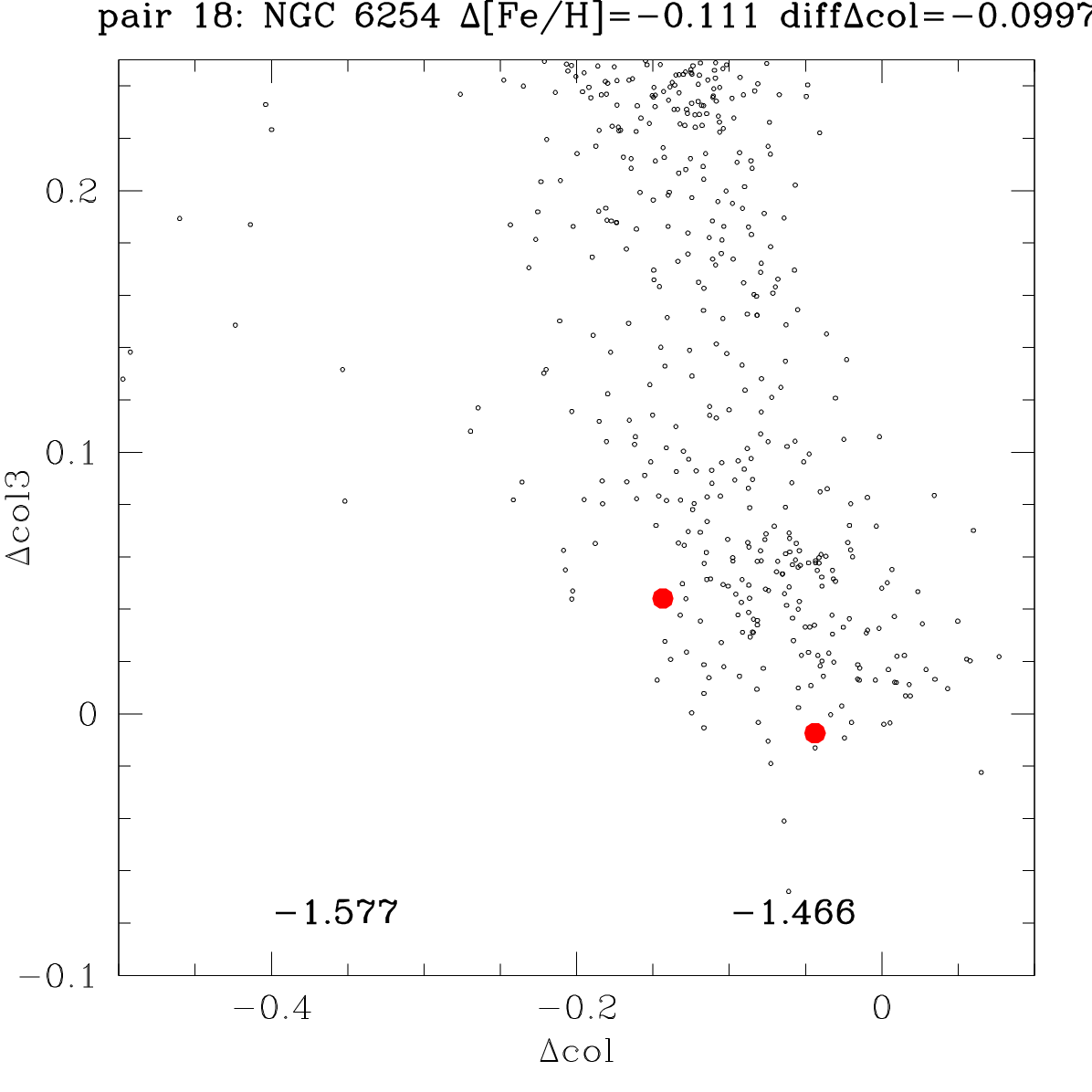}\includegraphics[scale=0.23]{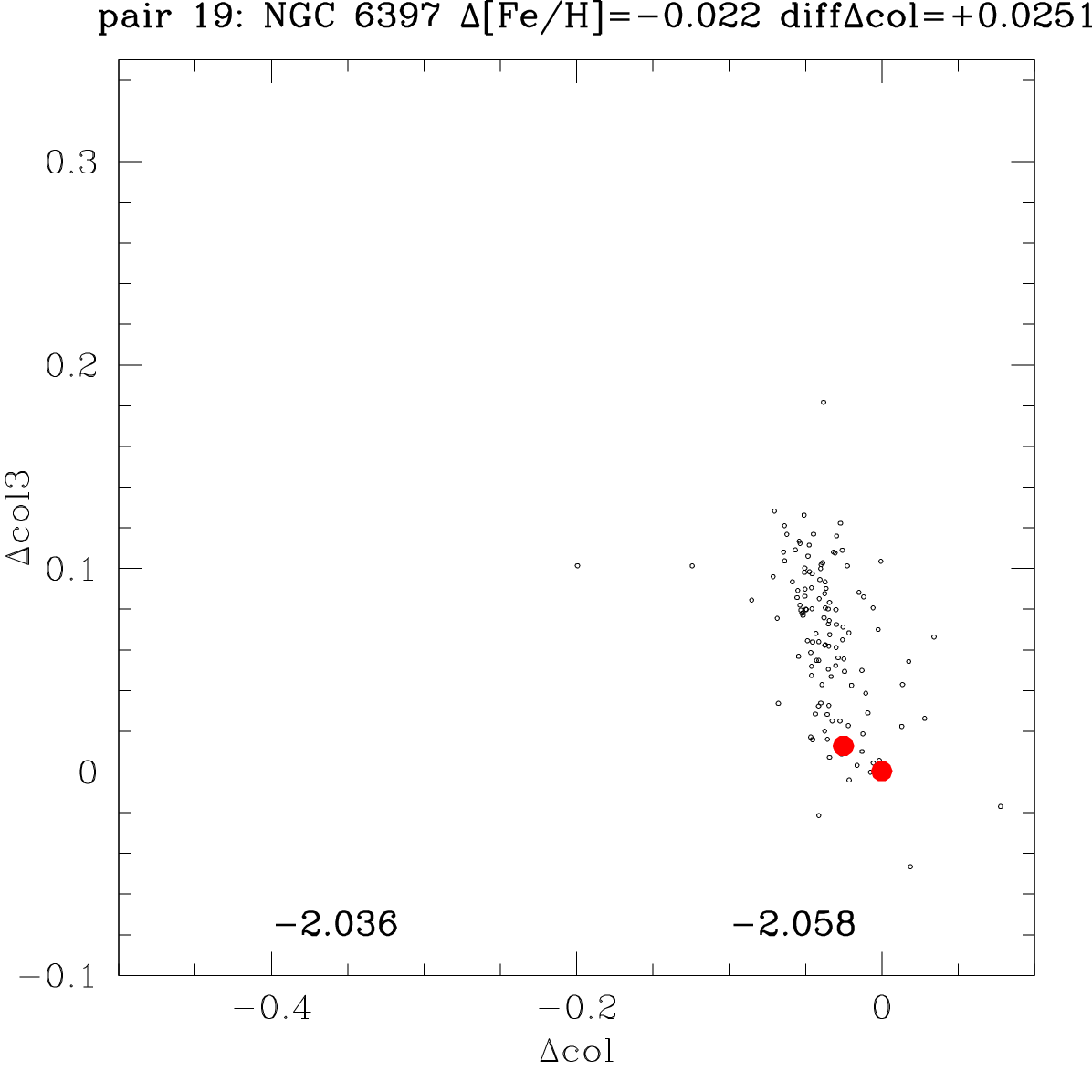}\includegraphics[scale=0.23]{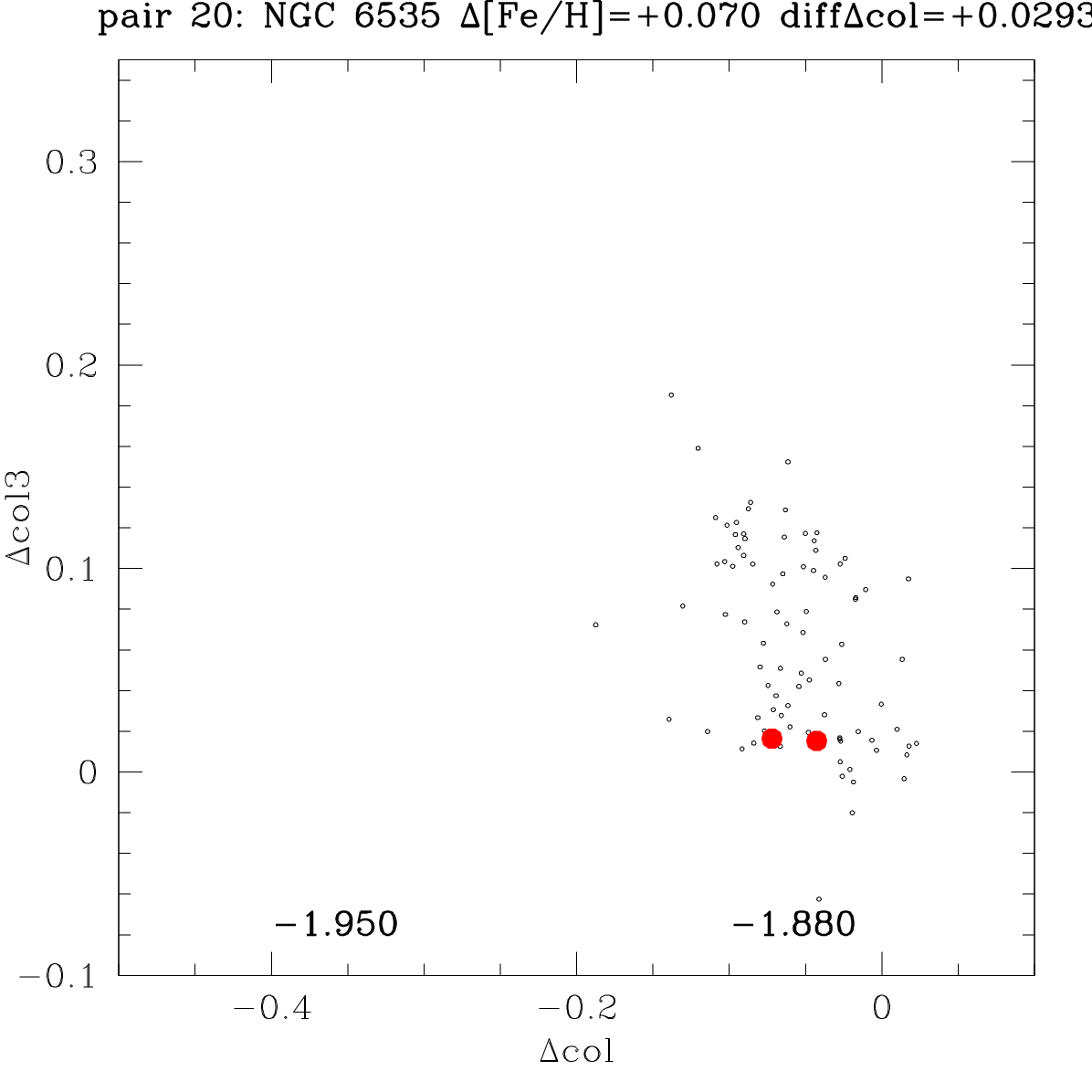}
\includegraphics[scale=0.23]{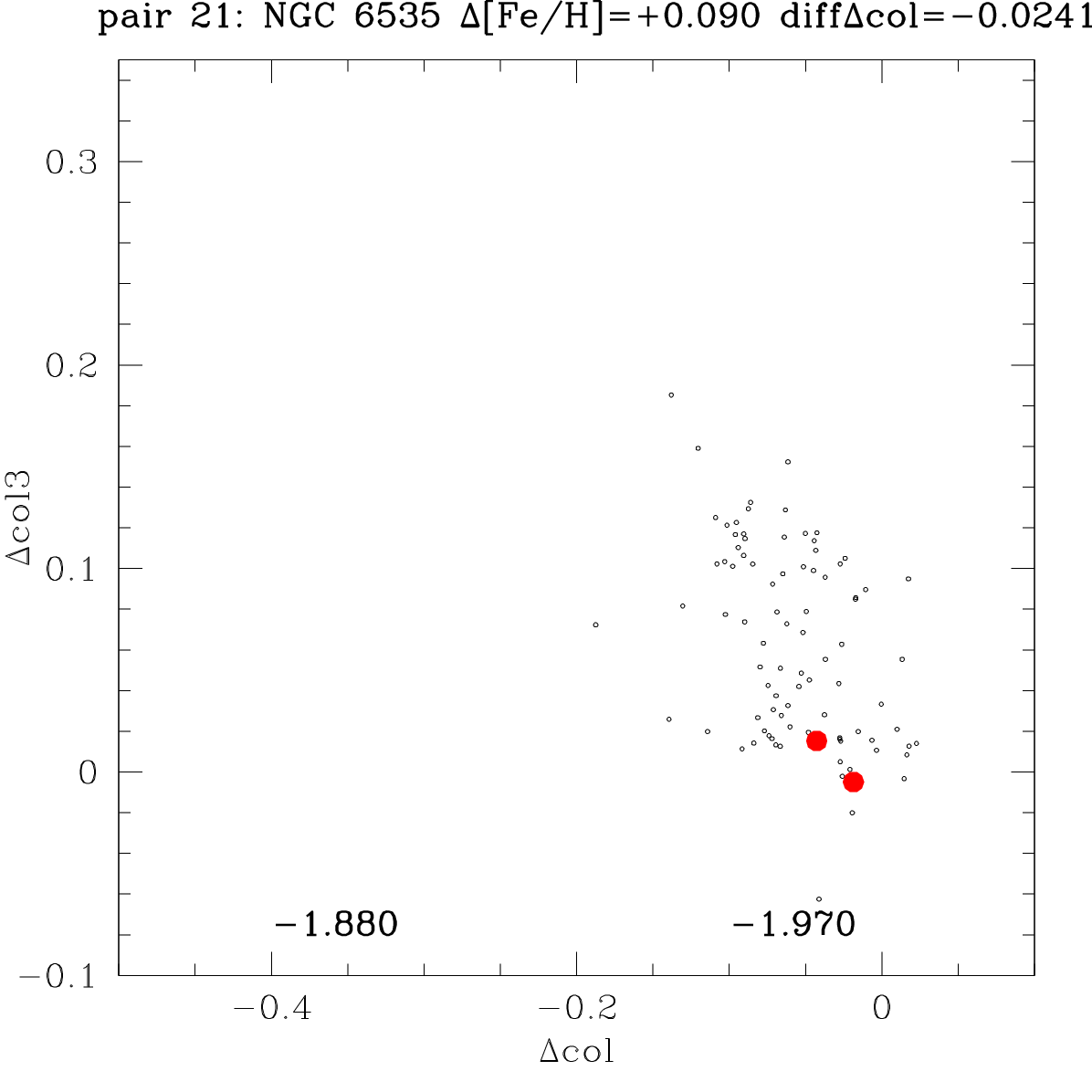}\includegraphics[scale=0.23]{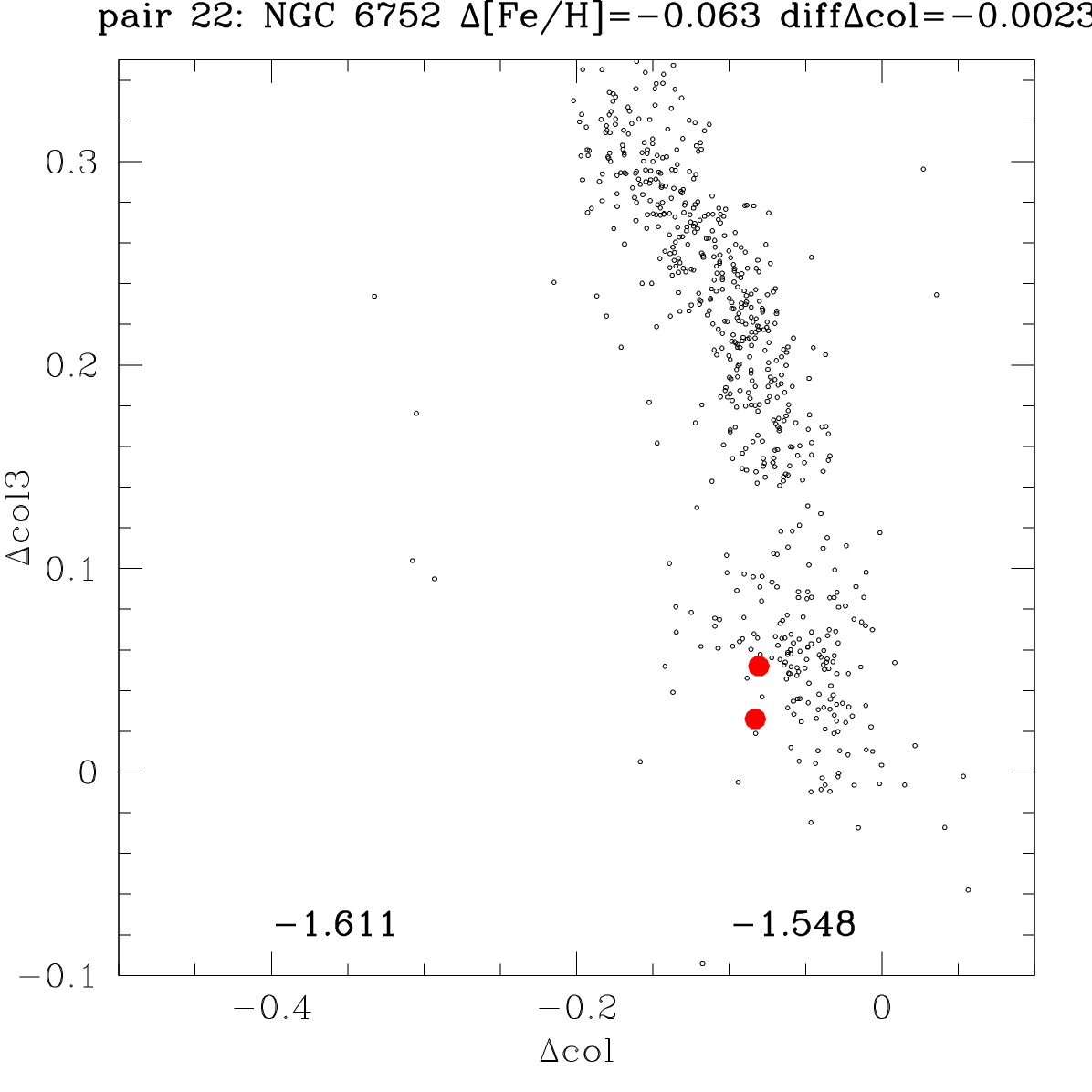}\includegraphics[scale=0.23]{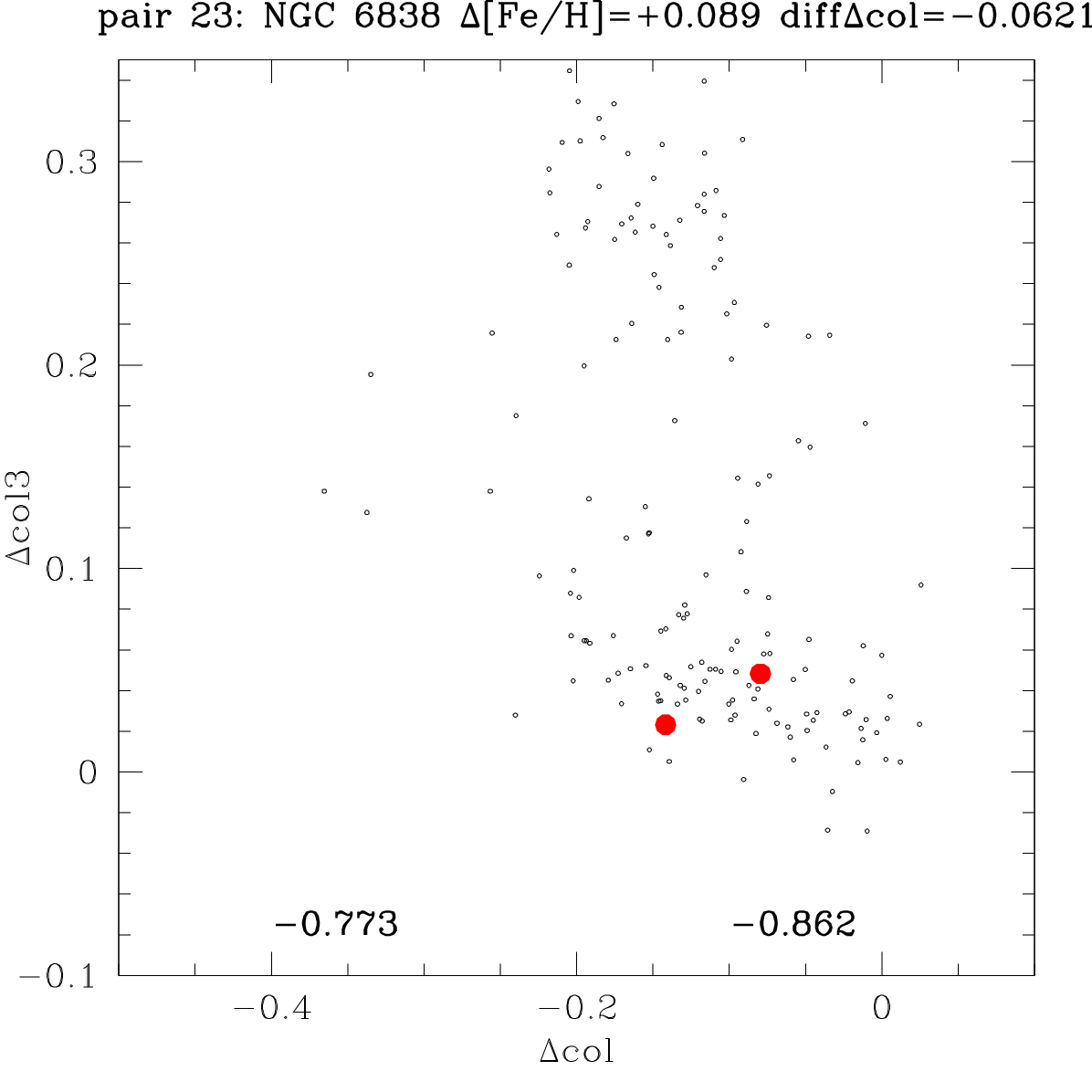}\includegraphics[scale=0.23]{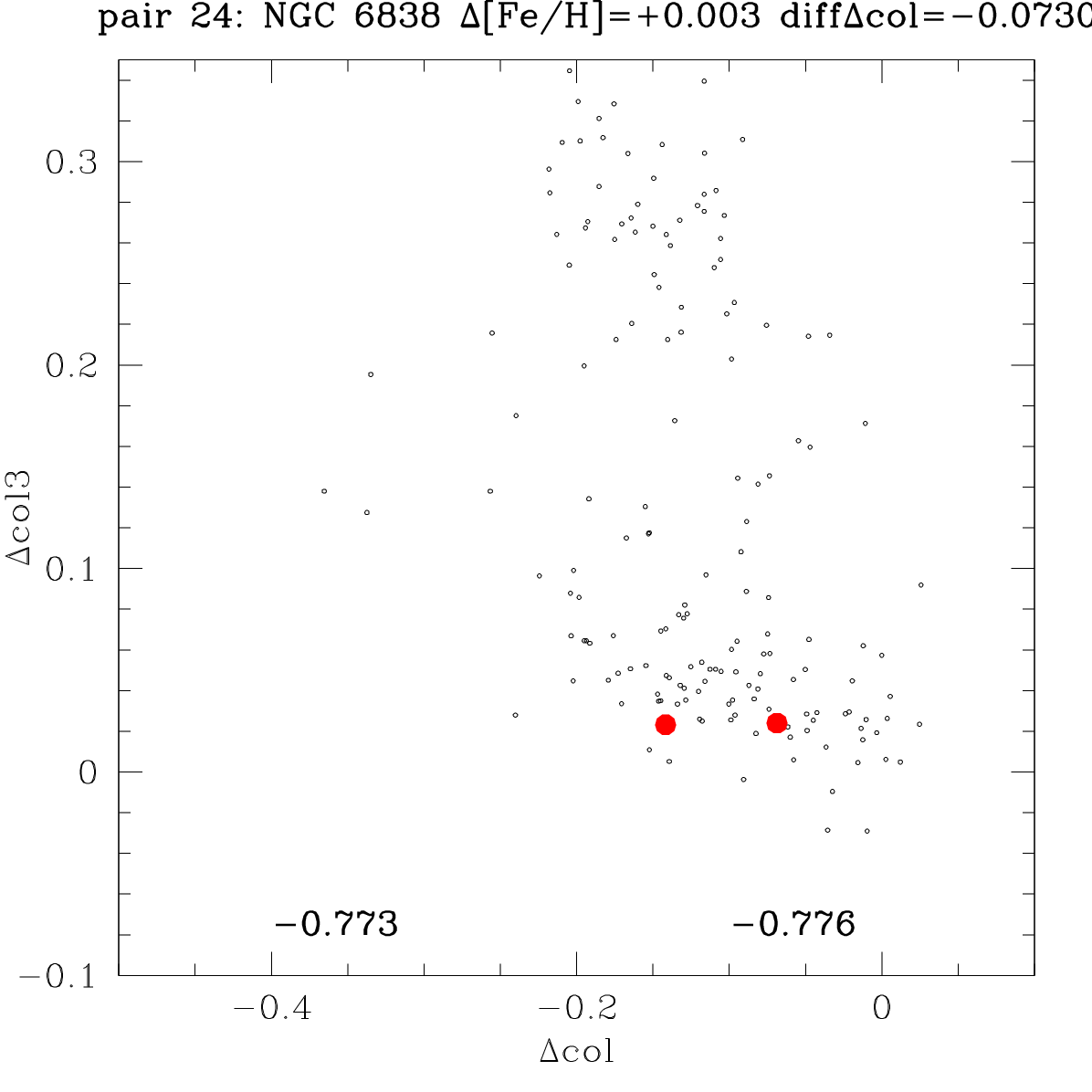}
\includegraphics[scale=0.23]{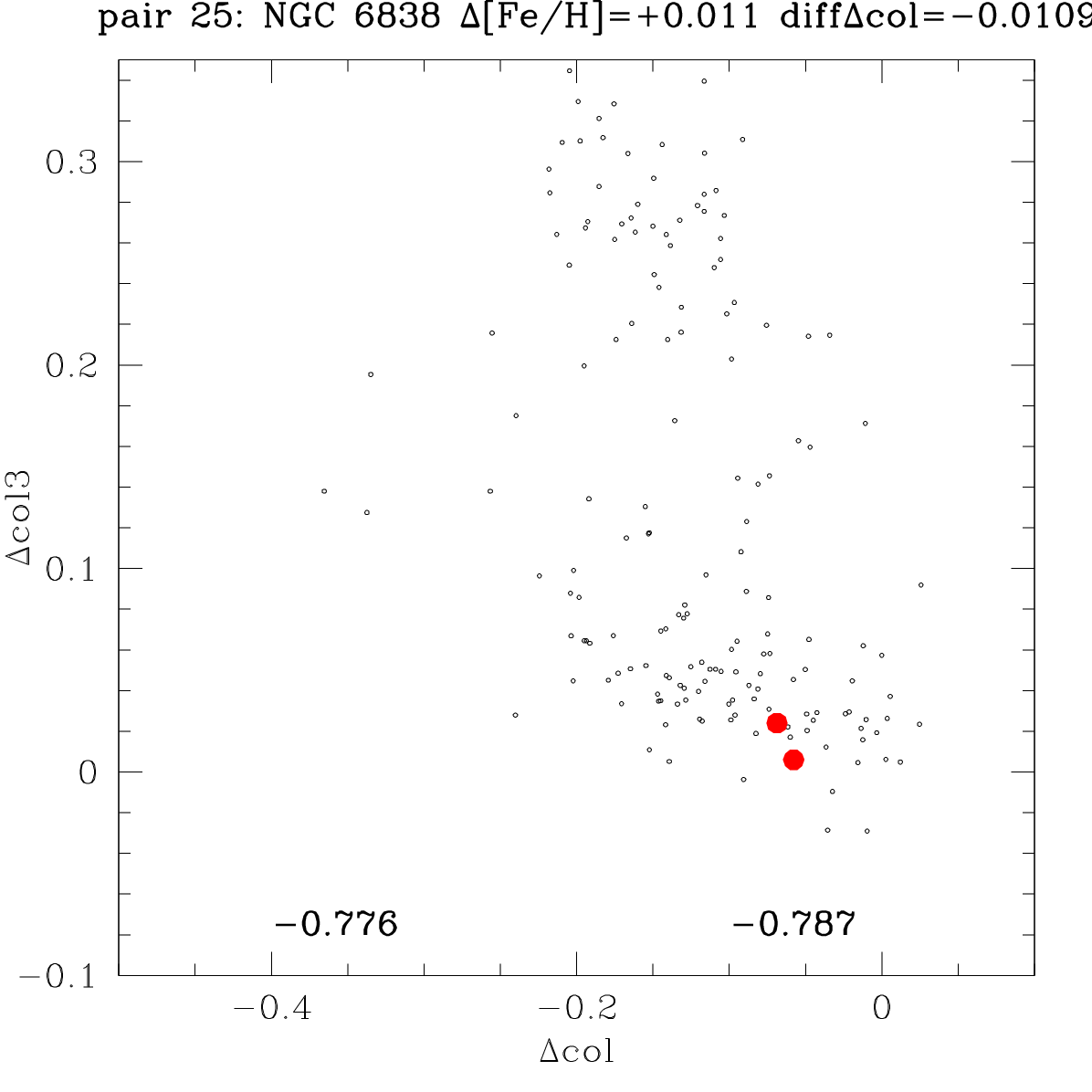}\includegraphics[scale=0.23]{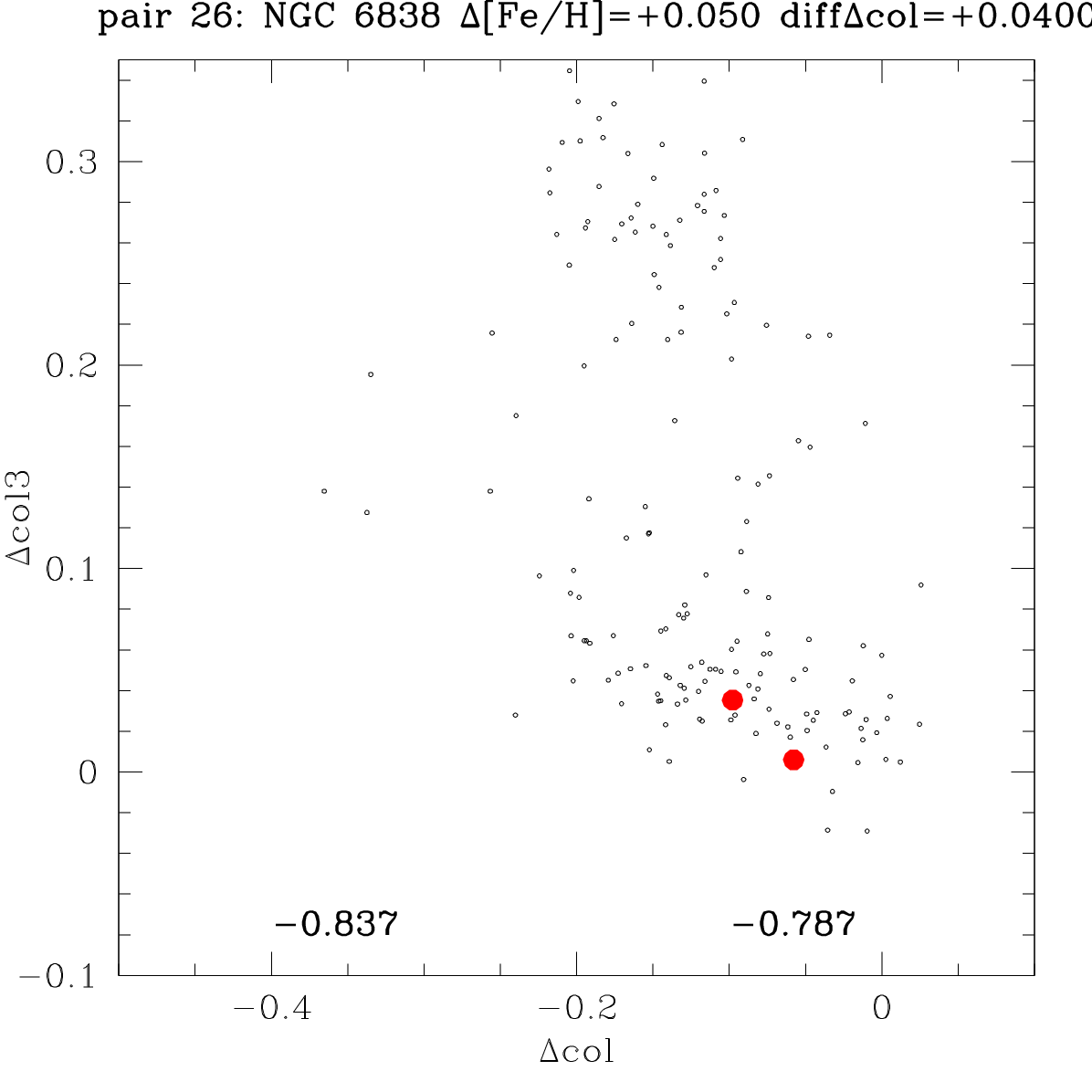}\includegraphics[scale=0.23]{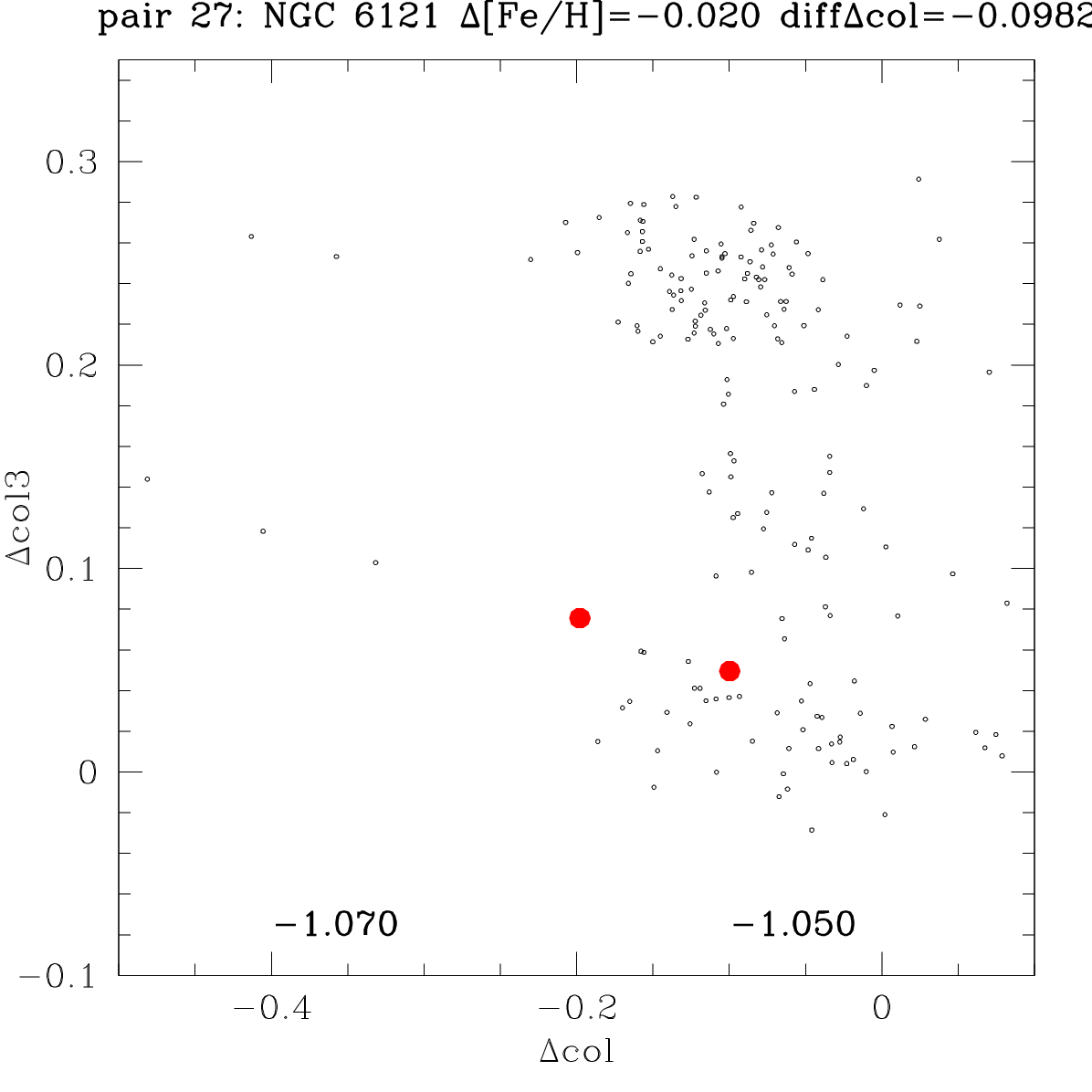}\includegraphics[scale=0.23]{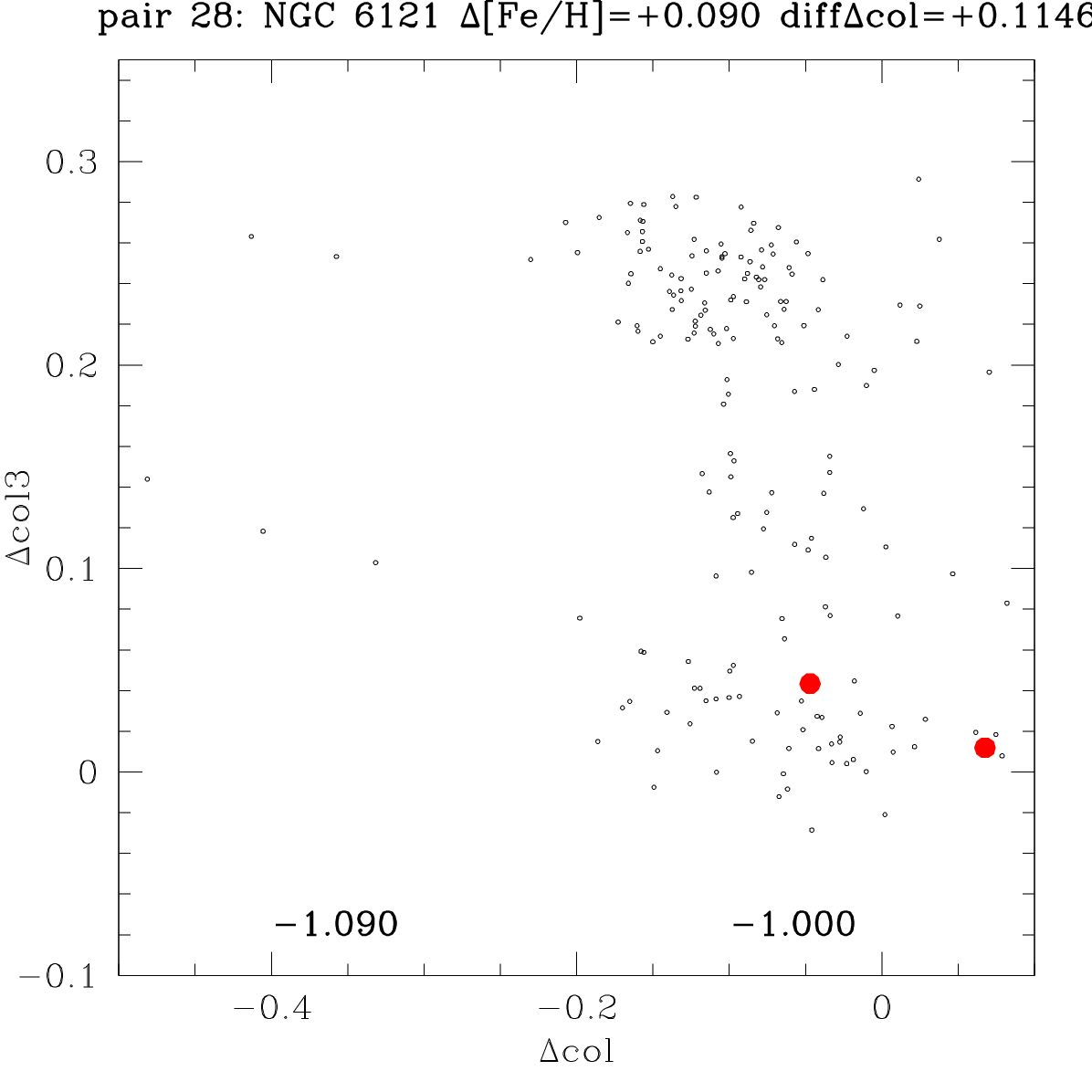}
\includegraphics[scale=0.23]{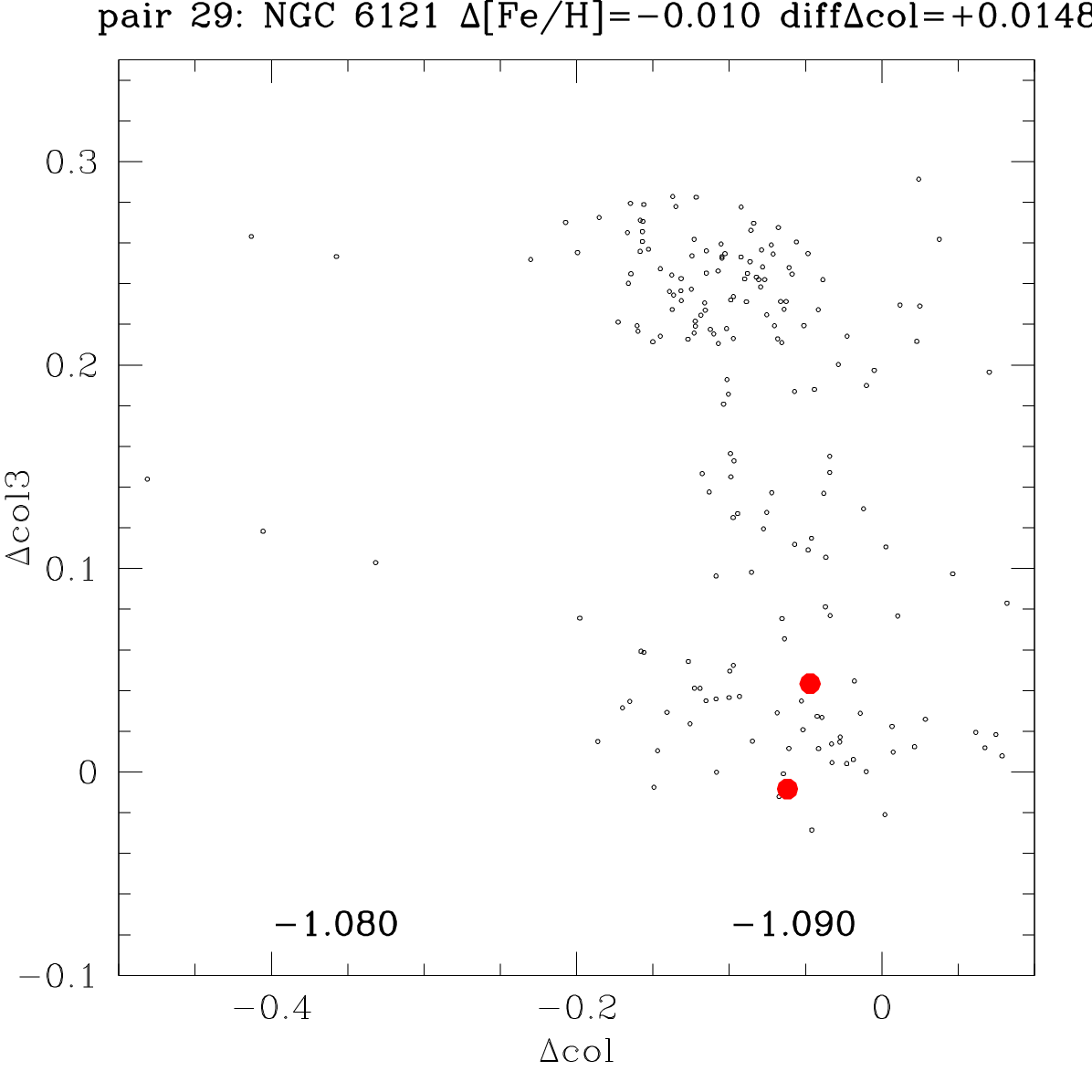}\includegraphics[scale=0.23]{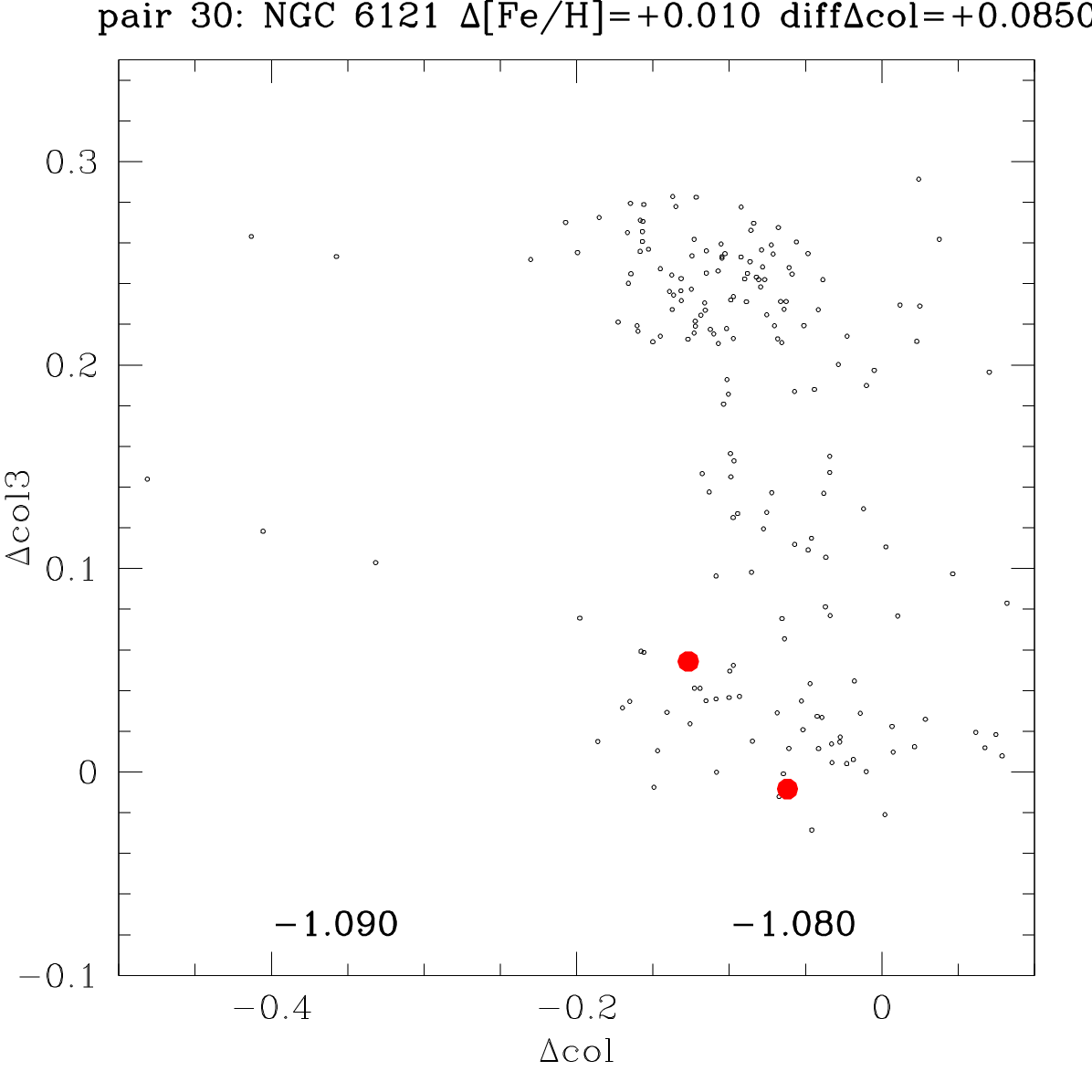}
\caption{Position of pairs of FG stars with similar atmospheric parameters
(red cicles) in NGC~6254, NGC~6397, NGC~6535, NGC~6838, and NGC~6121.}
\label{f:coppie2}
\end{figure*}

A popular technique to investigate possible metallicity spreads among FG stars
is the so called differential abundance analysis of stars with nearly identical
atmospheric parameters, since it permits to have very small uncertainties
associated to the derived abundances (see e.g., Lardo et al. 2023). The precise
abundances can be then related to the position of stars along $RG1$.

To have a better insight on the true link between the extension of $RG1$ and
iron spread we applied a similar method starting with the above large sample of
159 FG  stars. We further culled the sample by rejecting the stars found to
be misclassified as FG by photometry (CB24). Additionally, we excluded four
other stars classified as SG from their Na abundances and not recognized in
CB24. Furthermore, eight stars were conservatively rejected  due to their uncertain
position between $RG1$ and $RG2$. In each GC we then  searched and paired stars
with similar atmospheric parameters. Finally, we ended up with 46 genuine FG stars
in 12 GCs. We combined these stars in 30 pairs with a maximum difference in
effective temperature of 32 K (on average the difference in T$_{\rm eff}$ is 
$-4.6$ K, rms=16 K). The pairs are listed in Table~\ref{t:pairs}. For each paired
up star we report the identification of the pair, the position in the $RG1$ of
the PCM, the values of [Fe/H] and [Na/Fe], the star ID from Nardiello et al.
(2018), where the stellar coordinates can be found, the name of the GC with
its average metallicity, and the atmospheric parameters effective temperature, 
surface gravity, model abundance [A/H], and microturbulent velocity. References
for the abundance analysis are given in the last column. Most of the analysis
are from our homogeneous FLAMES survey (see Carretta et al. 2006), except for
NGC~6121, for which we used an independent source (Marino et al. 2008) as a
safety check.

The similarity of atmospheric parameters means that comparing the stars in each
couple we are doing an extensive differential analysis. If there is any
metallicity spread in any given GC, the assembled dataset of stars with nearly
identical parameters should reveal a correlation between [Fe/H] and position
along the $\Delta col$ in $RG1$.

The results of this exercise are plotted in Fig.~\ref{f:coppie1} and
Fig.~\ref{f:coppie2}. On top of each panel we report the GC name, the absolute
value of the metallicity difference of the two stars (the order of the stars in
each pair is not relevant) and their absolute  difference along the $\Delta col$
coordinate in the $RG1$ of the PCM (we are interested only in mapping the
distance between the stars). Moreover, the metallicities of the stars are
labelled at the bottom of each panel. The [Fe/H] value of the star with the
bluer $\Delta col$ value is always put on the left and the value for the redder
$\Delta col$ star on the right. Stars with very similar parameters may be used
to form more than a single pair.

The first clear evidence from these figures is that the star at bluer
$\Delta col$ is not always the most metal-poor (pairs 2, 5, 7, 11, 13, 14,
19, 21, 23, 24, 25, 29). In these cases, the bluer position is occupied by the 
more metal-rich star in the couple, albeit sometime by tiny amounts in [Fe/H].

Second, stars at the same (or very similar) position in $RG1$ have noticeably
distinct [Fe/H] values (pairs 1, 7, 9, 11, 15, 19, 20, 21, 22, 25). Vice versa,
in some other cases small amounts of difference in [Fe/H] (from 0.01 to less
than 0.05 dex) correspond to large distance in $\Delta col$ (pairs 2, 3, 4, 5,
12, 13, 17, 27, 30).

\begin{figure}
\centering
\includegraphics[scale=0.40]{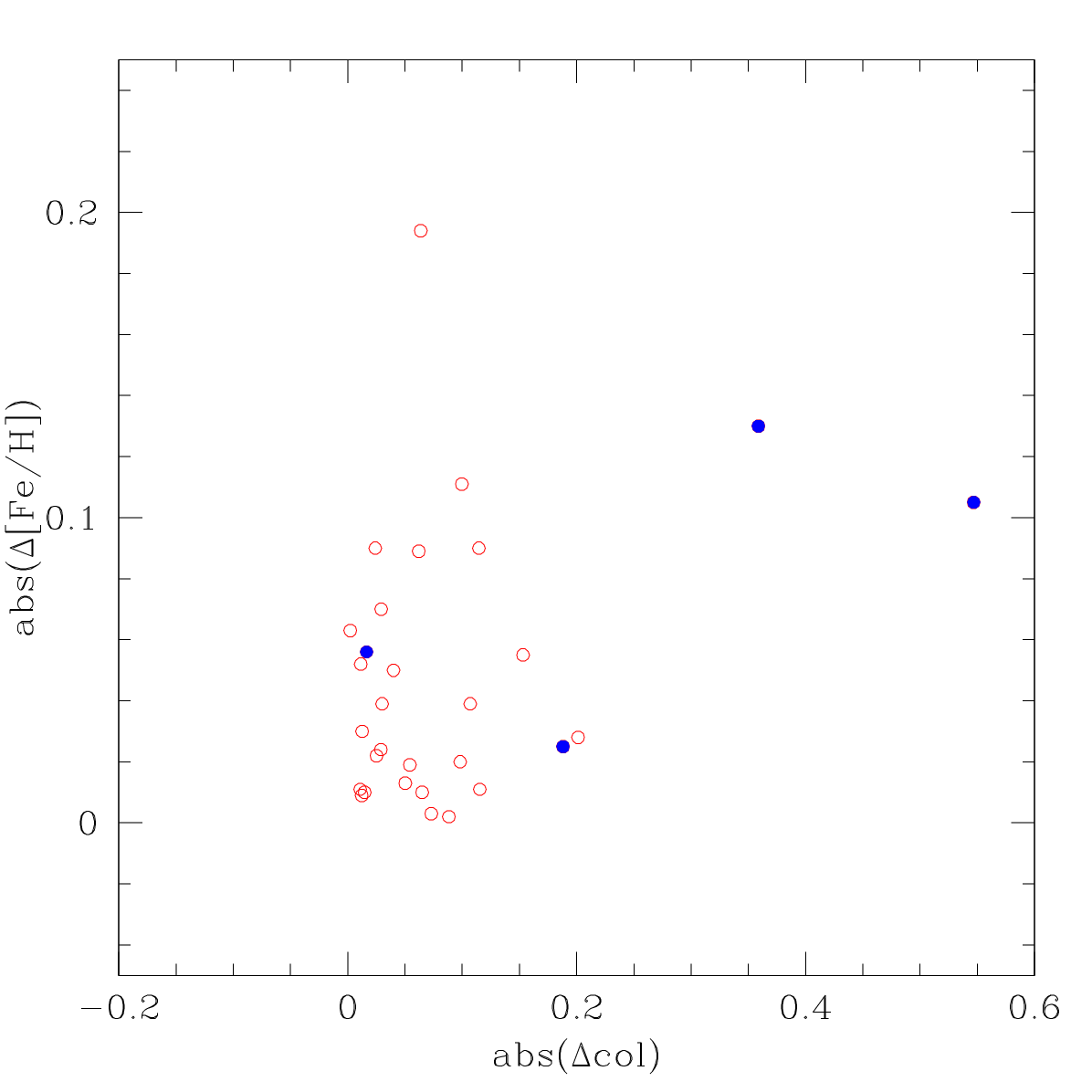}
\caption{Summary of the variations of [Fe/H] as a function of the
$\Delta col$ variation along the $RG1$ region on the PCM from 30 pairs of FG stars
with similar atmospheric parameters. The plotted values are the absolute 
differences in [Fe/H] and $\Delta col$ between the two stars in each pair.
Pairs formed with stars in NGC~3201 are plotted as filled blue circles.}
\label{f:diffcoppie}
\end{figure}

The relation between the displacement in [Fe/H] and in $\Delta col$ is
summarised in Fig.~\ref{f:diffcoppie}. As could be inferred from the above
discussion, there is no correlation or proportionality between these quantities.
Over 30 pairs of stars the mean value of the shift in $\Delta col$ is 0.090 mag
(rms=0.114 mag). To this nearly constant shift does correspond a large range in
metallicities, up to 0.2 dex, between stars in pairs, although the average
difference in [Fe/H] is not significant (0.049 dex, rms=0.045 dex).

Only in NGC~3201 large differences in [Fe/H] seem to be matched by large
differences along the length of the $RG1$. Even in this GC, however, the
apparent correlation is not statistically significant (two-tails probability
p=0.32, 4 data points).

On the other hand, from Fig.~\ref{f:coppie1} and Fig.~\ref{f:coppie2} it is
clear that most of the extended $RG1$ sequence in NGC~2808 is explained by
couple of stars with differences in [Fe/H] of 0.04, 0.01, and 0.05 dex only 
(pairs 2,3,4, at odds with the results by Lardo et al. 2023), whereas L22 claim
an unlikely large iron spread of 0.110 dex to explain the length of $RG1$ in
this GC.

The conclusion from these extensive differential analysis is that there seems to
be no simple or direct correspondence between metallicity spread and extension
of the $RG1$ region in the PCMs.

\section{Quantifying the extension of the $RG1$ sequences in the PCM}

Despite the results from the previous Section, a different location of stars in
$\Delta col$ on the $RG1$ in the PCMs according to the metallicity of the star
should be expected.  The colour $col=m_{F275W}-m_{F814W}$ (and, differentially,
the pseudo-colour  $\Delta col$) with its long baseline is sensitive mostly to
the effective temperature of stars (see e.g., CB24). Hence the spread in 
$\Delta col$ suggested the inference made by M17 and L22
of a metallicity spread, since the He content of a FG star is the one
established by the Big Bang nucleosynthesis, by definition.

For a given age, metal-poor stars are located on the blue side of the RGB in the
CMD, metal rich stars lie on the red side. In the same way, also when building a
PCM stars with lower metallicity are expected at bluer (more negative)  
$\Delta col$ values whereas metal rich stars can be expected to be at redder
(less negative) values of $\Delta col$ along $RG1$.
However, neither L22 nor M17 provide a precise quantity to express what they
mean with the statement that the extension of $RG1$ is too large with respect to
the errors. No precise measurements of this extent are given or published.

\begin{figure}
\centering
\includegraphics[scale=0.40]{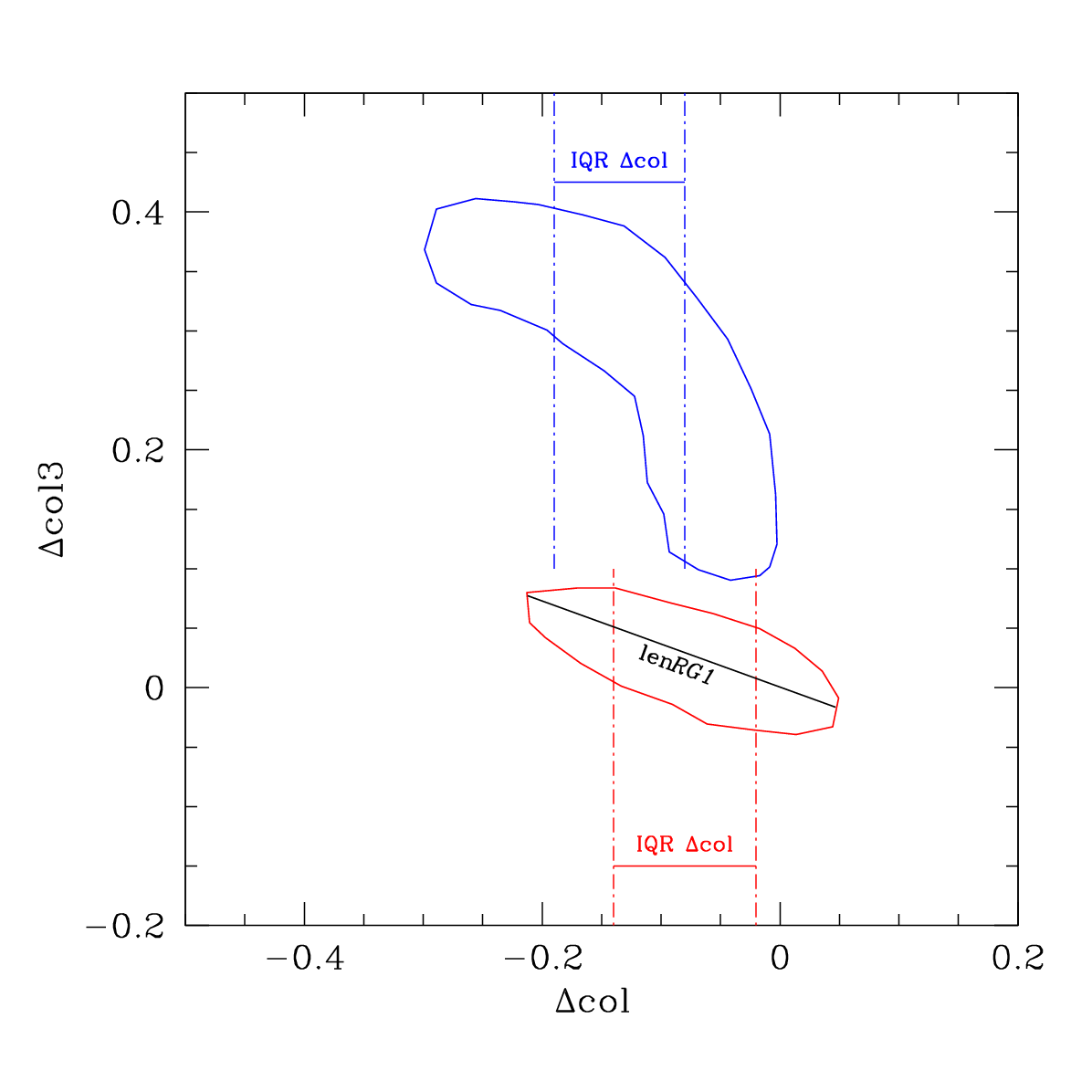}
\caption{Schematic plot of a PCM, with $RG1$ and $RG2$ indicated in red and blue,
respectively. The lines and segments show the different measures of their
extension in $\Delta col$, and along the elongated region of FG stars
enclosed by the ellipse (see text for a description).}
\label{f:sketch}
\end{figure}

We fix this shortcoming by providing for the first time the extension of $RG1$
in the PCMs of the GCs published in CB24. To quantify the sequence $RG1$ we used three kind
of indicators (see the sketch in Fig.~\ref{f:sketch}): 1) the interquartile
range iqrFG in  $\Delta col$ of the FG stars, to minimize the uncertainties
introduced by outliers; 2) the rms relative to the mean
$\Delta col$ value of the FG stars, rmsFG; and 3) the major axis of the ellipse
that best encloses the $RG1$ sequence, len$RG1$. 

\begin{table}
\centering
\caption{Extensions and spreads in $\Delta col$ for $RG1$ and $RG2$}
\setlength{\tabcolsep}{1.5mm}
\begin{tabular}{rccccccc}
\hline
GC  & len$RG1$ &rmsFG&  iqrFG &rmsSG & iqrSG & b/a & ang \\
\hline

 104 & 0.290 & 0.056 &  0.079 &0.054 &  0.071 &  0.60 & 34.99	\\
 288 & 0.142 & 0.061 &  0.037 &0.023 &  0.029 &  0.50 & 50.00	\\
 362 & 0.160 & 0.034 &  0.041 &0.047 &  0.056 &  0.50 & 28.96	\\
1851 & 0.186 & 0.038 &  0.049 &0.048 &  0.059 &  0.50 & 30.00	\\
2808 & 0.394 & 0.079 &  0.098 &0.113 &  0.175 &  0.35 & 16.38	\\
3201 & 0.356 & 0.084 &  0.103 &0.062 &  0.090 &  0.30 & 15.67	\\
4590 & 0.084 & 0.020 &  0.023 &0.019 &  0.026 &  0.35 & 20.00	\\
4833 & 0.244 & 0.053 &  0.065 &0.042 &  0.058 &  0.60 & 25.00	\\
5904 & 0.206 & 0.047 &  0.053 &0.046 &  0.062 &  0.35 & 30.00	\\
6093 & 0.222 & 0.045 &  0.053 &0.050 &  0.067 &  0.70 & 20.00	\\
6121 & 0.246 & 0.073 &  0.097 &0.056 &  0.064 &  0.30 &  7.51	\\
6205 & 0.166 & 0.036 &  0.043 &0.051 &  0.074 &  0.55 & 18.93	\\
6254 & 0.286 & 0.063 &  0.077 &0.044 &  0.061 &  0.45 & 28.00	\\
6397 & 0.058 & 0.026 &  0.025 &0.015 &  0.020 &  0.50 & 30.02	\\
6535 & 0.120 & 0.034 &  0.052 &0.036 &  0.051 &  0.55 & 19.99	\\
6752 & 0.138 & 0.034 &  0.035 &0.039 &  0.060 &  0.75 & 30.00	\\
6809 & 0.084 & 0.020 &  0.033 &0.023 &  0.032 &  0.70 & 40.00	\\
6838 & 0.300 & 0.062 &  0.095 &0.057 &  0.080 &  0.40 & 20.00	\\
7078 & 0.230 & 0.047 &  0.063 &0.056 &  0.059 &  0.30 &  9.49	\\
7099 & 0.088 & 0.021 &  0.027 &0.026 &  0.036 &  0.50 & 20.00	\\

\hline
\end{tabular}
\label{t:len}
\end{table}

More in detail, first we isolated FG and SG stars for each GC on the 
corresponding PCM. Conservatively, we dropped from the CB24 sample NGC~6388 and
NGC~6715 whose uncertainties due to the high reddening and to field
contamination may hamper somehow the results.
The measurements for all the other 20 GCs are listed in Table~\ref{t:len}.

The selection of stars to be assigned to $RG1$ and $RG2$ is made interactively
using TOPCAT (Taylor 2005); for most cases the selection, simply done 
visually, is straightforward, while in other  cases there is some ambiguity.
This is true also for the PCMs computed by M17, as clearly shown in their
plots.  We verified that these groups are properly located in the
$m_{F814W}-col$ and $m_{F814W}-col3$  CMD in the position where FG and SG stars
lie. Our selection is shown for all GCs in the figures in Appendix A, where FG
and SG stars are plotted in red and blue, respectively, whereas stars with an
uncertain attribution are in grey. Stars in this latter group are always a 
minority. In the worst case (NGC~2808) they are less than 15\% and they do not
alter our results, being clearly located slightly outside the tight sequences of
the RGB.

Once we have the samples of FG stars, we do a linear interpolation so we can 
measure the inclination of the FG distribution in the PCM. This is also the
inclination of the ellipse that best encloses the FG stars, apart from a few
slight adjustments made by eye. The values of the position angles are measured
clockwise starting from the x-axis of the PCM and listed in Table~\ref{t:len}.

\begin{figure}
\centering
\includegraphics[scale=0.40]{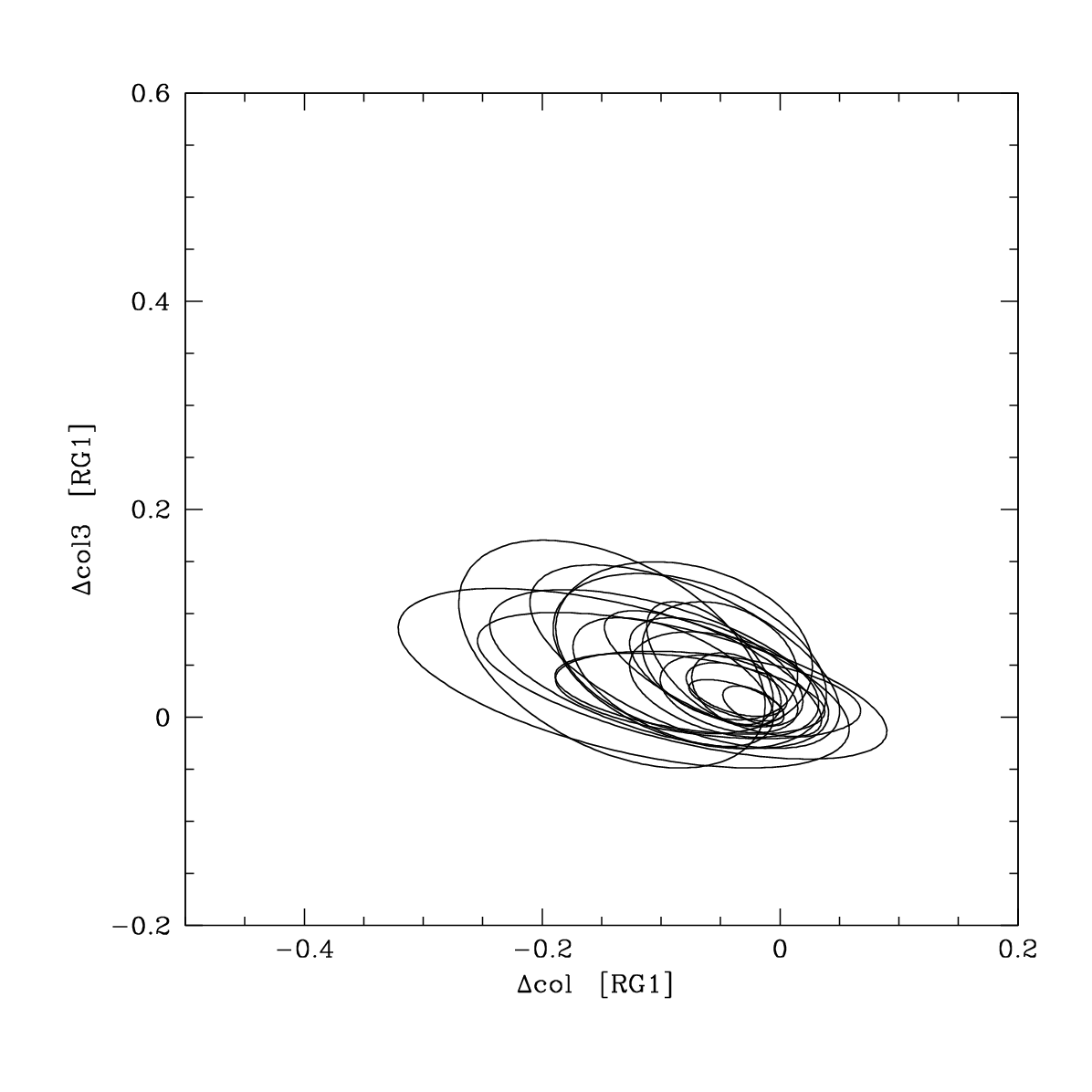}
\caption{Ellipses representing the $RG1$ region in all GCs considered here. The
major axis length, the axis ratio, and the inclination were derived as described
in the text.}
\label{f:ellipse}
\end{figure}

The orientation of the $RG1$ region is similar, but not equal among GCs (see
Fig.~\ref{f:ellipse}).  This is in contrast with the procedure by M17, who
adopted a line with a fixed inclination (an angle of  18 degrees with respect to
the x-axis in the PCM) to separate FG and SG stars.  Although the average
inclination in our sample of GCs (24.75 degrees, with an rms=10.11) seems not
very different from the value used by M17, the individual $RG1$ span a range
from the 7 degrees in NGC~6121 up to the steep 50 degrees of the FG stars
in NGC~288. 

\begin{figure*}
\centering
\includegraphics[scale=0.30]{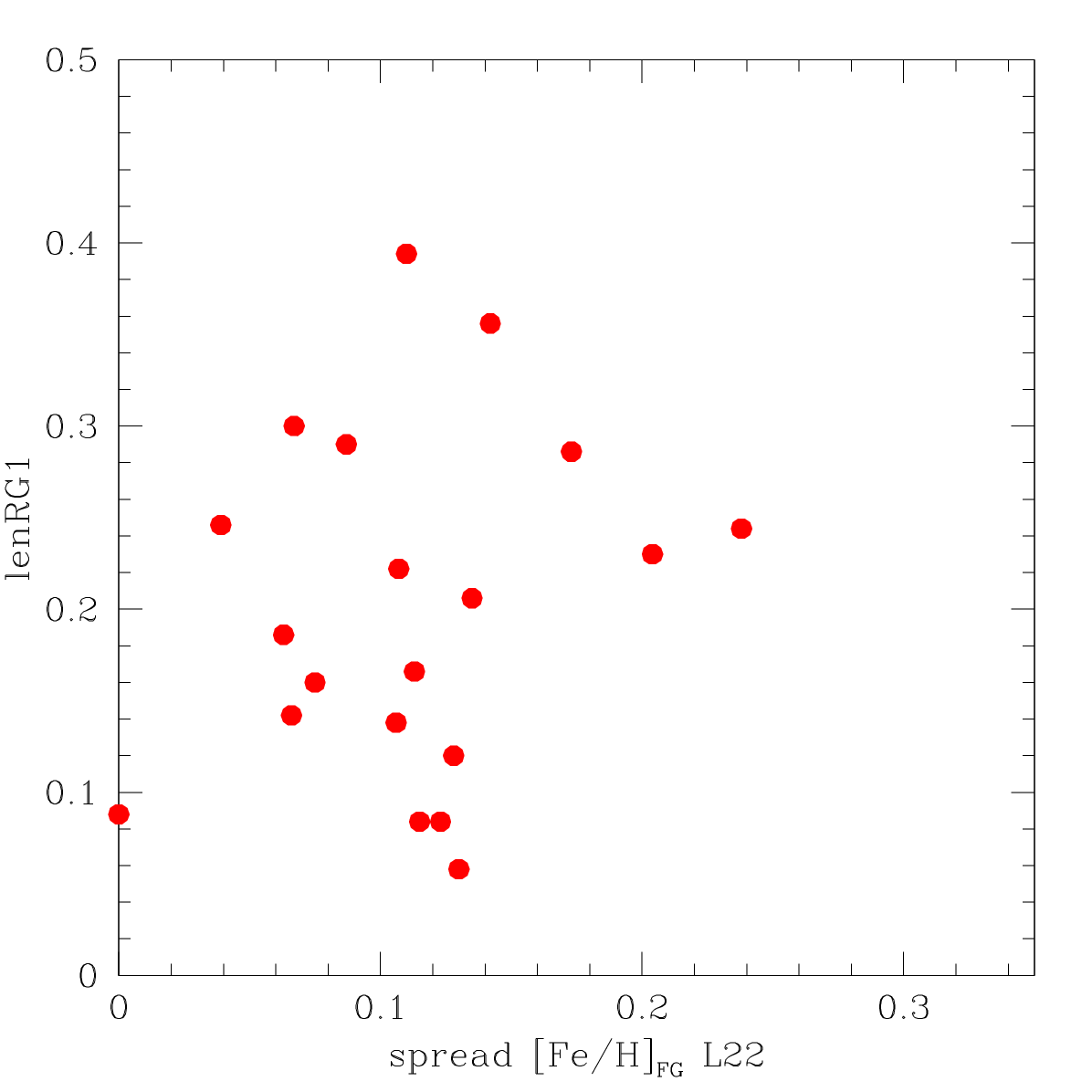}\includegraphics[scale=0.30]{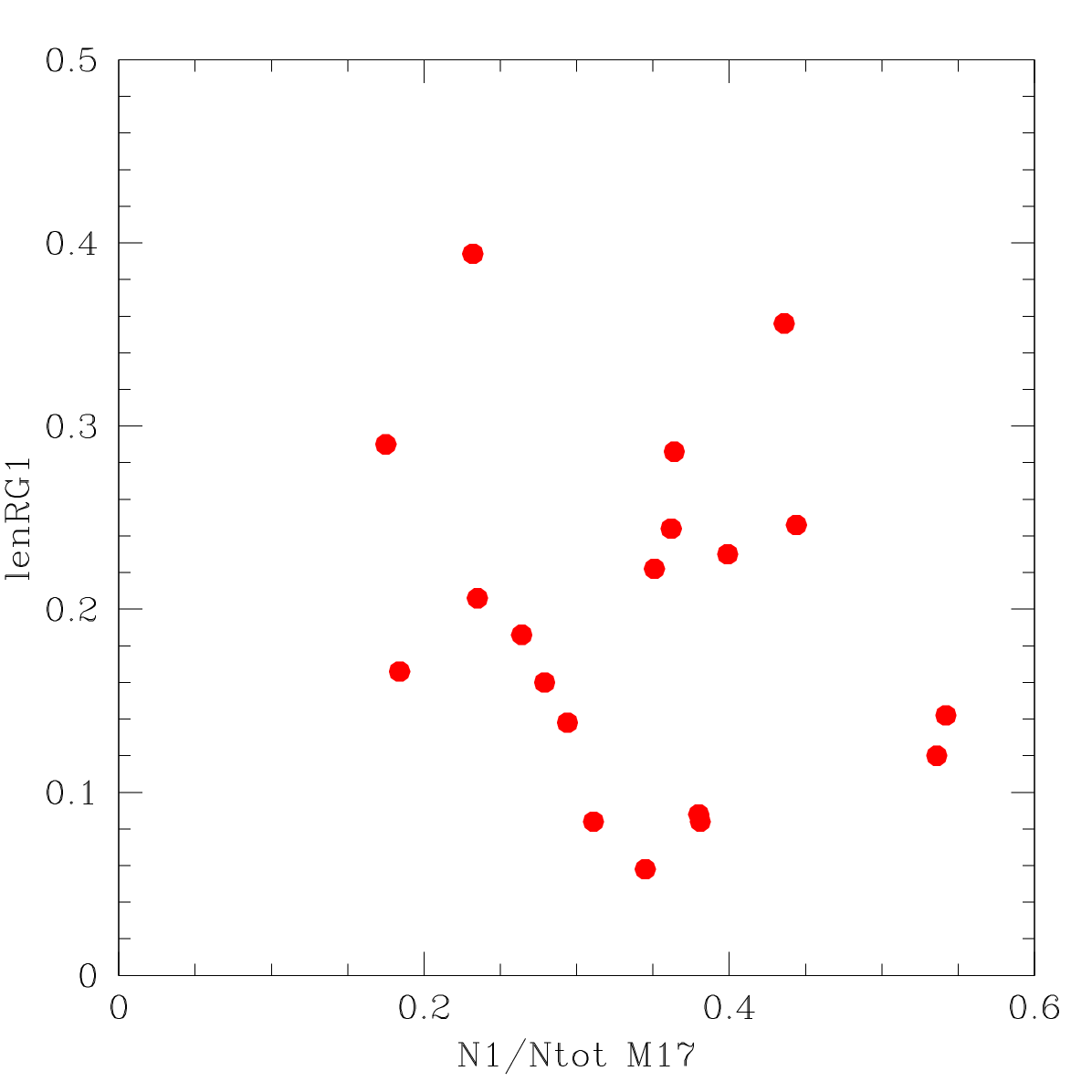}
\includegraphics[scale=0.30]{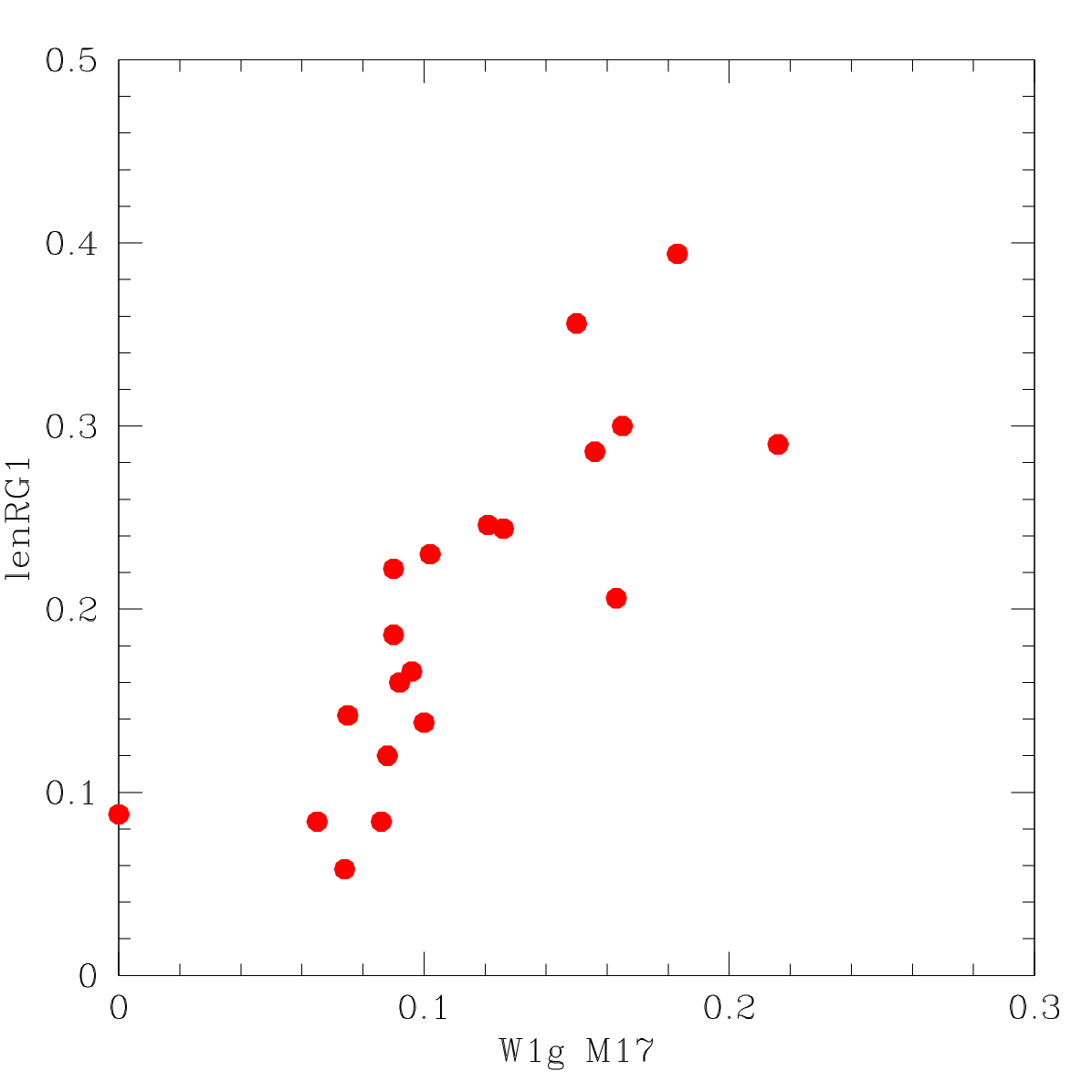}\includegraphics[scale=0.30]{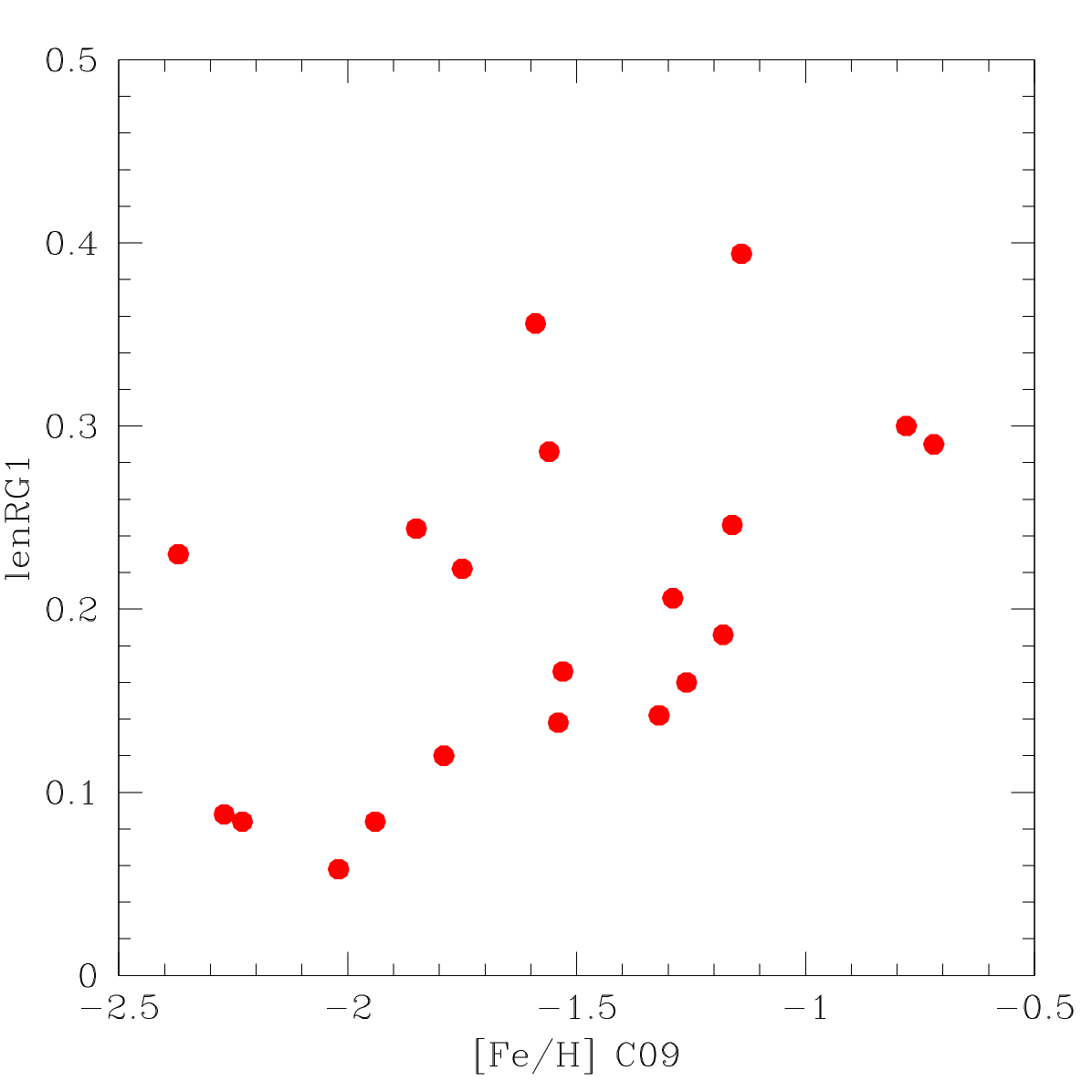}
\caption{Extension of the region $RG1$ plotted with respect to other quantities.
Upper left panel: Metallicity spread of FG stars by L22. Upper right panel:
Fraction of FG stars in each GC from M17. Lower left panel: Width of the FG
stars on the RGB in the $m_{F814W}-col$ CMD from M17. Lower right panel:
Metallicity of GCs on the metallicity scale by Carretta et al. (2009a).}
\label{f:lung}
\end{figure*}

\begin{table*}
\centering
\caption{Correlations of extensions and other parameters}
\begin{tabular}{rllll}
\hline

 & Extensions              &        &         &  \\
\hline
corr. & parameters         & $r_P$  &  prob.               & data source \\
 (1)  &   len$RG1$ - rmsFG & +0.900 & $<1.0\times 10^{-6}$ & this work \\
 (2)  &   len$RG1$ - iqrFG & +0.934 & $<1.0\times 10^{-6}$ & this work \\
 (3)  &   iqrFG  - rmsFG   & +0.907 & $<1.0\times 10^{-6}$ & this work \\
 (4)  &   iqrSG  - rmsSG   & +0.976 & $<1.0\times 10^{-6}$ & this work \\

\hline
  &   MP parameters            &        &         &  \\
\hline
corr. & parameters         & $r_P$  &  prob.             & data source other than this work\\
 (5)  &  len$RG1$ - N1/Ntot& -0.027 &  0.91              &  Milone et al. (2017) \\
 (6)  &	 len$RG1$ - W1g    & +0.815 &$1.2\times 10^{-5}$ &  Milone et al. (2017) \\
 (7)  &	 len$RG1$ - W2g    & +0.621 &$3.5\times 10^{-3}$ &  Milone et al. (2017) \\
 (8)  &	 W1g    - W2g      & +0.515 &$2.0\times 10^{-2}$ &  Milone et al. (2017) \\
 (9)  &	 iqrSG  - W1g      & +0.638 &$2.5\times 10^{-3}$ &  Milone et al. (2017) \\
(10)  &	 iqrSG  - W2g      & +0.858 &$1.0\times 10^{-6}$ &  Milone et al. (2017) \\
(11)  &  iqrSG  - iqrFG    & +0.719 &$3.5\times 10^{-4}$ &   \\

\hline
  &  GC parameters             &        &         &  \\
\hline
corr. & parameters                & $r_P$  &  prob.              & data source other than this work\\
(12)  &  len$RG1$ - Fe spread     & +0.205 &  0.39               &  L22 \\
(13)  &	 len$RG1$ - [Fe/H]        & +0.550 &  $1.2\times 10^{-2}$&  Carretta et al. (2009a) \\
(14)  &	 len$RG1$ - $\sigma_0$    & +0.043 &  0.86               &  Bailin and Von Klar (2022) \\
(15)  &	 len$RG1$ - $\sigma_{Fe}$ & -0.111 &  0.64               &  Willman and Strader (2012) \\
(16)  &	 len$RG1$ - age           & -0.349 &  0.13               &  Kruijssen et al. (2019) \\
(17)  &	 len$RG1$ - M$_V$         & -0.341 &  0.14               &  Harris (2010) \\
(18)  &	 len$RG1$ - M$_{ini}$     & +0.279 &  0.23               &  Baumgardt et al. (2019) \\
(19)  &	 len$RG1$ - iqr[O/Na]     & +0.234 &  0.32               &  Carretta et al. (2010c) \\
(20)  &	 iqrSG  - [Fe/H]          & +0.474 &  $3.5\times 10^{-2}$&  Carretta et al. (2009a) \\
(21)  &  iqrSG  - M$_{ini}$       & +0.443 &  0.050              &  Baumgardt et al. (2019) \\
(22)  &	 rmsSG  - M$_{ini}$       & +0.467 &  0.038              &  Baumgardt et al. (2019) \\
\hline

\end{tabular}
\label{t:corrtot}
\end{table*}

All the derived indicators iqrFG, rmsFG, and len$RG1$ are tightly correlated with
each other, as shown in the upper part of Table~\ref{t:corrtot} (correlations 1,
2, and 3), that is they measure essentially the same quantity and are equivalent
as true estimates of the extension of $RG1$ in the PCMs. In this table we list
for each correlation the involved parameters, the Pearson's correlation
coefficient for the linear regression and the two-tails probability testing the
null hypothesis that the observed values come from a population in which the
true correlation is zero.
We also measured the extension in $\Delta col$ of the $RG2$ sequence, again
taking into consideration the interquartile range iqrSG and the rmsSG values.
Due to the shape of the $RG2$ regions, it was not possible to use simple
geometric forms like the ellipse to measure their extension, anyway both iqrSG
and rmsSG are well suited to quantify the spread of SG stars in $\Delta col$
(correlation 4 in Table~\ref{t:corrtot}).

For geometric reasons we chose the major axis of the ellipse as our preferred
observable for the extension of the $RG1$ region since it reproduces the length
of the FG distribution exactly along the direction where it is more elongated. 
Notice that L22 and M17
only used the displacement in $\Delta col$ along the horizontal
axis, without considering the inclination of $RG1$. In other words, their
measurements are a scaled version of our measured extensions according to the
Pythagoras' theorem. We will use len$RG1$ to investigate the dependence of the
$RG1$ region on global GC parameters. On the other hand, to quantitatively
compare the properties of FG and SG stars we are necessarily forced to employ
the other quantities, iqr and rms values.

In most cases the values of iqrFG and iqrSG (as well as rmsFG and rmsSG) are not
dramatically different. This result is unexpected, since it is generally claimed
that the extension in $\Delta col$ is larger for FG than for SG stars (see e.g.,
L22 and M17). We note that the derivation of a metallicity spread in M17 and L22
is focused only on the FG spread or extension in $\Delta col$, whereas the
similar spread in SG stars is almost completely neglected, albeit it is
significant (see next Section). The cluster NGC~2808 is the most extreme case,
since its len$RG1$ is smaller than the extent of $RG2$. Excluding NGC~2808, the
average values become more comparable, as shown by the iqr values in
Table~\ref{t:len}.

The relation of the measured extent of $RG1$ with respect to other relevant
quantities is shown in Fig.~\ref{f:lung}. Statistical values for the
correlations (or the lack of any relation) regarding mutual parameters of the
MPs in GCs are reported in the middle part of Table~\ref{t:corrtot}. Possible
interesting correlations with global GC parameters have their statistical values
listed in the third part of Table~\ref{t:corrtot}.

\begin{figure*}
\centering
\includegraphics[bb=30 505 560 680, clip, scale=0.90]{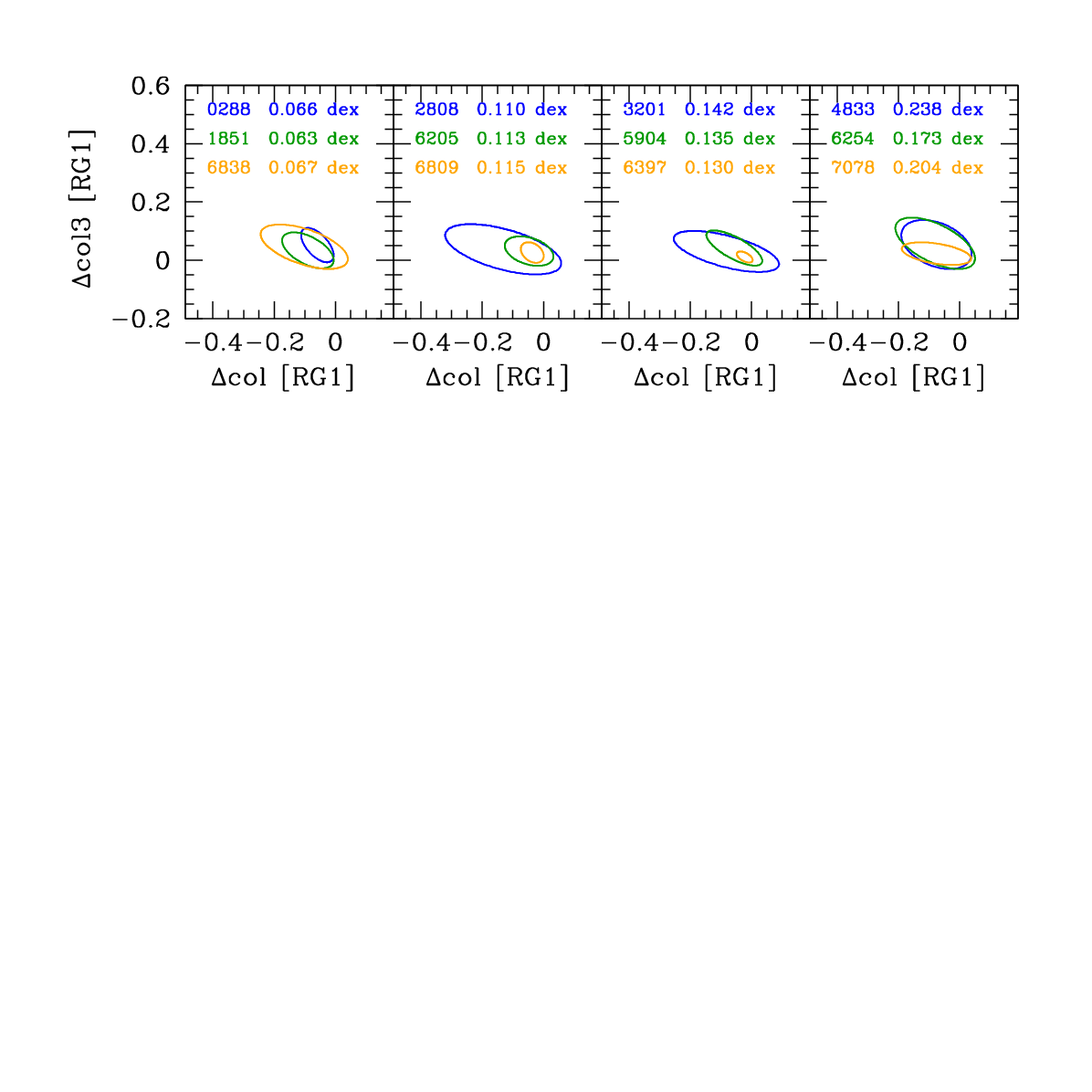}
\caption{Relation (or lack thereof) between the extension in $RG1$
(here describer by the major axis of the ellipses) and the spread in metallicity
proposed by L22. For clarity, we show four panels with only three cases with
similar spreads, as claimed by L22, in each one. The GCs and the metallicity
spreads are indicated in the plots with the same colour as the corresponding
ellipse. The extensions len$RG1$ are in Table~\ref{t:len}.}
\label{f:confelli}
\end{figure*}

The first consideration from Fig.~\ref{f:lung} (upper left panel) is that
there is no correlation between the extent of the $RG1$ sequence of FG stars and
the values of the metallicity spread as computed by L22 (see correlation 12 in
Tab.~\ref{t:corrtot}). This striking evidence is clearly visualised in
Fig.~\ref{f:confelli}. In the four panels of this figure we grouped GCs
according to the [Fe/H] spread of FG stars derived by L22, and for each group we
plotted the ellipses enclosing the $RG1$ sequences. As a matter of fact, 
metallicity spreads and extensions of FG stars in the PCM are unrelated.
Putative, almost identical FG [Fe/H] spreads correspond to wildly different
extensions of $RG1$.

Similarly, there is no correlation between len$RG1$ and
the fraction of FG stars in each GC estimated by M17 on the PCM (upper right
panel in Fig.~\ref{f:lung} and correlation 5 in Table~\ref{t:corrtot}). Thus, how
much the $RG1$ extends seems to be not dependent on the number of FG stars in the
GC, apparently requiring the involvement of some other parameters.

A robust correlation does actually exist between the extension len$RG1$ and the
width W1g of the FG stars on the RGB in  $col=m_{F275W}-m_{F814W}$ from M17
(lower left panel in Fig.~\ref{f:lung}). This width is the same used by L22 as
starting point to translate the width of the FG region into a metallicity
spread. Hence, the results of correlations (6) and (12) altogether would suggest
once more that the width W1g of FG stars used by L22 are sound, being wider as
the len$RG1$ is larger, but are somehow converted into unrealistic metallicity
spreads that, moreover, do not seems to have any correspondence with the
measured extension of $RG1$ in the PCM. This  would then suggests that is the
conversion of W1g into [Fe/H] values to be the source of the error.

Moreover, there is a positive and statistically significant correlation
between  the len$RG1$ and the GC global metallicity (lower right panel of
Fig.~\ref{f:lung}), on the metallicity scale by Carretta et al. (2009a). 
However, this correlation goes in the opposite direction with respect to the one
given by L22. In L22 larger metallicity spread are found for metal-poor GCs (see
their Figure 12 or their Table 4), whereas in Fig.~\ref{f:lung} the len$RG1$ is
larger in more metal-rich GCs.

Finally, len$RG1$ does not correlate with the GC intrinsic metallicity
spreads (correlations 14 and 15 in Table~\ref{t:corrtot}) or the age and mass
(either present-day mass or initial mass) of the GCs (correlations 16, 17, and
18). Not significant is also the relation of len$RG1$ with the extension of the
Na-O anticorrelation, the main spectroscopic signature of MPs (correlation 19).

\begin{figure}
\centering
\includegraphics[scale=0.40]{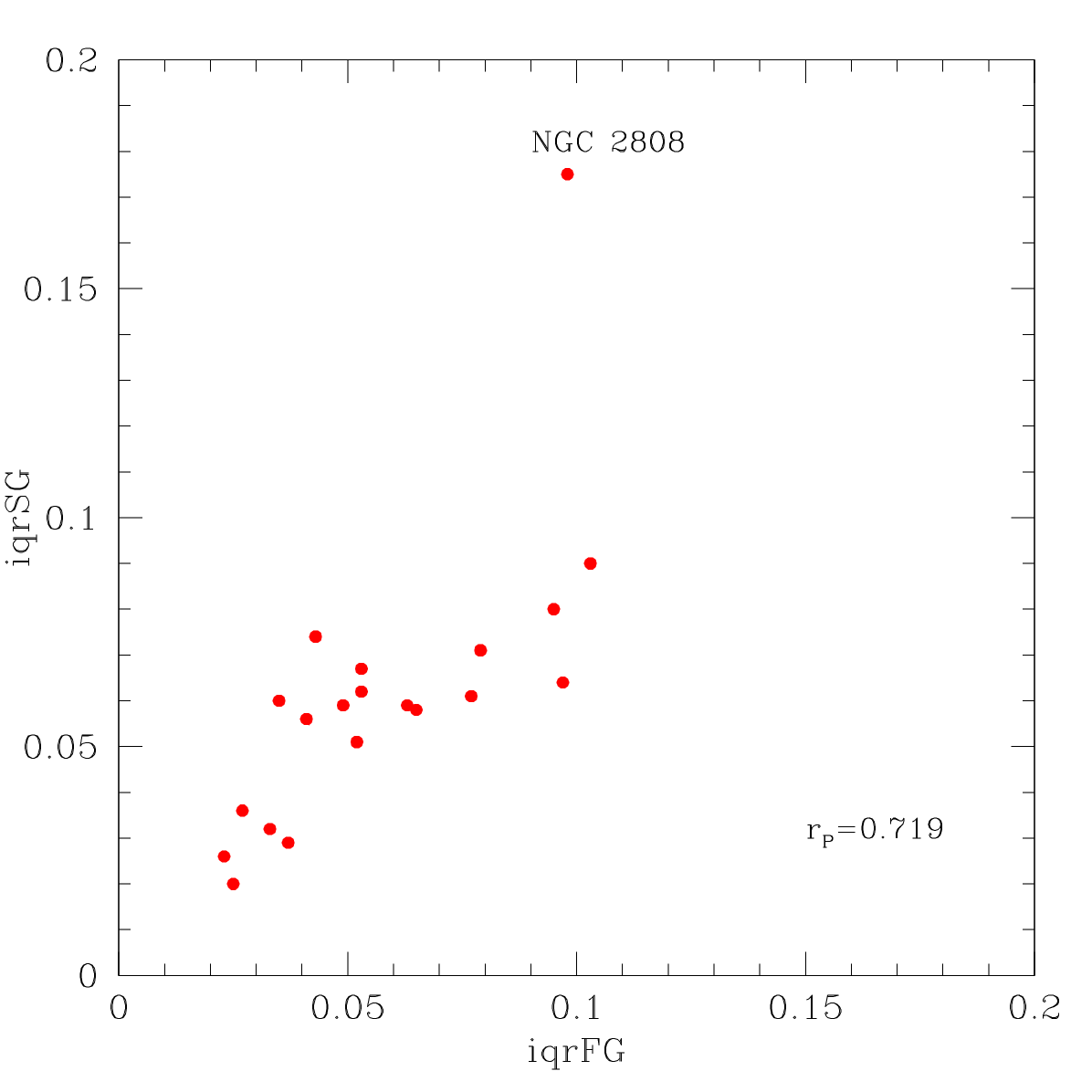}
\caption{The extension of the $RG2$ region as a function of the extension of the
sequence of FG stars, $RG1$. The Pearson's correlation coefficient is reported.}
\label{f:lungiqr}
\end{figure}

\begin{figure*}
\centering
\includegraphics[scale=0.8]{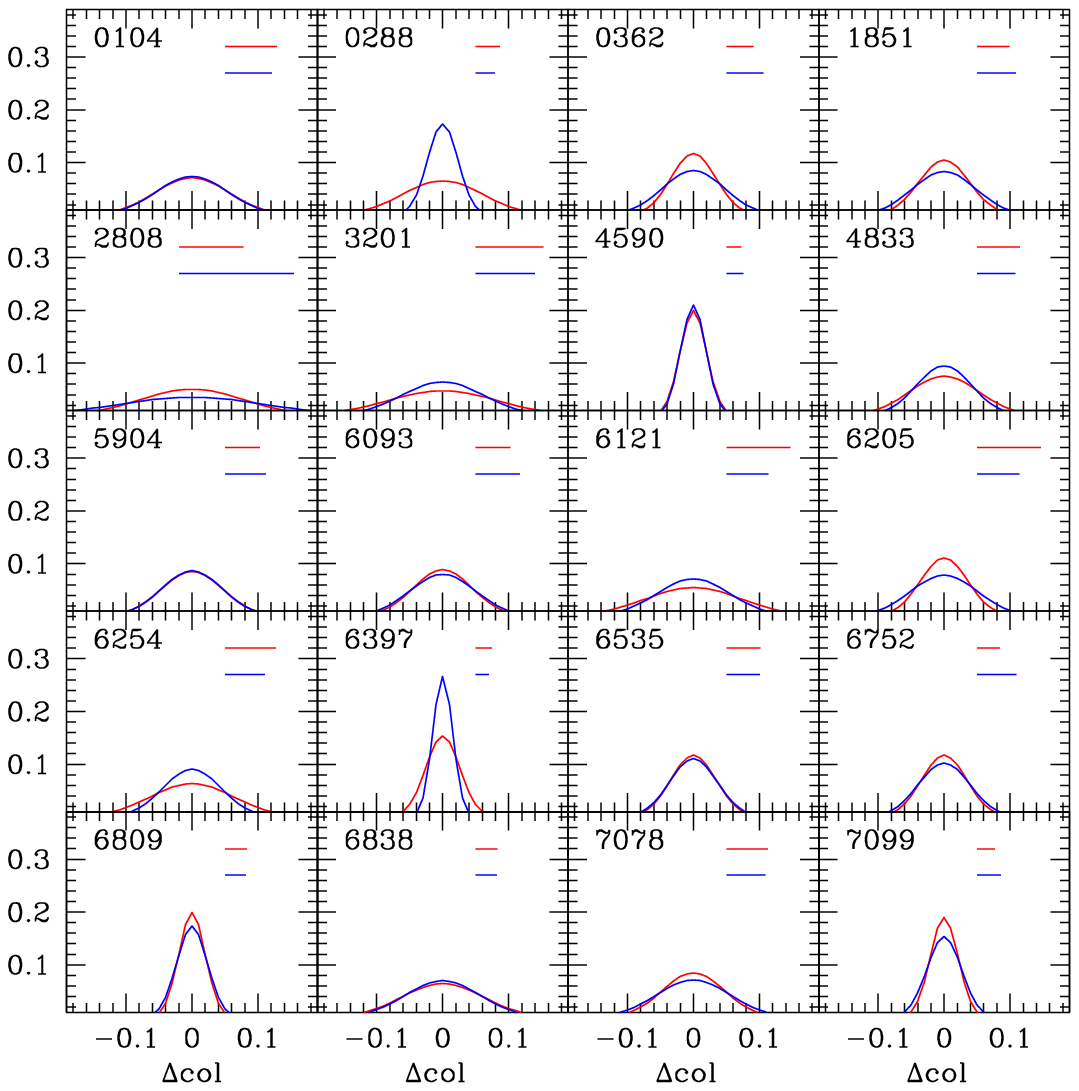}
\caption{Comparison between the distributions of $RG1$ (in red) and $RG2$ (in
blue) in $\Delta col$ in 20 GCs from CB24. In each panel the two 
gaussians with $\sigma$ equal to the rms are shown. The peaks have been shifted both to 
zero to get an immediate comparison. The horizontal lines in each panel indicate
the iqrFG and iqrSG (red and blue, respectively).}
\label{f:gaussianeFGSG}
\end{figure*}

A maybe unexpected result when quantifying the extension of the sequences in the
PCM is shown in Fig.~\ref{f:lungiqr}. We found that the extension of $RG1$ is
tightly correlated to the extent of the $RG2$ in $\Delta col$. In this figure we
also report the Pearson's correlation coefficient. The tight relation between
iqrFG and iqrSG is true regardless NGC~2808 is included (p$=3.5\times 10^{-4}$)
or not  (p$=5.9\times 10^{-5}$) in the sample. This result prompted us to have a
more careful look at the extension of the $RG2$ sequence and its relation to the
global cluster parameters, filling the void left in the M17 and L22 analysis who
only focus on the spread in $\Delta col$ of the FG stars.

\section{The elephant in the room: Spreads in the neglected SG stars in GCs}

The basic observation upon which a claim of a metallicity spread is based is the
large extension of the $RG1$ with respect to the observational errors, estimated
with Monte Carlo simulations.

The main effects of abundances on the shape of PCMs are clearly summarised in 
Marino et al. (2019: their figure 27). An enhancement of about 1.2 dex in [N/Fe]
explains the vertical extension of SG stars in the region $RG2$ on
the PCM along the $\Delta col3$ coordinate. A spread of iron affects the
position of FG stars along $RG1$. 

This simplified scheme, however, does not account for other features of the PCM.
The $RG1$ is not horizontal but 
tilted in the PCM, meaning that in principle it must also have a 
vertical component along $\Delta col3$, the coordinate mainly affected by
changes in the N content, the basic feature of the photometric classification of
stars into MPs in GCs.

On the other hand, beside a strong component in $\Delta col3$, the $RG2$ is
showing a noticeable component also along the  $\Delta col$. Hence, if the claim
by M17 and L22 of a link between the spread in $\Delta col$ and
iron is sound, there is no reason not to contemplate a metallicity spread also
among SG stars.

The comparison shown in Fig.~\ref{f:lungiqr} clearly indicates that as the
length of the FG sequence in $\Delta col$ increases, so does the extension of
the $RG2$ region populated by SG stars. The correlation has a high statistical
significance, hence if the displacement along $\Delta col$ has to be associated
to a metallicity spread, the logical inference would be to infer a spread in
[Fe/H] also in SG stars. From Fig.~\ref{f:lungiqr} the comparison between the
respective extensions of $RG1$ and $RG2$ is highly significant. The spread of $RG2$ along
the $\Delta col$ coordinate (supposedly related to the metallicity spread) is
not only well comparable to the measured len$RG1$ values, but in the case of
NGC~2808 is clearly much larger than len$RG1$ itself.

A more thorough comparison can be seen in Fig.~\ref{f:gaussianeFGSG}. In this
figure we show the distributions of $RG1$ and $RG2$ in $\Delta col$ for the 20
GCs from CB24. In each panel we show the gaussian with
$\sigma$ equal to the rms; the gaussian peaks have been shifted to zero using
the mean values, to get an immediate comparison between $RG1$ and $RG2$
distributions (in red and blue, respectively). To complete the comparison, we
plot also lines indicating the iqr values for FG and SG stars, using the same
colours.

\begin{figure}
\centering
\includegraphics[scale=0.40]{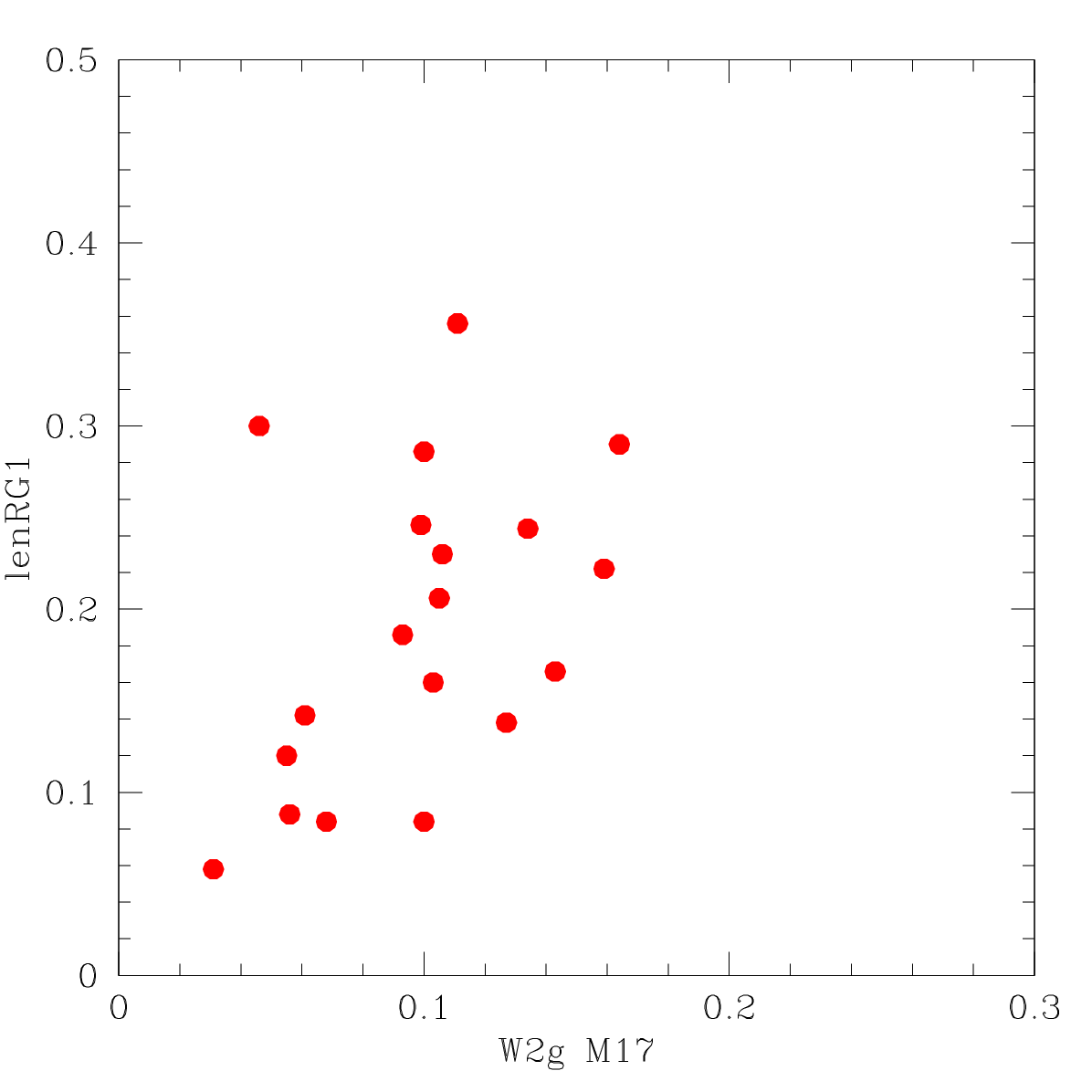}
\caption{The extension of the RG1 region as a function of the width of the SG
stars (from M17) along the RGB in 20 GCs.}
\label{f:lung4a}
\end{figure}

To introduce further complexity, in Fig.~\ref{f:lung4a} the extension of the FG
sequence is found to be correlated to the width of the SG stars on the CMD, with
high statistical significance (see correlation 7 in Tab.~\ref{t:corrtot}). At
face value, this means that, were the extension of $RG1$ be really related only
to the metallicity spread, the same would be required also to explain  the
spread of the SG stars on the CMD. Clearly, the scenario depicted by L22 by
neglecting the polluted stars in their speculations is too simplistic.

\begin{figure*}
\centering
\includegraphics[scale=0.30]{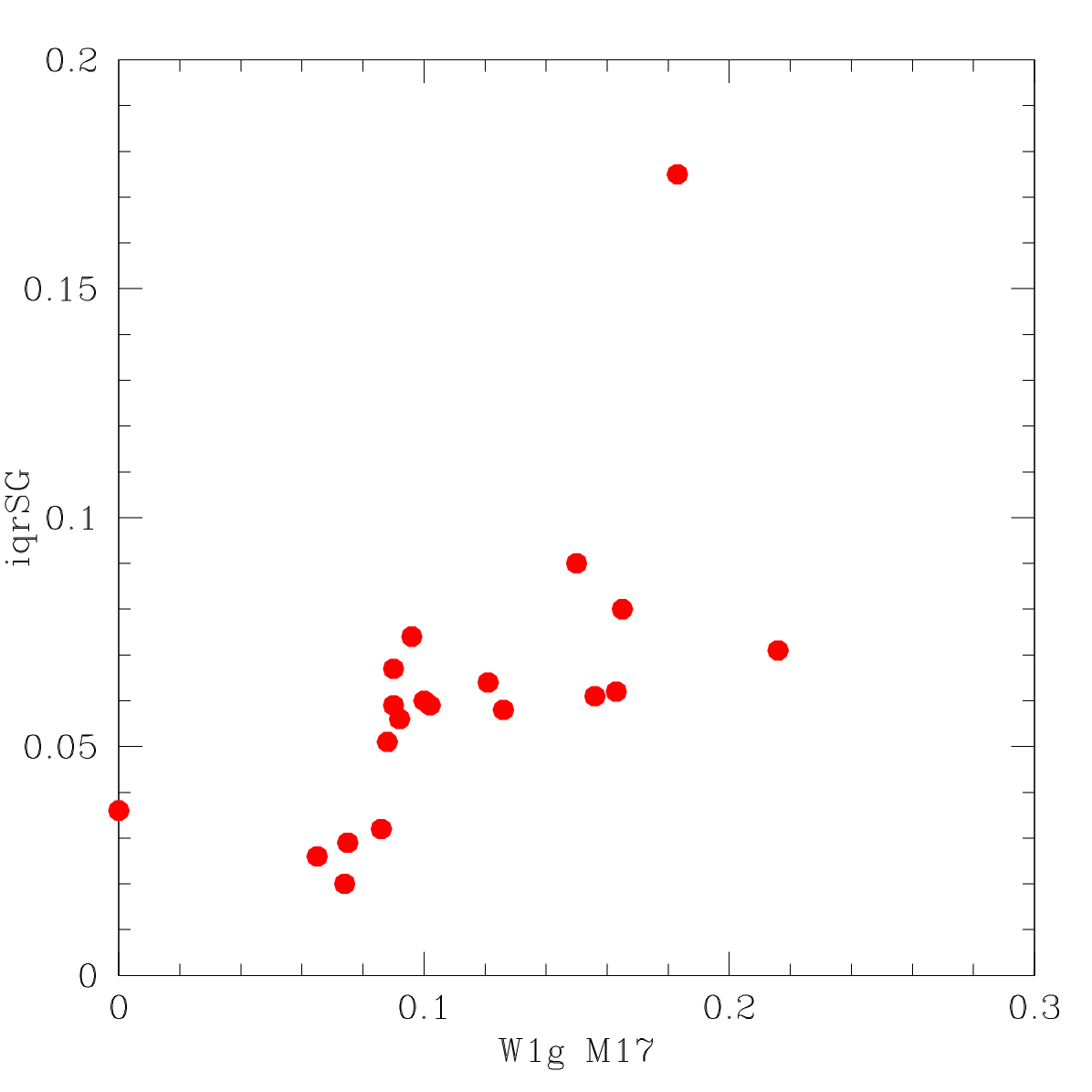}\includegraphics[scale=0.30]{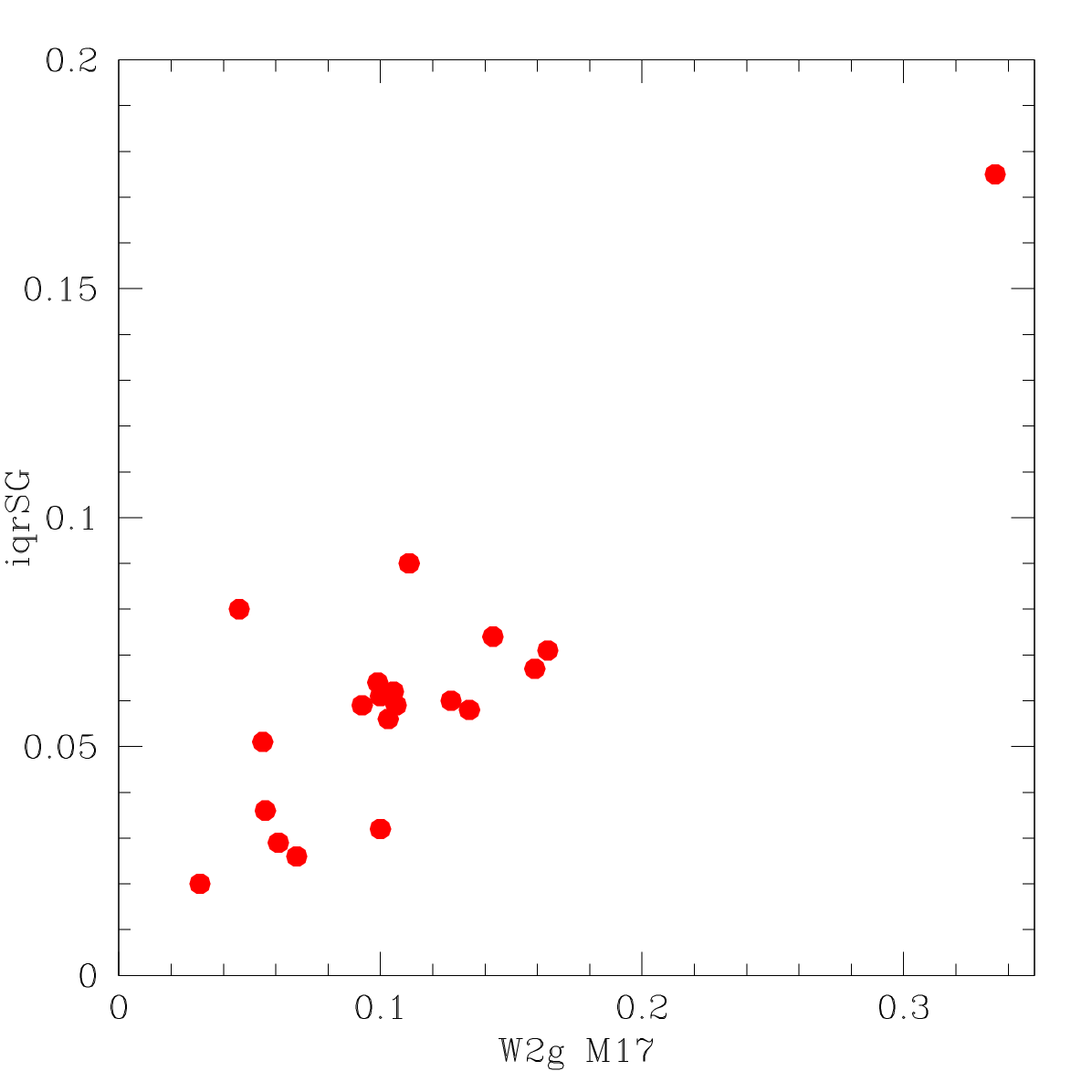}\includegraphics[scale=0.30]{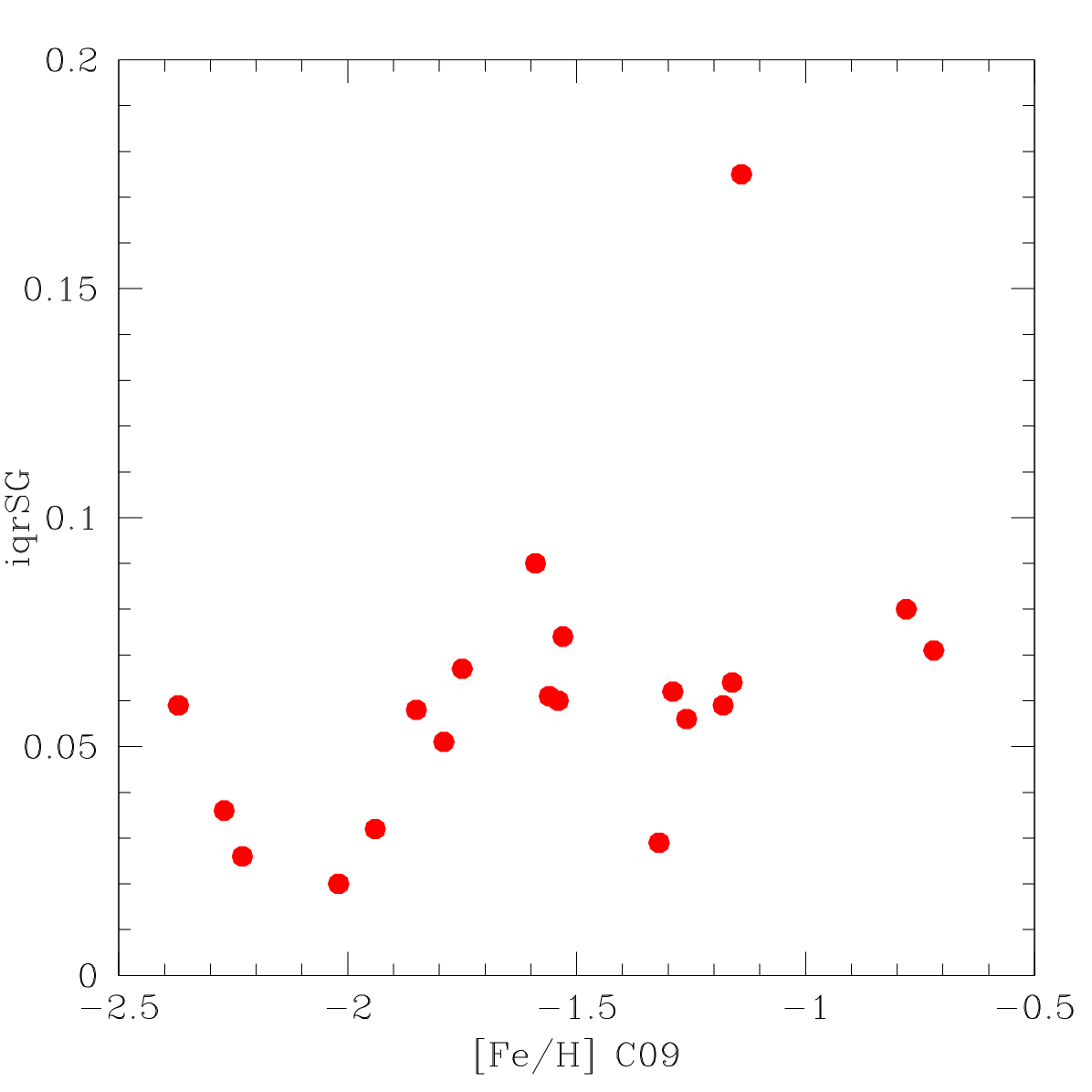}
\caption{Extension of the $RG2$ region of SG stars as a function of the width
of FG stars on the RGB (left panel), the width of SG stars (middle panel), and
metallicity [Fe/H] (right panel) in 20 GCs.}
\label{f:lungSG1}
\end{figure*}

The extensions of the $RG2$ region along the $\Delta col$ coordinate
in the PCM is correlated to the width W2g of SG stars on the RGB but also to the
width W1g of FG stars (left and middle panels in Fig.~\ref{f:lungSG1}). A
statistically significant dependence on the GC metallicity is detected for iqrSG
(right panel in Fig.~\ref{f:lungSG1}).

\begin{figure*}
\centering
\includegraphics[scale=0.30]{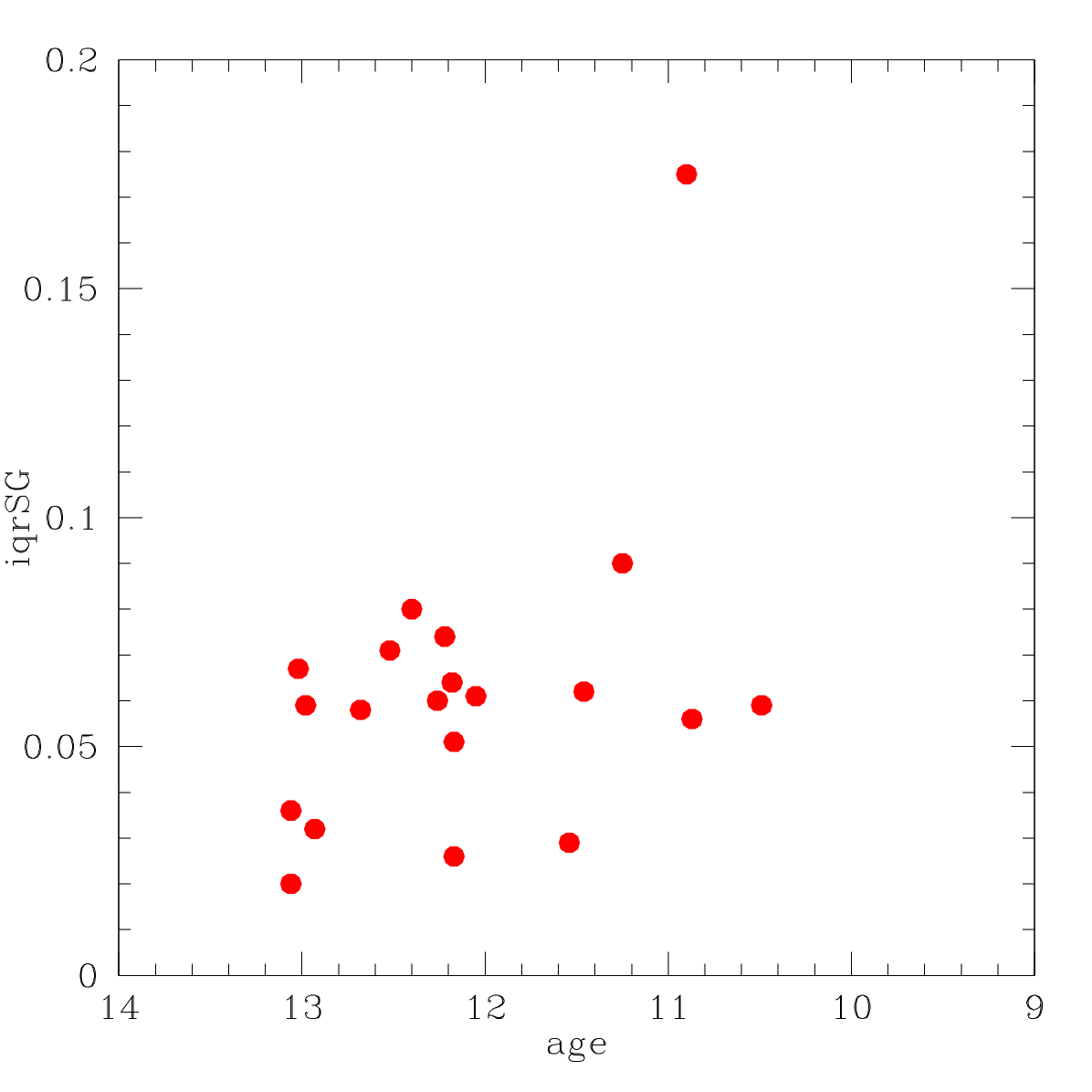}\includegraphics[scale=0.30]{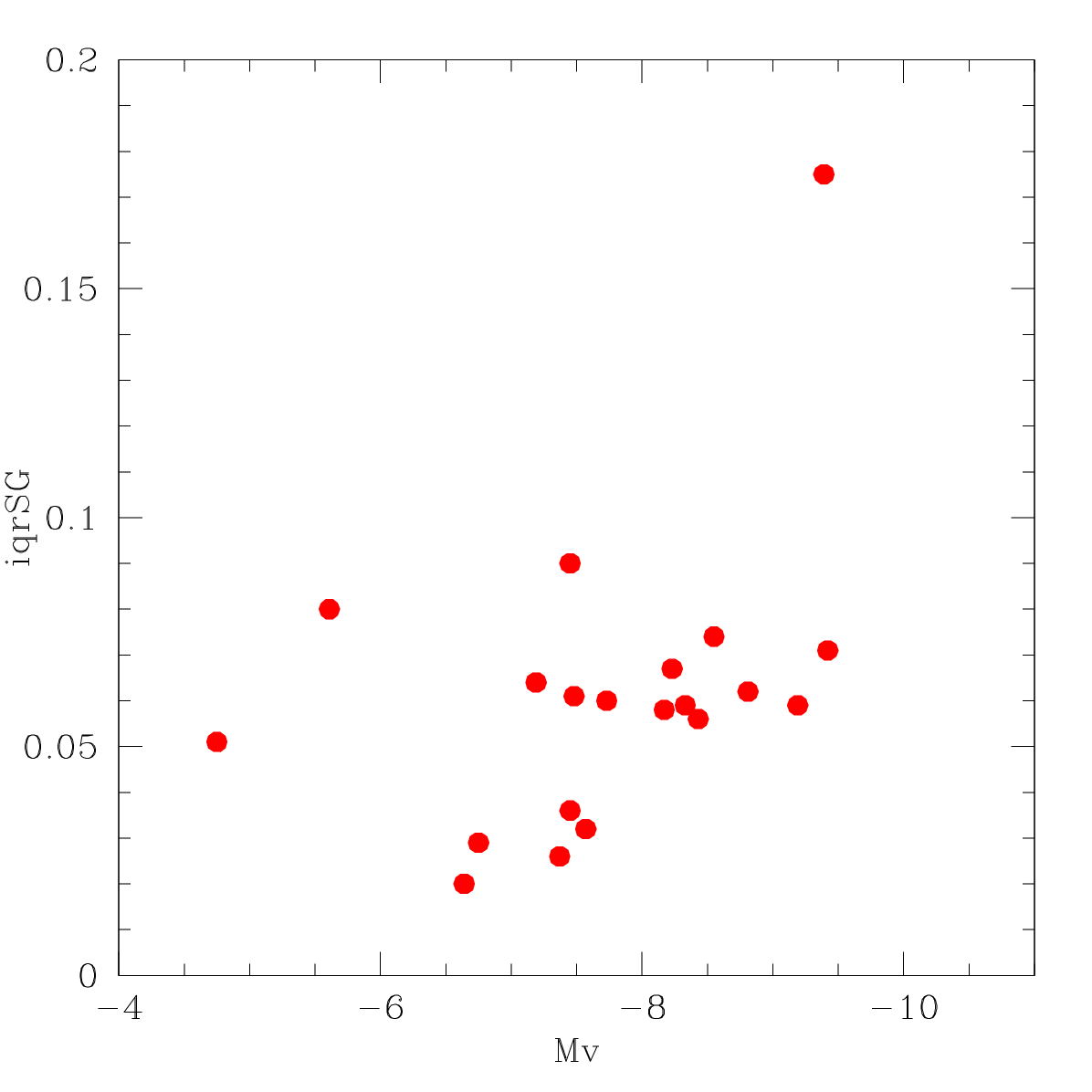}
\includegraphics[scale=0.30]{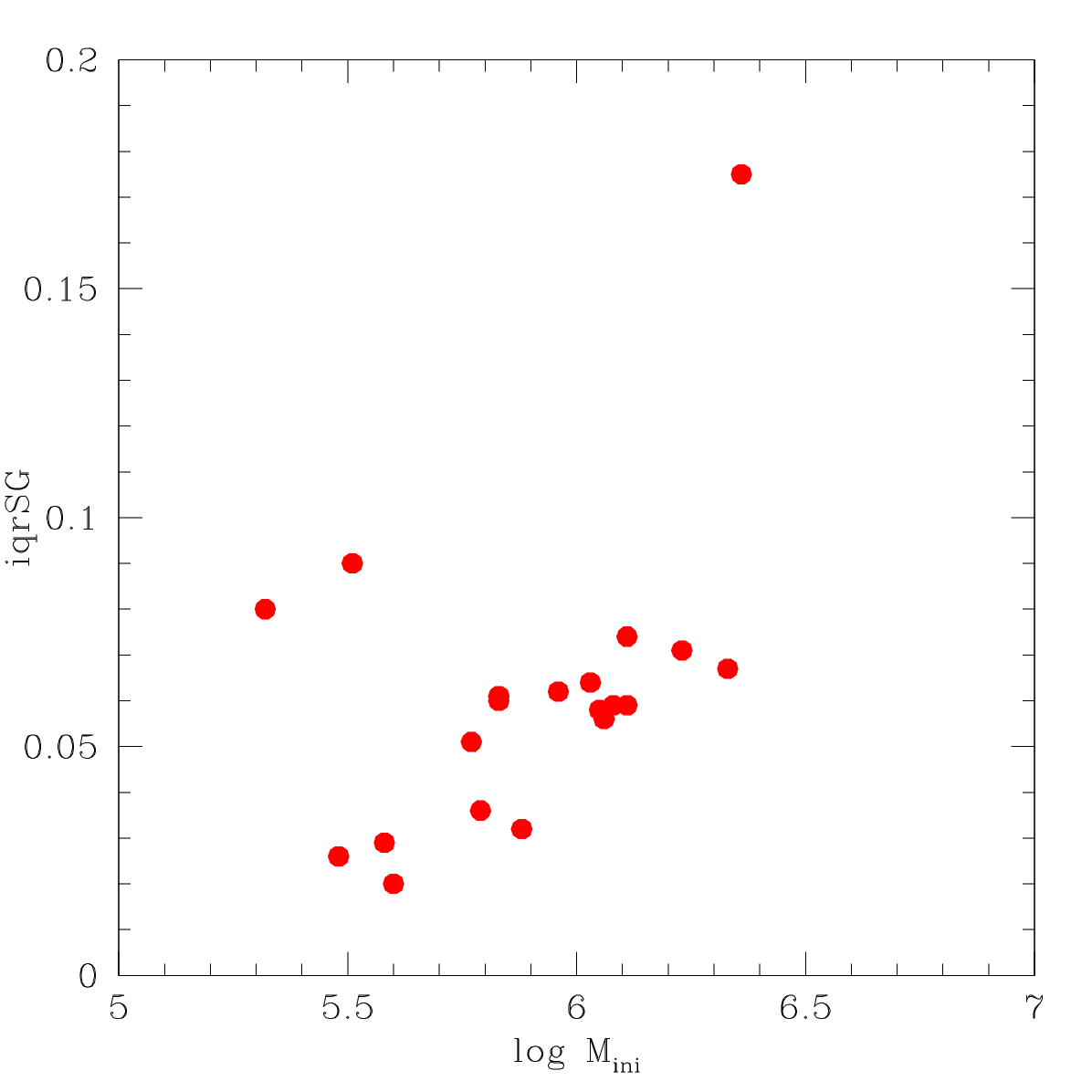}\includegraphics[scale=0.30]{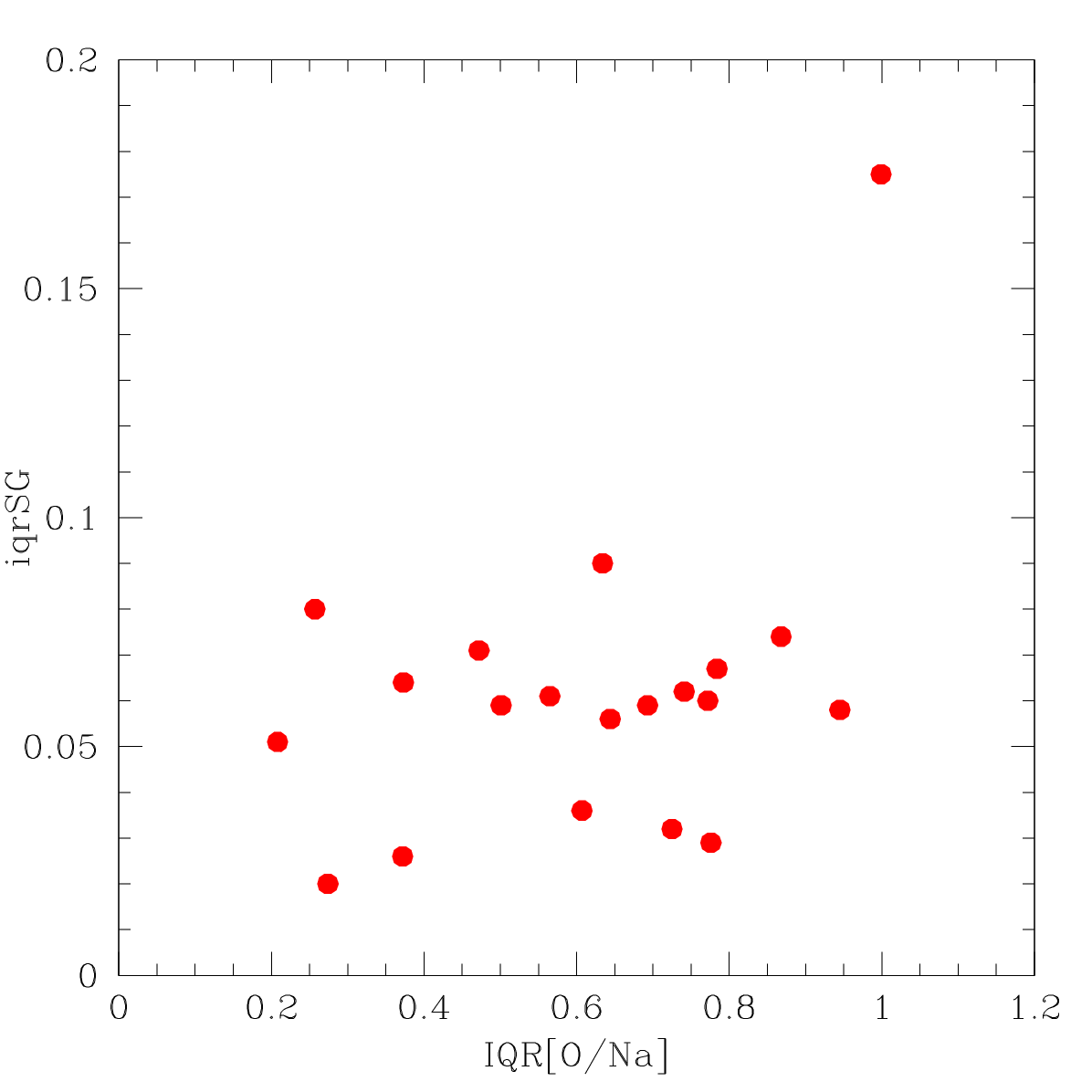}
\caption{Extension of the RG2 region of SG stars in 20 GCs as a function of 
the GC ages (upper left panel), total absolute magnitude (upper right panel), 
initial GC mass (lower left panel), and interquartile range of the [O/Na] ratio
(lower right panel). References for data sources are listed in
Table~\ref{t:corrtot}.}
\label{f:lungSG2}
\end{figure*}

The extension of $RG2$ in $\Delta col$ is plotted as a function of a few other 
global parameters of GCs in Fig.~\ref{f:lungSG2}. The correlation with the
cluster age is formally not significant (p=0.055), but this result is borderline,
since had we used the quantity rmsSG, instead, the correlation would have become significant
(p=0.049), although barely. No clear dependence from the present-day mass of GCs
is detected, whereas we found that the extension of $RG2$ seems to depend on the
initial mass of GCs (lower left panel in Fig.~\ref{f:lungSG2} and correlations 21
and 22 in Table~\ref{t:corrtot}). Further studies on the spread in 
(pseudo-)colours and chemical abundances as their cause are clearly called for
also concerning the interpretation of the properties of SG stars.

\section{Summary}

In the present paper we addressed the issue of the metallicity spreads claimed
to be hosted among FG stars in GCs.
We investigated the large metallicity spreads in FG GC stars found by L22,
surprising in stellar systems formed without an extended star formation. The 
explanation provided by L22 is based on inhomogeneous mixing during the primary
star formation. Afterward, when mixing of polluter matter and pristine gas was
used to form the following population of SG, low mass stars, a more thorough
homogenisation is predicted to occur, effectively removing any detectable
difference in [Fe/H].

We used a large sample of FG stars with spectroscopic abundances and the only
available published set of PCMs to match the $\Delta col$ position of
stars along the $RG1$ sequences. Selecting stars with very similar atmospheric
parameters we found no clear correlation between differences in [Fe/H] and
displacement in $\Delta col$. Over 30 pairs of stars in 12 GCs, the average
shift in $\Delta col$ (0.090 mag, rms=0.114 mag), which is not statistically
significant, is not proportional to the differences in metallicity between the
stars in each pair.

For the first time we quantified the real extension of the $RG1$ sequence of FG
stars on the PCM. We measured the interquartile range  of the extension in
$\Delta col$ (to alleviate the impact of outliers), the spread in $\Delta col$,
rmsFG, and the major axis of the ellipse enclosing the FG sequence, len$RG1$.
All these parameters are perfectly consistent with each other.
Using these quantitative measurements 

\begin{figure}
\centering
\includegraphics[scale=0.40]{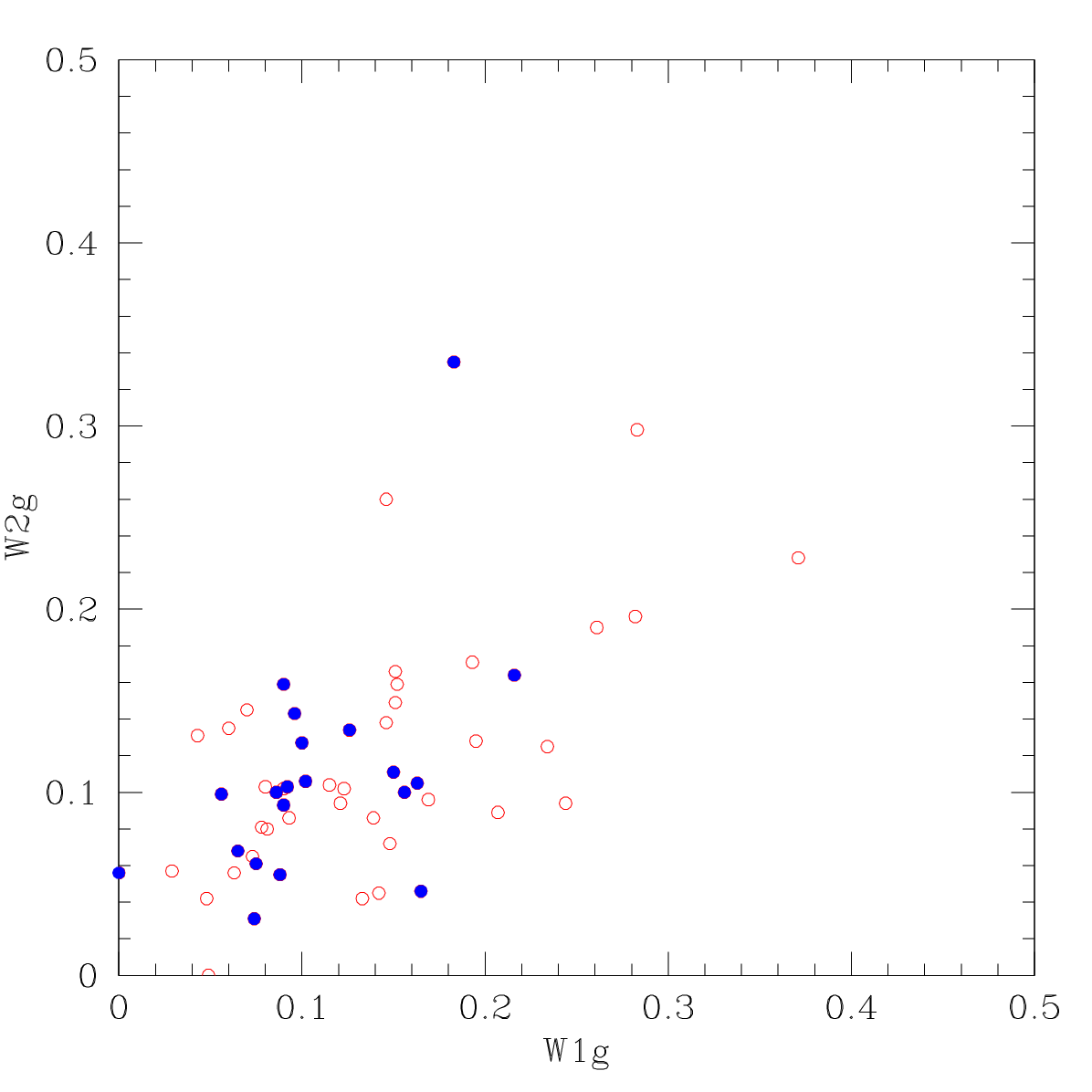}
\caption{Correlation of the width W1g of FG stars in $col$ and the width W2g of
SG stars for the 56 GCs in M17. The GCs in the subsample by CB24 utilised in the
present paper are plotted as filled blue symbols.}
\label{f:ww}
\end{figure}

\begin{enumerate}
\item we did not find any relation between the
metallicity spread among FG stars as derived by L22 and the true extension of
the region populated by FG stars in the PCM. Globular clusters with very
different $RG1$ extensions have identical spreads in [Fe/H] according to L22
(see Fig.~\ref{f:confelli});

\item the len$RG1$ values have no
significant correspondence also with any estimate of the intrinsic metallicity
spread of the GCs as a whole available in the literature. Instead we found that
the actual extent of the $RG1$ region in $\Delta col$ increases as the average
metallicity [Fe/H] of the GCs increases (Fig.~\ref{f:lung});

\item the true extension of the $RG1$ sequence, len$RG1$, is correlated to the
width W1g of the FG stars on the RGB in the $m_{F814W}$ versus $col$ plane 
(Fig.~\ref{f:lung}), but it is also correlated to the width W2g of the SG stars
(Fig.~\ref{f:lung4a}). Both these relations have high statistical significance; 

\item the len$RG1$ is very well correlated to the analogous extension 
in $\Delta col$ of the SG sequence. This result should not come unexpected,
since the spread of FG stars in $col$ on the RGB is complementary to the spread
of SG stars. The more extended is the location of SG stars, the less is the one
of FG stars, and vice versa. When the distance from the fiducial red and blue
envelopes of the RGB are used to produce the PCM, the procedure likely provides
the statistical significant correlation observed of len$RG1$ to the extent in
$\Delta col$ of the $RG2$. This occurrence, neglected by M17, is clearly shown in Fig.~\ref{f:ww}, where we plotted the
correlation of the widths W1g and W2g from M17. In the figure, GCs analysed in
the present paper are depicted as filled symbols, and the correlation is highly
significant (correlation 8 in Table~\ref{t:corrtot}). However, considering the
whole sample of 56 GCs tabulated in M17, the correlation of W1g with W2g is even
more significant (Pearson's correlation coefficient r=0.652, 
$p<1.0\times 10^{-6}$), a fact not mentioned either in M17 or in L22;

\item the extension in $\Delta col$ of the $RG2$ presents significant relations with
the cluster average metallicity [Fe/H] and the total GC mass, either the
present-day mass or the initial mass before any dynamical evolution onset. The
correlation between the extension of $RG2$ and the GC age is more uncertain,
being barely statistically significant.

\end{enumerate}

The implications of our findings challenge the scenario
depicted by L22 and M17. If the extension of $RG1$ has to
be related to a metallicity spread, then the same must be true also for SG
stars. We should observe a metallicity spread in any population of GC stars,
occurrence which is neither mentioned by L22, M17 nor
observed in most GCs. Vice versa, the tight correlation between the spread in
$\Delta col$ in $RG1$ and $RG2$ implies that if no metallicity spread is present
among SG stars, then it hardly should be existing among FG stars.
The good agreement between spreads in $\Delta col$ in $RG1$ and $RG2$,
highlighted for the first time thanks to quantitative measurements of their
extension in the PCMs, indicates that no significant difference is present
concerning the homogeneity of the intracluster medium from which the primary
star formation formed the FG stars and the following phase of formation of SG
stars, invalidating the scenario proposed by L22.

Finally, since the extension of $RG1$ and $RG2$ increases as the width of FG and
SG stars increases on the RGB, this probably means that W1g and W2g are
reasonable estimates. It is likely the conversion of these width into
spread of  [Fe/H] to be the problem.

\begin{acknowledgements}
We thank the referee for careful and useful comments and suggestions.
This research has made use of the VizieR catalogue access tool, CDS, 
 Strasbourg, France (DOI: 10.26093/cds/vizier). The original description of the
 VizieR service was published in 2000, A\&AS 143, 23. Use of the NASA's
Astrophysical Data System, and TOPCAT (Taylor 2005) are also acknowledged.
\end{acknowledgements}

\begin{appendix}
\onecolumn
\section{Selected samples of FG and SG stars in the program GCs}

We present here the selection of FG and SG stars (i.e., the populations of the
$RG1$ and $RG2$ regions, respectively) for the 20 GCs discussed in the present
paper. We used the PCMs derived by CB24 and isolated $RG1$ and $RG2$ stars
manually, using TOPCAT; the result of the selection is shown in the following
figures (from Fig.~\ref{f:app1} to Fig.~\ref{f:app5}), where FG stars are
coloured in red and SG stars are coloured in blue, respectively. Stars falling
far from the main distributions in the CMDs, and consequently in the PCMs, 
were ignored; they are always a
minority and are indicated in grey in the plots. In each plot we also show the
ellipse enclosing the $RG1$ (in black) with the major axis measuring the actual
extension of $RG1$ drawn in green.
Information on the extension of the $RG1$ and its inclination is given in
Table~4 in the main text.

\begin{figure*}[b!]
\centering
\includegraphics[scale=0.9]{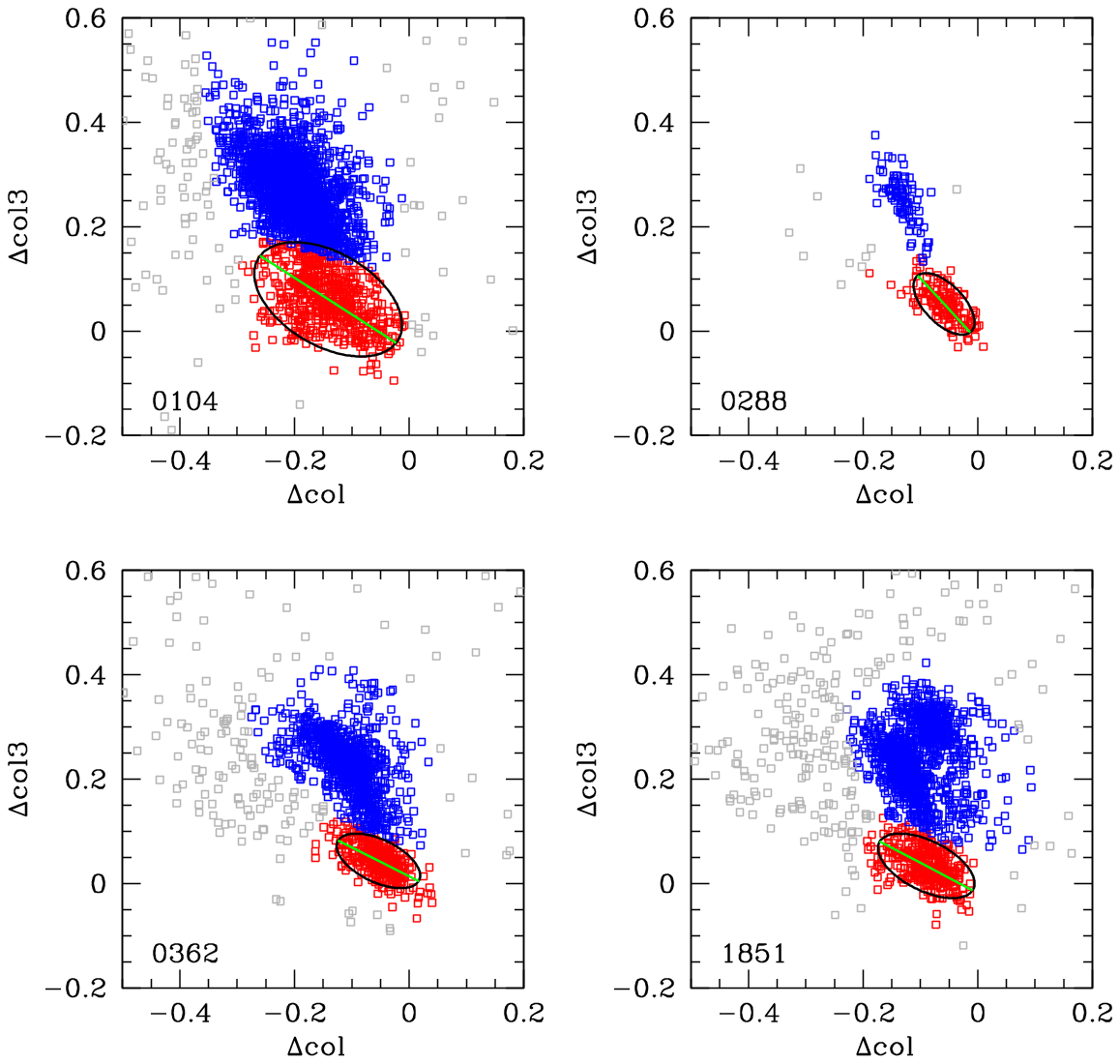}
\caption{Separation of FG and SG stars (in red and blue, respectively) in the
PCMs of the four clusters NGC~0104, NGC~0288, NGC~0362, and NGC~1851. The
ellipses indicate the $RG1$ region.}
\label{f:app1}
\end{figure*}

\newpage
\begin{figure*}
\centering
\includegraphics[scale=0.9]{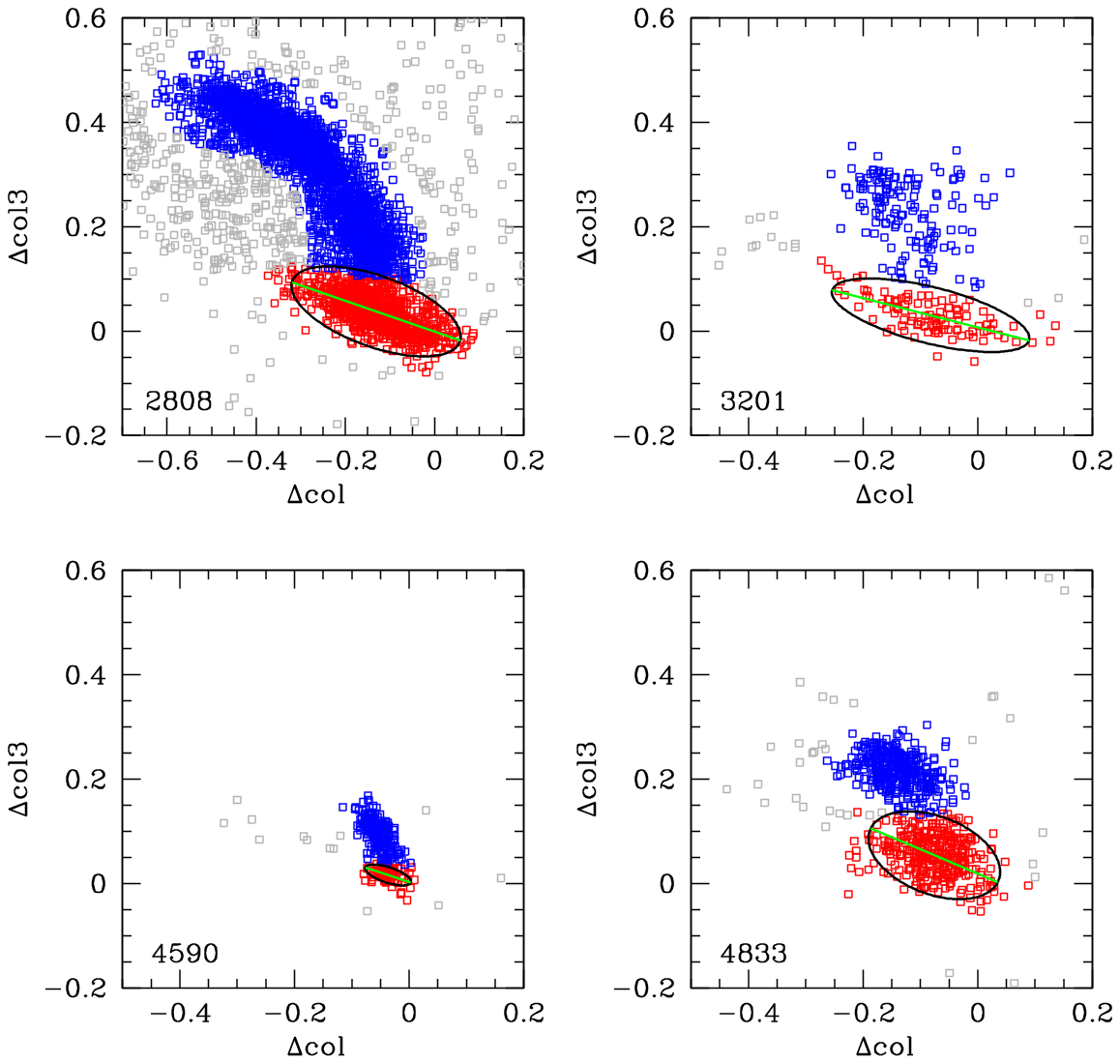}
\caption{The same, for NGC~2808 (notice the slightly different x-axis scale),
NCGC~3201, NGC~4590, and NGC~4833.}
\label{f:app2}
\end{figure*}

\newpage
\begin{figure*}
\centering
\includegraphics[scale=0.9]{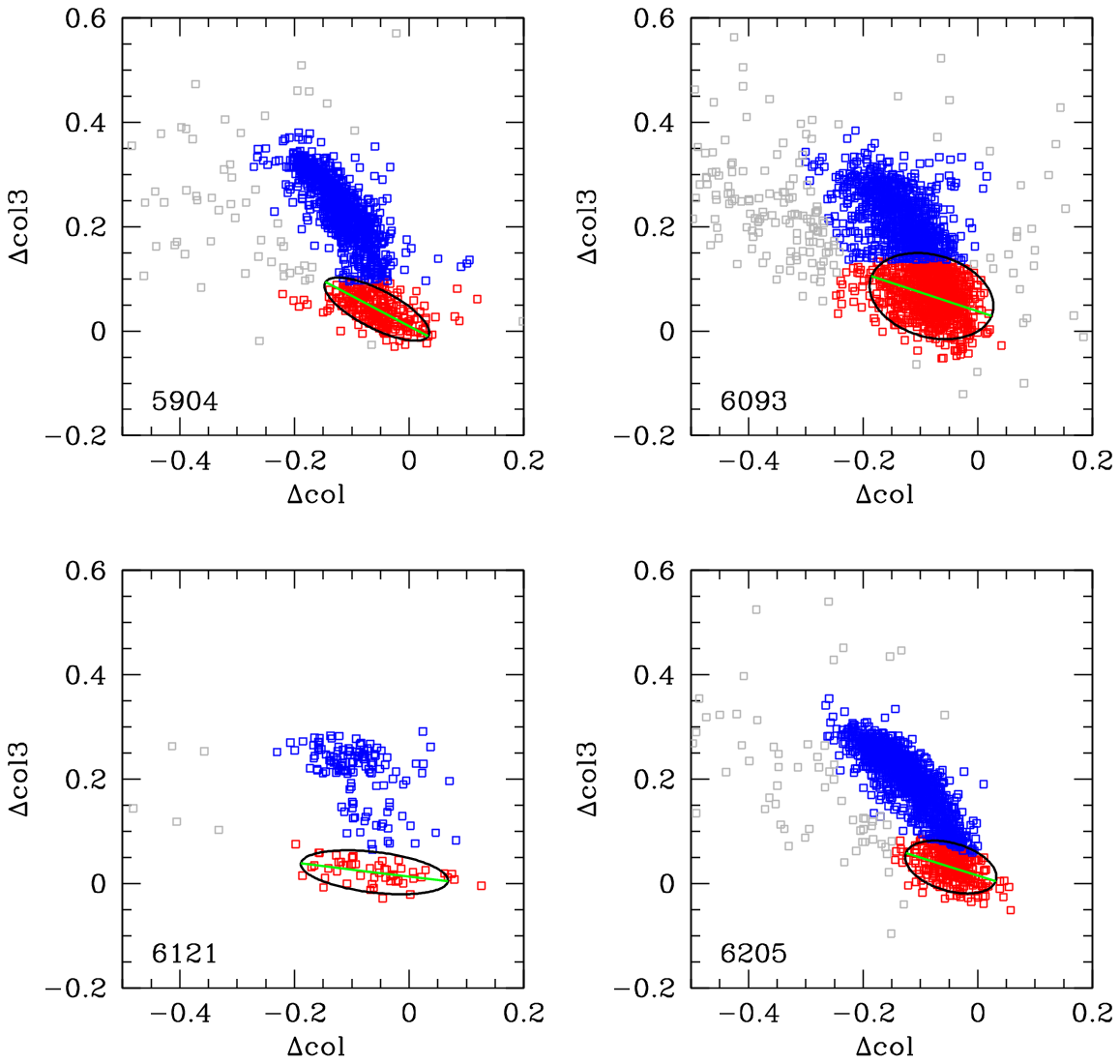}
\caption{The same, for NGC~5904, NCGC~6093, NGC~6121, and NGC~6205.}
\label{f:app3}
\end{figure*}

\newpage
\begin{figure*}
\centering
\includegraphics[scale=0.9]{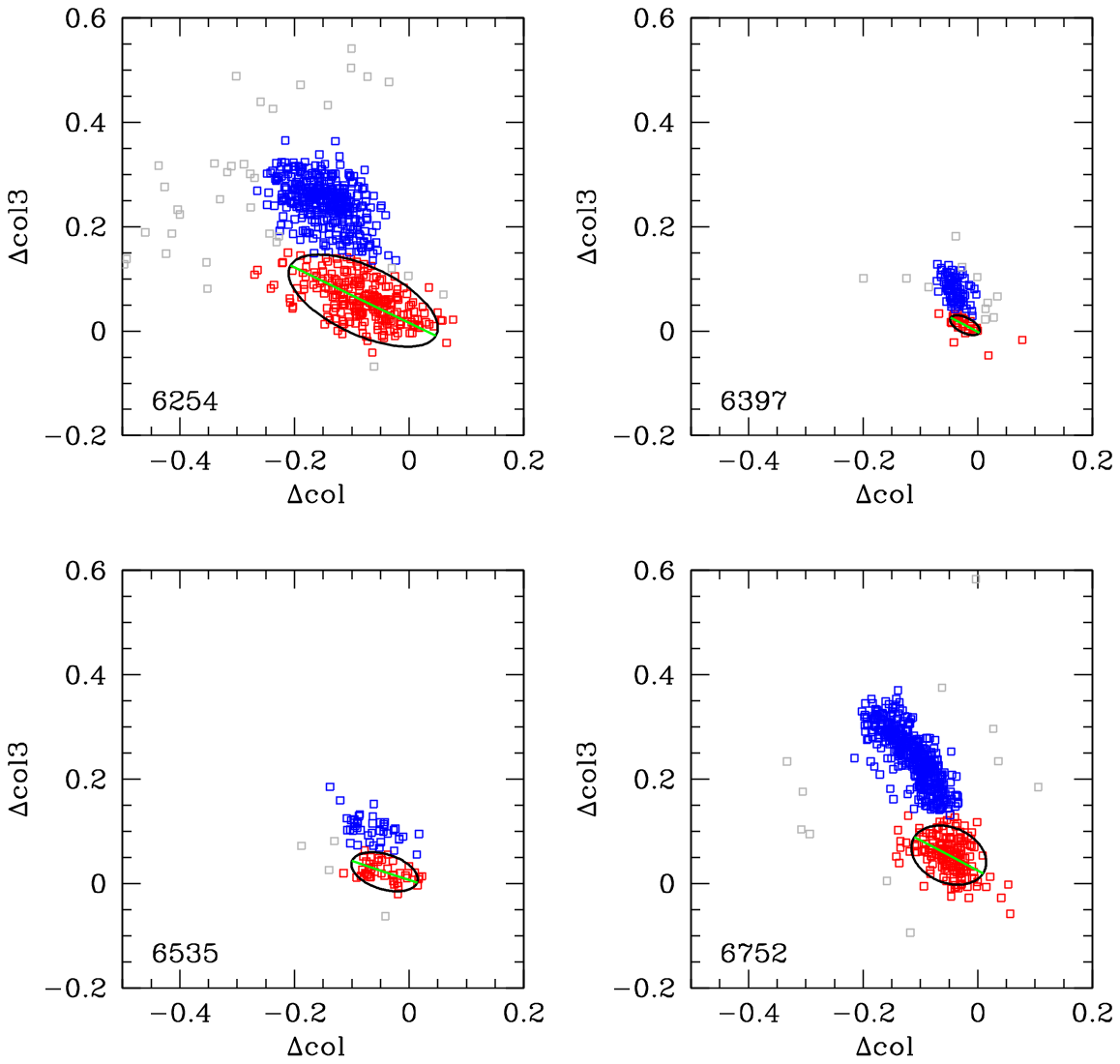}
\caption{The same, for NGC~6254, NCGC~6397, NGC~6535, and NGC~6752.}
\label{f:app4}
\end{figure*}

\newpage
\begin{figure*}
\centering
\includegraphics[scale=0.9]{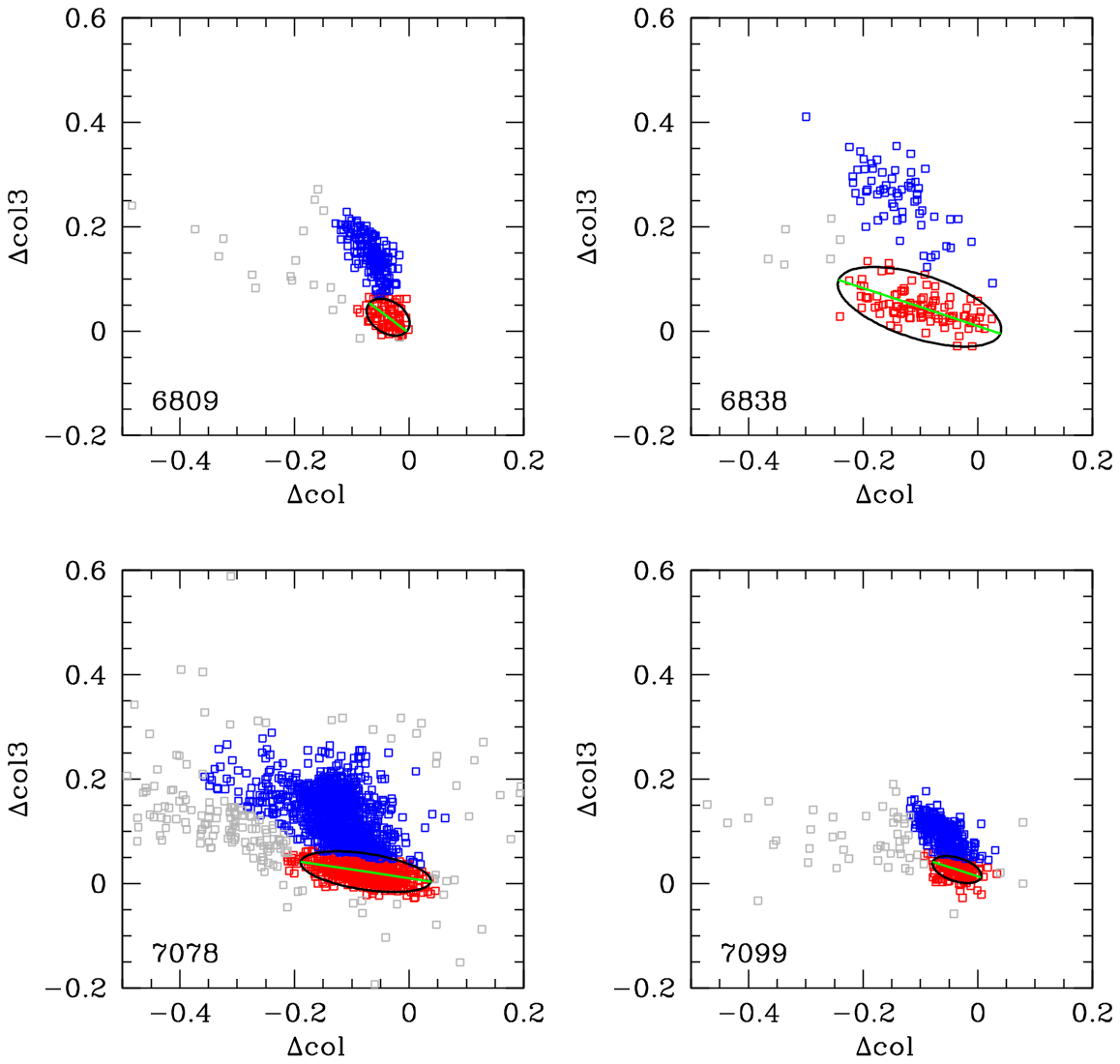}
\caption{The same, for NGC~6809, NCGC~6838, NGC~7078, and NGC~7099.}
\label{f:app5}
\end{figure*}

\end{appendix}

\begin{thebibliography}{}

\bibitem[]{} Bailin, J. 2019, ApJS, 245, 5
\bibitem[]{} Bailin, J., von Klar, R. 2022, ApJ, 925, 36
\bibitem[]{} Bastian, N., Lardo, C. 2018, ARA\&A, 56, 83
\bibitem[]{} Battaglia, G., Tolstoy, E., Helmi, A. et al. 2006, A\&A, 459, 423 
\bibitem[]{} Baumgardt, H., Hilker, M., Sollima, A., Bellini, A. 2019, MNRAS,
  482, 5138 
\bibitem[]{} Bekki, K., Freeman, K.C. 2003, MNRAS, 346, L11 
\bibitem[]{} Boberg, O.M., Friel, E.D., Vesperini, E. 2016, ApJ, 824, 5 
\bibitem[]{} Bragaglia, A., Carretta, E., D'Orazi, V. et al. 2017, A\&A, 607, A44 
\bibitem[]{} Calamida, A., Bono, G., Stetson, P.B. et al. 2007, ApJ, 670, 400 
\bibitem[]{} Carretta, E. 2015, ApJ, 810, 148 
\bibitem[]{} Carretta, E., Bragaglia, A. 2024, A\&A, 690, A158 
\bibitem[]{} Carretta, E., Gratton, R.G., Clementini, G., Fusi Pecci, F. 2000,
  ApJ, 533, 215
\bibitem[]{} Carretta, E., Bragaglia, A., Gratton R.G., et al. 2006, A\&A, 450, 523 
\bibitem[]{} Carretta, E., Bragaglia, A., Gratton, R.G., D'Orazi, V., Lucatello,
 S. 2009a, A\&A, 508, 695 
\bibitem[]{}  Carretta, E., Bragaglia, A., Gratton, R.G. et al. 2009b, 
  A\&A, 505, 117 
\bibitem[]{} Carretta, E., Bragaglia, A., Gratton, R.G., Lucatello, S. 2009c, 
 A\&A, 505, 139 
\bibitem[]{} Carretta, E., Bragaglia, A., Gratton, R.G. et al. 2010, ApJ, 714, L7 
\bibitem[]{} Carretta, E., Bragaglia, A., Gratton, R.G., D'Orazi, V., 
  Lucatello, S. 2011a, A\&A, 535, 121 
\bibitem[]{} Carretta, E., Lucatello, S., Gratton, R.G., Bragaglia, A., D'Orazi,
  V. 2011b, A\&A, 533, 69 
\bibitem[]{} Carretta, E., Bragaglia, A., Gratton, R.G. et al. 2014, A\&A, 564, A60 
\bibitem[]{} Carretta, E., Bragaglia, A., Gratton, R.G. et al. 2015, A\&A, 578, A116 
\bibitem[]{} Cohen, J.G., Melendez, J. 2005, AJ, 129, 303 
\bibitem[]{} Dopita, M.A., Smith, G.H. 1986, ApJ, 304, 283
\bibitem[]{} Dotter, A., Chaboyer, B., Jevremovic, D. et al. 2007, AJ, 134, 376
\bibitem[]{} Fanelli, C., Origlia, L., Rich, R.M. et al. 2024, A\&A, 690, A139
\bibitem[]{} Gaia Collaboration, Brown, A.~G.~A., Vallenari, A., et al. 2018a,
  A\&A, 616, A1.
\bibitem[]{} Gaia Collaboration, Helmi, A., van Leeuwen, F., et al. 2018b, A\&A,
  616, A12.
\bibitem[]{} Gaia Collaboration, Vallenari, A., Brown, A.G.A.,  et al. 2023,
  A\&A, 674, A1.
\bibitem[]{} Gratton, R.G., Bonifacio, P., Bragaglia, A., et al. 2001, 
 A\&A, 369, 87 
\bibitem[]{} Gratton, R.G., Sneden, C., Carretta, E. 2004, ARA\&A, 42, 385
\bibitem[]{} Gratton, R.G., Carretta, E., Bragaglia, A. 2012, A\&ARv, 20, 50 
\bibitem[]{} Gratton, R.G., Bragaglia, A., Carretta, E., D'Orazi, V., 
  Lucatello, S., Sollima, A. 2019, A\&ARv, 27, 8
\bibitem[]{} Harris, W.E. 2010, ArXiv, 1012.3224
\bibitem[]{} Johnson, C.I., Rich, M.R., Pilachowski, C.A. et al. 2015, AJ, 150, 63 
\bibitem[]{} Johnson, C.I., Calamida, A., Kader, J.A. et al. 2023, AJ, 163, 3
\bibitem[]{} Kirby, E.N., Guhathakurta, P., Simon, J.D. et al. 2010, ApJS, 191
  352 
\bibitem[]{} Kirby, E.N., Lanfranchi, G.A., Simon, J.D., Cohen, J.G., 
  Guhathakurta, P. 2011, ApJ, 727, 78 
\bibitem[]{} Kruijssen, J.M.D. , Pfeffer, J.L., Reina-Campos, M., Crain, R.A.
 Bastian, N. 2019, MNRAS, 486, 3180 
\bibitem[]{} Langer, G.E., Hoffman, R., Sneden, C. 1993, PASP, 105, 301 
\bibitem[]{} Lardo, C., Bellazzini, M., Pancino, E. et al. 2011, A\&A, 525, A114 
\bibitem[]{} Lardo, C., Salaris, M., Cassisi, et al. 2023, A\&A, 669, A19
\bibitem[]{} Lee, J.-W., Lee, J., Kang, Y.-W. et al. 2009, ApJ, 695, L78 
\bibitem[]{} Legnardi, M.V., Milone, A.P., Armillotta, L. et al. 2022, MNRAS,
  513, 735
\bibitem[]{} Legnardi, M.V., Milone, A.P., Cordoni, G. et al. 2024, A\&A,
  687, A160
\bibitem[]{} Leitinger, E., Baumgardt, H., Cabrera-Ziri, I., Hilker, M.,
  Pancino, E. 2023, MNRAS, 520, 1456  
\bibitem[]{} Marino, A.F., Villanova, S., Piotto, G. et al. 2008, A\&A, 490, 625
\bibitem[]{} Marino, A.F., Milone, A.P., Sills, A. et al. 2019, ApJ, 887, 91
\bibitem[]{} Milone, A.P., Piotto, G., Renzini, A. et al. 2017, MNRAS, 464,
  3636 
\bibitem[]{} Nardiello, D., Libralato, M., Piotto, G. et al. 2018, MNRAS, 481,
  3382 
\bibitem[]{} Salaris, M., Cassisi, S., Weis, A. 2002, PASP, 114, 375
\bibitem[]{} Sbordone, L., Salaris, M., Weiss, A., Cassisi, S. 2011, A\&A, 534,
  A9 
\bibitem[]{} Siegel, M.H., Dotter, A., Majewski, S.R. et al. 2007, ApJ, 667, L57 
\bibitem[]{} Sneden, C., Kraft, R.P., Guhathakurta, P., Peterson, R.C.,
  Fulbright, J.P. 2004, AJ, 127, 2162 
\bibitem[]{} Stetson, P.B., Pancino, E., Zocchi, A., Sanna, N., Monelli, M.
  2019, MNRAS, 485, 3042
\bibitem[]{} Taylor, M.B. 2005, Astronomical Data Analysis Software and Systems
  XIV, 347, 29
\bibitem[]{} VandenBerg, D.A., Brogaard, K., Leaman, R., Casagrande, L. 2013,
  ApJ, 775, 134 
\bibitem[]{} Willman, B., Strader, J. 2012, AJ, 144, 76


\end{thebibliography}
\end{document}